%% file: onshellmethods.tex
\documentclass[11pt,a4paper]{article}

\DeclareMathAlphabet{\mathpzc}{OT1}{pzc}{m}{it}
\DeclareFontFamily{OT1}{pzc}{}
\DeclareFontShape{OT1}{pzc}{m}{it}{<-> s * [0.900] pzcmi7t}{}
\DeclareMathAlphabet{\mathpzc}{OT1}{pzc}{m}{it}

\usepackage[dvips]{graphicx}
\usepackage[dvips,dvipsnames]{color}
\usepackage{jheppub}
\usepackage[dvips,dvipsnames]{xcolor}
\usepackage{hyperref}
\usepackage[dvips]{graphicx}
\usepackage{amsfonts}
\usepackage{latexsym,amsmath,amssymb,amsthm,graphics,stmaryrd}
\usepackage{array}
\usepackage{multirow}
\usepackage{pstricks}
\usepackage{pst-node}
\usepackage{pst-coil}
\usepackage{bm}
\usepackage{dsfont}
\usepackage[bf]{caption}
\usepackage{subcaption}
\usepackage{fancyhdr}

\usepackage{tabularx}
\usepackage{longtable}
\usepackage{multirow}

\usepackage{xkeyval}

\usepackage{afterpage} 

\allowdisplaybreaks

\makeatletter
\DeclareRobustCommand*{\bfseries}{%
\not@math@alphabet\bfseries\mathbf
\fontseries\bfdefault\selectfont
\boldmath
}
\makeatother

\def\LT@makecaption#1#2#3{%
\LT@mcol\LT@cols c{\hbox to\z@{\hss\parbox[t]\LTcapwidth{%

     \sbox\@tempboxa{#1{#2: }#3}%
    \ifdim\wd\@tempboxa>\hsize
       #1{#2: }#3%
    \else
       \hbox to\hsize{\hfil\box\@tempboxa\hfil}%
     \fi
     \endgraf\vskip\baselineskip}%
   \hss}}}

\usepackage[textsize=tiny,textwidth=2.25cm]{todonotes}

\usepackage{comment}

\numberwithin{equation}{section}

\usepackage{feynmp}

\newlength{\eqoff}

\newlength{\unit}
\psset{xunit=\unit,yunit=\unit,runit=\unit}
\newlength{\linew}
\psset{linewidth=\linew}
\input{input4mod}

\definecolor{labelcolor}{rgb}{1,0,0}

\input{picturemacro}

\input{Macro-Dhritiman}

\newcommand{\col}{~,}
\newcommand{\pnt}{~.}
\newcommand{\AdS}{\text{AdS}}

\newcommand{\YM}{{\scriptscriptstyle\text{YM}}}
\newcommand{\E}{{\scriptscriptstyle\text{E}}}

\newcommand{\BPS}{{\scriptscriptstyle\text{BPS}}}

\newcommand{\ren}{{\text{R}}}

\newcommand{\bare}{{\text{B}}}

\newcommand{\teps}{\varepsilon}

\newcommand{\peps}{\epsilon}

\newcommand{\e}{\operatorname{e}}

\newcommand{\ncsusyF}{{\hat{\cal F}}}

\newcommand{\susyF}{{\cal F}}

\newcommand{\bosoF}{{F}}

\newcommand{\ratioF}{{f}}

\newcommand{\de}{\operatorname{d}\!}

\newcommand{\parderiv}[2][]{\frac{\partial #1}{\partial #2}}

\newcommand{\unitmatrix}{\mathds{1}}

\newcommand{\tracematrix}{\mathds{T}}

\newlength{\neglength}
\newlength{\diameter}

\DeclareMathOperator{\tr}{tr}

\DeclareMathOperator{\T}{T}

\DeclareMathOperator{\im}{Im}

\begin{document}
\begin{fmffile}{graphs1}
\initialisepictures

\fmfcmd{%
style_def phantom_cross expr p =
    save amid,ang;
    amid=.5*length p;
    ang= angle direction amid of p;
    draw ((polycross 4) scaled 8 rotated ang) shifted point amid of p;
enddef;
}


\title{Cutting through form factors and cross sections of non-protected operators
in $\mathcal{N}=4$ SYM}
\author[1,2]{Dhritiman Nandan,}
\author[1,2]{Christoph Sieg,}
\author[1,2]{Matthias Wilhelm,}
\author[1]{and Gang Yang}

\affiliation[1]{
  Institut f\"ur Physik}
\affiliation[2]{  
  Institut f\"ur Mathematik \\ 
  Humboldt-Universit\"at zu Berlin\\
  IRIS Geb\"aude, Zum Gro\ss{}en Windkanal 6, 12489 Berlin
	}

\emailAdd{dhritiman@physik.hu-berlin.de}
\emailAdd{csieg@physik.hu-berlin.de}
\emailAdd{mwilhelm@physik.hu-berlin.de}
\emailAdd{gang.yang@physik.hu-berlin.de}


\abstract{
We study the form factors of the Konishi operator, the prime example of non-protected operators in $\mathcal{N}=4$ SYM theory, via the on-shell unitarity method. 
Since the Konishi operator is not protected by supersymmetry, its form factors
share many features with amplitudes in QCD, such as the occurrence of rational terms
and of UV divergences that require renormalization.
A subtle point is that this operator depends on the spacetime dimension. 
This requires a modification when calculating its form factors via the on-shell
unitarity method.
We derive a rigorous prescription that implements this modification to all loop orders
and obtain the two-point form factor up to two-loop order and the three-point form factor to one-loop order.
From these form factors, we construct an IR-finite cross-section-type quantity, 
namely the inclusive decay rate of the (off-shell) Konishi operator to any final (on-shell) state. Via the optical theorem, it is connected to the imaginary part of the two-point correlation function.
We extract the Konishi anomalous dimension up to two-loop order from it.

}

\flushbottom

\maketitle

\input{introduction}

\input{csnutshell}

\input{FFactor}

\input{subtleties}

\input{crosssection}

\input{conclusion}

\input{acknowledgements}

\appendix

\input{appendices}

\input{feyndiag}

\footnotesize
\bibliographystyle{utcaps}
\bibliography{references}

\end{fmffile}
\end{document}

%% file: input4mod.tex
\usepackage{rotating}
\newlength{\arlength}
\newlength{\arheight}

\newcommand{\svertex}[2]{%
\fmfiequ{#1}{point length(#2)/2 of (#2)}
}

\newcommand{\chionetwo}[1][black]{%
\fmftop{v1}
\fmfbottom{v4}
\fmfforce{(0.125w,h)}{v1}
\fmfforce{(0.125w,0)}{v4}
\fmffixed{(0.25w,0)}{v1,v2}
\fmffixed{(0.25w,0)}{v2,v3}
\fmffixed{(0.25w,0)}{v4,v5}
\fmffixed{(0.25w,0)}{v5,v6}
\fmffixed{(0,whatever)}{vc1,vc3}
\fmffixed{(0,whatever)}{vc2,vc4}
\fmf{plain,tension=0.5,right=0.25}{v1,vc1}
\fmf{plain,tension=0.5,left=0.25}{v2,vc1}
\fmf{phantom,tension=0.5,right=0.25}{v2,vc2}
\fmf{plain,tension=0.5,left=0.25}{v3,vc2}
\fmf{plain,tension=0.5,left=0.25}{v4,vc3}
\fmf{phantom,tension=0.5,right=0.25}{v5,vc3}
\fmf{plain,tension=0.5,left=0.25}{v5,vc4}
\fmf{plain,tension=0.5,right=0.25}{v6,vc4}
\fmf{plain,tension=1.25,left=0}{vc1,vc3}
\fmf{plain,tension=1.25,left=0}{vc2,vc4}
\fmffreeze
\fmf{plain,tension=1,left=0}{vc2,vc3}
\fmf{plain,tension=0.5,right=0,width=1mm}{v4,v6}
\fmffreeze
\fmfposition
\fmfipath{p[]}
\fmfipair{vd[],vm[],vu[]}
\fmfiset{p1}{vpath(__v1,__vc1)}
\fmfiset{p2}{vpath(__v2,__vc1)}
\fmfiset{p6}{vpath(__v3,__vc2)}
\fmfiset{p4}{vpath(__v4,__vc3)}
\fmfiset{p8}{vpath(__v5,__vc4)}
\fmfiset{p9}{vpath(__v6,__vc4)}
\fmfiset{p3}{vpath(__vc1,__vc3)}
\fmfiset{p7}{vpath(__vc2,__vc4)}
\fmfiset{p5}{vpath(__vc2,__vc3)}
\svertex{vm1}{p1}
\svertex{vm2}{p2}
\svertex{vm3}{p3}
\svertex{vm4}{p4}
\svertex{vm5}{p5}
\svertex{vm6}{p6}
\svertex{vm7}{p7}
\svertex{vm8}{p8}
\svertex{vm9}{p9}
}

\newcommand{\chionetwoone}[1][black]{%
\fmftop{v1}
\fmfbottom{v4}
\fmfforce{(0.125w,h)}{v1}
\fmfforce{(0.125w,0)}{v4}
\fmffixed{(0.25w,0)}{v1,v2}
\fmffixed{(0.25w,0)}{v2,v3}
\fmffixed{(0.25w,0)}{v4,v5}
\fmffixed{(0.25w,0)}{v5,v6}
\fmffixed{(0,whatever)}{vc1,vc3}
\fmffixed{(0,whatever)}{vb2,vb4}
\fmffixed{(0,whatever)}{vc1,vb1}
\fmffixed{(0,whatever)}{vc1,vb3}
\fmffixed{(whatever,0)}{vb1,vb2}
\fmffixed{(whatever,0)}{vb3,vb4}
\fmf{plain,tension=0.5,right=0.25}{v1,vc1}
\fmf{plain,tension=0.5,left=0.25}{v2,vc1}
\fmf{phantom,tension=0.5,right=0.25}{v2,vb2}
\fmf{plain,tension=0.5,left=0.25}{v3,vb2}
\fmf{plain,tension=0.5,left=0.25}{v4,vc3}
\fmf{plain,tension=0.5,right=0.25}{v5,vc3}
\fmf{phantom,tension=0.5,left=0.25}{v5,vb4}
\fmf{plain,tension=0.5,right=0.25}{v6,vb4}
\fmf{plain,tension=2,left=0}{vc1,vb1}
\fmf{plain,tension=2,left=0}{vb1,vb3}
\fmf{plain,tension=2,left=0}{vb3,vc3}
\fmf{plain,tension=2,left=0}{vb2,vb4}
\fmffreeze
\fmf{plain,tension=2,left=0}{vb1,vb2}
\fmf{plain,tension=2,left=0}{vb3,vb4}
\fmf{plain,tension=0.5,right=0,width=1mm}{v4,v6}
\fmffreeze
\fmfposition
\fmfipath{p[]}
\fmfipair{vd[],vm[],vu[]}
\fmfiset{p1}{vpath(__v1,__vc1)}
\fmfiset{p2}{vpath(__v2,__vc1)}
\fmfiset{p3}{vpath(__vc1,__vb1)}
\fmfiset{p4}{vpath(__vb1,__vb3)}
\fmfiset{p5}{vpath(__vb1,__vb2)}
\fmfiset{p6}{vpath(__v3,__vb2)}
\fmfiset{p7}{vpath(__vb2,__vb4)}
\fmfiset{p8}{vpath(__vb3,__vb4)}
\fmfiset{p9}{vpath(__v6,__vb4)}
\fmfiset{p10}{vpath(__vb3,__vc3)}
\fmfiset{p11}{vpath(__v4,__vc3)}
\fmfiset{p12}{vpath(__v5,__vc3)}
\svertex{vm1}{p1}
\svertex{vm2}{p2}
\svertex{vm3}{p3}
\svertex{vm4}{p4}
\svertex{vm5}{p5}
\svertex{vm6}{p6}
\svertex{vm7}{p7}
\svertex{vm8}{p8}
\svertex{vm9}{p9}
\svertex{vm10}{p10}
\svertex{vm11}{p11}
\svertex{vm12}{p12}
}

\newcommand{\chionetwothree}[1][black]{%
\fmftop{v1}
\fmfbottom{v5}
\fmfforce{(0.125w,h)}{v1}
\fmfforce{(0.125w,0)}{v5}
\fmffixed{(0.25w,0)}{v1,v2}
\fmffixed{(0.25w,0)}{v2,v3}
\fmffixed{(0.25w,0)}{v3,v4}
\fmffixed{(0.25w,0)}{v5,v6}
\fmffixed{(0.25w,0)}{v6,v7}
\fmffixed{(0.25w,0)}{v7,v8}
\fmffixed{(0,whatever)}{vc1,vc4}
\fmffixed{(0,whatever)}{vc2,vc5}
\fmffixed{(0,whatever)}{vc3,vc6}

\fmf{plain,tension=0.5,right=0.25}{v1,vc1}
\fmf{plain,tension=0.5,left=0.25}{v2,vc1}
\fmf{phantom,tension=0.5,right=0.25}{v2,vc2}
\fmf{plain,tension=0.5,left=0.25}{v3,vc2}
\fmf{phantom,tension=0.5,right=0.25}{v3,vc3}
\fmf{plain,tension=0.5,left=0.25}{v4,vc3}
\fmf{plain,tension=0.5,left=0.25}{v5,vc4}
\fmf{phantom,tension=0.5,right=0.25}{v6,vc4}
\fmf{plain,tension=0.5,left=0.25}{v6,vc5}
\fmf{phantom,tension=0.5,right=0.25}{v7,vc5}
\fmf{plain,tension=0.5,left=0.25}{v7,vc6}
\fmf{plain,tension=0.5,right=0.25}{v8,vc6}
\fmf{plain,tension=1.25,left=0}{vc1,vc4}
\fmf{plain,tension=1.25,left=0}{vc2,vc5}
\fmf{plain,tension=1.25,left=0}{vc3,vc6}
\fmffreeze
\fmf{plain,tension=1,left=0}{vc4,vc2}
\fmf{plain,tension=1,left=0}{vc5,vc3}
\fmf{plain,tension=0.5,right=0,width=1mm}{v5,v8}
\fmffreeze
\fmfposition
}

\newcommand{\chitwoonethree}[1][black]{%
\fmftop{v1}
\fmfbottom{v5}
\fmfforce{(0.125w,h)}{v1}
\fmfforce{(0.125w,0)}{v5}
\fmffixed{(0.25w,0)}{v1,v2}
\fmffixed{(0.25w,0)}{v2,v3}
\fmffixed{(0.25w,0)}{v3,v4}
\fmffixed{(0.25w,0)}{v5,v6}
\fmffixed{(0.25w,0)}{v6,v7}
\fmffixed{(0.25w,0)}{v7,v8}
\fmffixed{(whatever,0.5h)}{v5,vc1}
\fmffixed{(0,whatever)}{vc1,vc4}
\fmffixed{(0,whatever)}{vc2,vc5}
\fmffixed{(0,whatever)}{vc3,vc6}
\fmffixed{(whatever,0)}{vc1,vc3}
\fmffixed{(whatever,0)}{vc3,vc5}

\fmf{plain,tension=0.5,right=0.125}{v1,vc1}
\fmf{phantom,tension=0.5,left=0.25}{v2,vc1}
\fmf{plain,tension=0.5,right=0.25}{v2,vc2}
\fmf{plain,tension=0.5,left=0.25}{v3,vc2}
\fmf{phantom,tension=0.5,right=0.25}{v3,vc3}
\fmf{plain,tension=0.5,left=0.125}{v4,vc3}
\fmf{plain,tension=0.5,left=0.25}{v5,vc4}
\fmf{plain,tension=0.5,right=0.25}{v6,vc4}
\fmf{phantom,tension=0.5,left=0.25}{v6,vc5}
\fmf{phantom,tension=0.5,right=0.25}{v7,vc5}
\fmf{plain,tension=0.5,left=0.25}{v7,vc6}
\fmf{plain,tension=0.5,right=0.25}{v8,vc6}
\fmf{plain,tension=1.25,left=0}{vc1,vc4}
\fmf{plain,tension=1.25,left=0}{vc2,vc5}
\fmf{plain,tension=1.25,left=0}{vc3,vc6}
\fmffreeze
\fmf{plain,tension=1,left=0}{vc1,vc5}
\fmf{plain,tension=1,left=0}{vc5,vc3}
\fmf{plain,tension=0.5,right=0,width=1mm}{v5,v8}
\fmffreeze
\fmfposition
}

\newcommand{\chionethreetwo}[1][black]{%
\fmftop{v1}
\fmfbottom{v5}
\fmfforce{(0.125w,h)}{v1}
\fmfforce{(0.125w,0)}{v5}
\fmffixed{(0.25w,0)}{v1,v2}
\fmffixed{(0.25w,0)}{v2,v3}
\fmffixed{(0.25w,0)}{v3,v4}
\fmffixed{(0.25w,0)}{v5,v6}
\fmffixed{(0.25w,0)}{v6,v7}
\fmffixed{(0.25w,0)}{v7,v8}
\fmffixed{(whatever,0.5h)}{v5,vc2}
\fmffixed{(0,whatever)}{vc1,vc4}
\fmffixed{(0,whatever)}{vc2,vc5}
\fmffixed{(0,whatever)}{vc3,vc6}
\fmffixed{(whatever,0)}{vc2,vc4}
\fmffixed{(whatever,0)}{vc4,vc6}
\fmf{plain,tension=0.5,right=0.25}{v1,vc1}
\fmf{plain,tension=0.5,left=0.25}{v2,vc1}
\fmf{phantom,tension=0.5,right=0.25}{v2,vc2}
\fmf{phantom,tension=0.5,left=0.25}{v3,vc2}
\fmf{plain,tension=0.5,right=0.25}{v3,vc3}
\fmf{plain,tension=0.5,left=0.25}{v4,vc3}
\fmf{plain,tension=0.5,left=0.125}{v5,vc4}
\fmf{phantom,tension=0.5,right=0.25}{v6,vc4}
\fmf{plain,tension=0.5,left=0.25}{v6,vc5}
\fmf{plain,tension=0.5,right=0.25}{v7,vc5}
\fmf{phantom,tension=0.5,left=0.25}{v7,vc6}
\fmf{plain,tension=0.5,right=0.125}{v8,vc6}
\fmf{plain,tension=1.25,left=0}{vc1,vc4}
\fmf{plain,tension=1.25,left=0}{vc2,vc5}
\fmf{plain,tension=1.25,left=0}{vc3,vc6}
\fmffreeze
\fmf{plain,tension=1,left=0}{vc2,vc4}
\fmf{plain,tension=1,left=0}{vc2,vc6}
\fmf{plain,tension=0.5,right=0,width=1mm}{v5,v8}
\fmffreeze
\fmfposition
}

\newcommand{\chionetwothreetwo}[1][black]{%
\fmftop{v1}
\fmfbottom{v5}
\fmfforce{(0.125w,h)}{v1}
\fmfforce{(0.125w,0)}{v5}
\fmffixed{(0.25w,0)}{v1,v2}
\fmffixed{(0.25w,0)}{v2,v3}
\fmffixed{(0.25w,0)}{v3,v4}
\fmffixed{(0.25w,0)}{v5,v6}
\fmffixed{(0.25w,0)}{v6,v7}
\fmffixed{(0.25w,0)}{v7,v8}
\fmffixed{(0,whatever)}{va1,va2}
\fmffixed{(whatever,0)}{va1,vc1}
\fmffixed{(whatever,0)}{va2,vb3}
\fmffixed{(0,whatever)}{vc1,vc3}
\fmffixed{(0,whatever)}{vb2,vb4}
\fmffixed{(0,whatever)}{vc1,vb1}
\fmffixed{(0,whatever)}{vc1,vb3}
\fmffixed{(whatever,0)}{vb1,vb2}
\fmffixed{(whatever,0)}{vb3,vb4}
\fmf{phantom,tension=0.5,right=0.25}{v2,vc1}
\fmf{plain,tension=0.5,left=0.25}{v3,vc1}
\fmf{phantom,tension=0.5,right=0.25}{v3,vb2}
\fmf{plain,tension=0.5,left=0.25}{v4,vb2}
\fmf{plain,tension=0.5,left=0.25}{v6,vc3}
\fmf{plain,tension=0.5,right=0.25}{v7,vc3}
\fmf{phantom,tension=0.5,left=0.25}{v7,vb4}
\fmf{plain,tension=0.5,right=0.25}{v8,vb4}
\fmf{plain,tension=2,left=0}{vc1,vb1}
\fmf{plain,tension=2,left=0}{vb1,vb3}
\fmf{plain,tension=2,left=0}{vb3,vc3}
\fmf{plain,tension=2,left=0}{vb2,vb4}
\fmffreeze
\fmf{plain,tension=0.5,right=0.25}{v1,va1}
\fmf{plain,tension=0.5,left=0.25}{v2,va1}
\fmf{plain,tension=1}{va1,va2}
\fmf{plain,tension=0}{va2,vc1}
\fmf{plain,tension=0.5,left=0.25}{v5,va2}
\fmf{phantom,tension=0.5,right=0.25}{v6,va2}
\fmf{plain,tension=1,left=0}{vb1,vb2}
\fmf{plain,tension=1,left=0}{vb3,vb4}
\fmf{plain,tension=0.5,right=0,width=1mm}{v5,v8}
}

\newcommand{\chitwothreetwoone}[1][black]{%
\fmftop{v1}
\fmfbottom{v5}
\fmfforce{(0.125w,h)}{v1}
\fmfforce{(0.125w,0)}{v5}
\fmffixed{(0.25w,0)}{v1,v2}
\fmffixed{(0.25w,0)}{v2,v3}
\fmffixed{(0.25w,0)}{v3,v4}
\fmffixed{(0.25w,0)}{v5,v6}
\fmffixed{(0.25w,0)}{v6,v7}
\fmffixed{(0.25w,0)}{v7,v8}
\fmffixed{(0,whatever)}{va1,va2}
\fmffixed{(whatever,0)}{va2,vc3}
\fmffixed{(whatever,0)}{va1,vb1}
\fmffixed{(0,whatever)}{vc1,vc3}
\fmffixed{(0,whatever)}{vb2,vb4}
\fmffixed{(0,whatever)}{vc1,vb1}
\fmffixed{(0,whatever)}{vc1,vb3}
\fmffixed{(whatever,0)}{vb1,vb2}
\fmffixed{(whatever,0)}{vb3,vb4}
\fmf{plain,tension=0.5,right=0.25}{v2,vc1}
\fmf{plain,tension=0.5,left=0.25}{v3,vc1}
\fmf{phantom,tension=0.5,right=0.25}{v3,vb2}
\fmf{plain,tension=0.5,left=0.25}{v4,vb2}
\fmf{phantom,tension=0.5,left=0.25}{v6,vc3}
\fmf{plain,tension=0.5,right=0.25}{v7,vc3}
\fmf{phantom,tension=0.5,left=0.25}{v7,vb4}
\fmf{plain,tension=0.5,right=0.25}{v8,vb4}
\fmf{plain,tension=2,left=0}{vc1,vb1}
\fmf{plain,tension=2,left=0}{vb1,vb3}
\fmf{plain,tension=2,left=0}{vb3,vc3}
\fmf{plain,tension=2,left=0}{vb2,vb4}
\fmffreeze
\fmf{plain,tension=0.5,right=0.25}{v1,va1}
\fmf{phantom,tension=0.5,left=0.25}{v2,va1}
\fmf{plain,tension=1}{va1,va2}
\fmf{plain,tension=0}{va1,vc3}
\fmf{plain,tension=0.5,left=0.25}{v5,va2}
\fmf{plain,tension=0.5,right=0.25}{v6,va2}
\fmf{plain,tension=1,left=0}{vb1,vb2}
\fmf{plain,tension=1,left=0}{vb3,vb4}
\fmf{plain,tension=0.5,right=0,width=1mm}{v5,v8}
}

\newcommand{\chitwoonetwothree}[1][black]{%
\fmftop{v1}
\fmfbottom{v5}
\fmfforce{(0.125w,h)}{v1}
\fmfforce{(0.125w,0)}{v5}
\fmffixed{(0.25w,0)}{v1,v2}
\fmffixed{(0.25w,0)}{v2,v3}
\fmffixed{(0.25w,0)}{v3,v4}
\fmffixed{(0.25w,0)}{v5,v6}
\fmffixed{(0.25w,0)}{v6,v7}
\fmffixed{(0.25w,0)}{v7,v8}
\fmffixed{(0,whatever)}{va1,va2}
\fmffixed{(whatever,0)}{va2,vc3}
\fmffixed{(whatever,0)}{va1,vb1}
\fmffixed{(0,whatever)}{vc1,vc3}
\fmffixed{(0,whatever)}{vb2,vb4}
\fmffixed{(0,whatever)}{vc1,vb1}
\fmffixed{(0,whatever)}{vc1,vb3}
\fmffixed{(whatever,0)}{vb1,vb2}
\fmffixed{(whatever,0)}{vb3,vb4}
\fmf{plain,tension=0.5,right=0.25}{v2,vc1}
\fmf{plain,tension=0.5,left=0.25}{v3,vc1}
\fmf{plain,tension=0.5,right=0.25}{v1,vb2}
\fmf{phantom,tension=0.5,left=0.25}{v2,vb2}
\fmf{plain,tension=0.5,left=0.25}{v6,vc3}
\fmf{phantom,tension=0.5,right=0.25}{v7,vc3}
\fmf{plain,tension=0.5,left=0.25}{v5,vb4}
\fmf{phantom,tension=0.5,right=0.25}{v6,vb4}
\fmf{plain,tension=2,left=0}{vc1,vb1}
\fmf{plain,tension=2,left=0}{vb1,vb3}
\fmf{plain,tension=2,left=0}{vb3,vc3}
\fmf{plain,tension=2,left=0}{vb2,vb4}
\fmffreeze
\fmf{phantom,tension=0.5,right=0.25}{v3,va1}
\fmf{plain,tension=0.5,left=0.25}{v4,va1}
\fmf{plain,tension=1}{va1,va2}
\fmf{plain,tension=0}{va1,vc3}
\fmf{plain,tension=0.5,left=0.25}{v7,va2}
\fmf{plain,tension=0.5,right=0.25}{v8,va2}
\fmf{plain,tension=1,left=0}{vb1,vb2}
\fmf{plain,tension=1,left=0}{vb3,vb4}
\fmf{plain,tension=0.5,right=0,width=1mm}{v5,v8}
}

\newcommand{\chithreeonetwoone}[1][black]{%
\fmftop{v1}
\fmfbottom{v5}
\fmfforce{(0.125w,h)}{v1}
\fmfforce{(0.125w,0)}{v5}
\fmffixed{(0.25w,0)}{v1,v2}
\fmffixed{(0.25w,0)}{v2,v3}
\fmffixed{(0.25w,0)}{v3,v4}
\fmffixed{(0.25w,0)}{v5,v6}
\fmffixed{(0.25w,0)}{v6,v7}
\fmffixed{(0.25w,0)}{v7,v8}
\fmffixed{(0,whatever)}{va1,va2}
\fmffixed{(whatever,0)}{va1,vc1}
\fmffixed{(whatever,0)}{va2,vb1}
\fmffixed{(0,whatever)}{vc1,vc3}
\fmffixed{(0,whatever)}{vb2,vb4}
\fmffixed{(0,whatever)}{vc1,vb1}
\fmffixed{(0,whatever)}{vc1,vb3}
\fmffixed{(whatever,0)}{vb1,vb2}
\fmffixed{(whatever,0)}{vb3,vb4}
\fmf{plain,tension=0.5,right=0.25}{v1,vc1}
\fmf{plain,tension=0.5,left=0.25}{v2,vc1}
\fmf{phantom,tension=0.5,right=0.25}{v2,vb2}
\fmf{phantom,tension=0.5,left=0.25}{v3,vb2}
\fmf{plain,tension=0.5,left=0.25}{v5,vc3}
\fmf{plain,tension=0.5,right=0.25}{v6,vc3}
\fmf{phantom,tension=0.5,left=0.125}{v6,vb4}
\fmf{plain,tension=0.5,right=0.125}{v7,vb4}
\fmf{plain,tension=2,left=0}{vc1,vb1}
\fmf{plain,tension=2,left=0}{vb1,vb3}
\fmf{plain,tension=2,left=0}{vb3,vc3}
\fmf{plain,tension=2,left=0}{vb2,vb4}
\fmffreeze
\fmf{plain,tension=0.5,right=0.25}{v3,va1}
\fmf{plain,tension=0.5,left=0.25}{v4,va1}
\fmf{plain,tension=1}{va1,va2}
\fmf{plain,tension=0}{va2,vb2}
\fmf{phantom,tension=0.5,left=0.125}{v7,va2}
\fmf{plain,tension=0.5,right=0.125}{v8,va2}
\fmf{plain,tension=1,left=0}{vb1,vb2}
\fmf{plain,tension=1,left=0}{vb3,vb4}
\fmf{plain,tension=0.5,right=0,width=1mm}{v5,v8}
}

\newcommand{\chionethreetwothree}[1][black]{%
\fmftop{v1}
\fmfbottom{v5}
\fmfforce{(0.125w,h)}{v1}
\fmfforce{(0.125w,0)}{v5}
\fmffixed{(0.25w,0)}{v1,v2}
\fmffixed{(0.25w,0)}{v2,v3}
\fmffixed{(0.25w,0)}{v3,v4}
\fmffixed{(0.25w,0)}{v5,v6}
\fmffixed{(0.25w,0)}{v6,v7}
\fmffixed{(0.25w,0)}{v7,v8}
\fmffixed{(0,whatever)}{va1,va2}
\fmffixed{(whatever,0)}{va1,vc1}
\fmffixed{(whatever,0)}{va2,vb1}
\fmffixed{(0,whatever)}{vc1,vc3}
\fmffixed{(0,whatever)}{vb2,vb4}
\fmffixed{(0,whatever)}{vc1,vb1}
\fmffixed{(0,whatever)}{vc1,vb3}
\fmffixed{(whatever,0)}{vb1,vb2}
\fmffixed{(whatever,0)}{vb3,vb4}
\fmf{plain,tension=0.5,right=0.25}{v3,vc1}
\fmf{plain,tension=0.5,left=0.25}{v4,vc1}
\fmf{phantom,tension=0.5,right=0.25}{v2,vb2}
\fmf{phantom,tension=0.5,left=0.25}{v3,vb2}
\fmf{plain,tension=0.5,left=0.25}{v7,vc3}
\fmf{plain,tension=0.5,right=0.25}{v8,vc3}
\fmf{plain,tension=0.5,left=0.125}{v6,vb4}
\fmf{phantom,tension=0.5,right=0.125}{v7,vb4}
\fmf{plain,tension=2,left=0}{vc1,vb1}
\fmf{plain,tension=2,left=0}{vb1,vb3}
\fmf{plain,tension=2,left=0}{vb3,vc3}
\fmf{plain,tension=2,left=0}{vb2,vb4}
\fmffreeze
\fmf{plain,tension=0.5,right=0.25}{v1,va1}
\fmf{plain,tension=0.5,left=0.25}{v2,va1}
\fmf{plain,tension=1}{va1,va2}
\fmf{plain,tension=0}{va2,vb2}
\fmf{plain,tension=0.5,left=0.125}{v5,va2}
\fmf{phantom,tension=0.5,right=0.125}{v6,va2}
\fmf{plain,tension=1,left=0}{vb1,vb2}
\fmf{plain,tension=1,left=0}{vb3,vb4}
\fmf{plain,tension=0.5,right=0,width=1mm}{v5,v8}
}

\newcommand{\chionetwoonethree}[1][black]{%
\fmftop{v1}
\fmfbottom{v5}
\fmfforce{(0.125w,h)}{v1}
\fmfforce{(0.125w,0)}{v5}
\fmffixed{(0.25w,0)}{v1,v2}
\fmffixed{(0.25w,0)}{v2,v3}
\fmffixed{(0.25w,0)}{v3,v4}
\fmffixed{(0.25w,0)}{v5,v6}
\fmffixed{(0.25w,0)}{v6,v7}
\fmffixed{(0.25w,0)}{v7,v8}
\fmffixed{(0,whatever)}{va1,va2}
\fmffixed{(whatever,0)}{va1,vb3}
\fmffixed{(whatever,0)}{va2,vc3}
\fmffixed{(0,whatever)}{vc1,vc3}
\fmffixed{(0,whatever)}{vb2,vb4}
\fmffixed{(0,whatever)}{vc1,vb1}
\fmffixed{(0,whatever)}{vc1,vb3}
\fmffixed{(whatever,0)}{vb1,vb2}
\fmffixed{(whatever,0)}{vb3,vb4}
\fmf{plain,tension=0.5,right=0.25}{v1,vc1}
\fmf{plain,tension=0.5,left=0.25}{v2,vc1}
\fmf{phantom,tension=0.5,right=0.125}{v2,vb2}
\fmf{plain,tension=0.5,left=0.125}{v3,vb2}
\fmf{plain,tension=0.5,left=0.25}{v5,vc3}
\fmf{plain,tension=0.5,right=0.25}{v6,vc3}
\fmf{phantom,tension=0.5,left=0.25}{v6,vb4}
\fmf{phantom,tension=0.5,right=0.25}{v7,vb4}
\fmf{plain,tension=2,left=0}{vc1,vb1}
\fmf{plain,tension=2,left=0}{vb1,vb3}
\fmf{plain,tension=2,left=0}{vb3,vc3}
\fmf{plain,tension=2,left=0}{vb2,vb4}
\fmffreeze
\fmf{phantom,tension=0.5,right=0.125}{v3,va1}
\fmf{plain,tension=0.5,left=0.125}{v4,va1}
\fmf{plain,tension=1}{va1,va2}
\fmf{plain,tension=0}{va1,vb4}
\fmf{plain,tension=0.5,left=0.25}{v7,va2}
\fmf{plain,tension=0.5,right=0.25}{v8,va2}
\fmf{plain,tension=1,left=0}{vb1,vb2}
\fmf{plain,tension=1,left=0}{vb3,vb4}
\fmf{plain,tension=0.5,right=0,width=1mm}{v5,v8}
}

\newcommand{\chionetwoonetwoone}[1][black]{%
\fmftop{v1}
\fmfbottom{v4}
\fmfforce{(0.125w,h)}{v1}
\fmfforce{(0.125w,0)}{v4}
\fmffixed{(0.25w,0)}{v1,v2}
\fmffixed{(0.25w,0)}{v2,v3}
\fmffixed{(0.25w,0)}{v4,v5}
\fmffixed{(0.25w,0)}{v5,v6}
\fmffixed{(0,whatever)}{vc1,vc3}
\fmffixed{(0,whatever)}{vb2,vb4}
\fmffixed{(0,whatever)}{vc2,vb2}
\fmffixed{(0,whatever)}{vc4,vb4}
\fmffixed{(0,whatever)}{vc1,vb1}
\fmffixed{(0,whatever)}{vc1,vb3}
\fmffixed{(whatever,0)}{vc1,vc2}
\fmffixed{(whatever,0)}{vb1,vb2}
\fmffixed{(whatever,0)}{vb3,vb4}
\fmffixed{(whatever,0)}{vc3,vc4}
\fmf{plain,tension=0.5,right=0.25}{v1,vc1}
\fmf{plain,tension=0.5,left=0.25}{v2,vc1}
\fmf{phantom,tension=0.5,right=0.25}{v2,vc2}
\fmf{plain,tension=0.5,left=0.25}{v3,vc2}
\fmf{plain,tension=0.5,left=0.25}{v4,vc3}
\fmf{phantom,tension=0.5,right=0.25}{v5,vc3}
\fmf{plain,tension=0.5,left=0.25}{v5,vc4}
\fmf{plain,tension=0.5,right=0.25}{v6,vc4}
\fmf{plain,tension=2,left=0}{vc1,vb1}
\fmf{plain,tension=2,left=0}{vb1,vb3}
\fmf{plain,tension=2,left=0}{vb3,vc3}
\fmf{plain,tension=2,left=0}{vc2,vb2}
\fmf{plain,tension=2,left=0}{vb2,vb4}
\fmf{plain,tension=2,left=0}{vb4,vc4}
\fmffreeze
\fmf{plain,tension=2,left=0}{vb1,vc2}
\fmf{plain,tension=2,left=0}{vb3,vb2}
\fmf{plain,tension=2,left=0}{vc3,vb4}
\fmf{plain,tension=0.5,right=0,width=1mm}{v4,v6}
}

\newcommand{\oneloopfftriang}[1][]{%
\settoheight{\eqoff}{$\times$}%
\setlength{\eqoff}{0.5\eqoff}%
\addtolength{\eqoff}{-8.5\unitlength}%
\raisebox{\eqoff}{%
\fmfframe(1,1)(4,1){%
\begin{fmfchar*}(15,15)
\fmfleft{vin}
\fmfright{vr}
\fmfforce{(0,0.5h)}{vin}
\fmfforce{(w,h)}{v1}
\fmfforce{(w,0.5h)}{v2}
\fmfforce{(w,0)}{v3}
\fmfpoly{phantom}{v3,v1,vo}
\fmf{phantom}{vo,vl1}
\fmf{phantom}{vl1,vr1}
\fmf{phantom,tension=3}{vr1,v1}
\fmf{phantom}{vo,vl3}
\fmf{phantom}{vl3,vr3}
\fmf{phantom,tension=3}{vr3,v3}
\fmffreeze
{#1}
\fmf{double}{vin,vo}
\end{fmfchar*}}}}

\newcommand{\twoloopfftriang}[1][]{%
\settoheight{\eqoff}{$\times$}%
\setlength{\eqoff}{0.5\eqoff}%
\addtolength{\eqoff}{-6\unitlength}%
\raisebox{\eqoff}{%
\fmfframe(1,1)(1,1){%
\begin{fmfchar*}(10,10)
\fmfleft{vin}
\fmfright{vr}
\fmfforce{(0,0.5h)}{vin}
\fmfforce{(w,h)}{v1}
\fmfforce{(w,0)}{v2}
\fmfpoly{phantom}{v2,v1,vo}
\fmf{phantom}{vo,vl1}
\fmf{phantom}{vl1,vr1}
\fmf{phantom,tension=3}{vr1,v1}
\fmf{phantom}{vo,vl2}
\fmf{phantom}{vl2,vr2}
\fmf{phantom,tension=3}{vr2,v2}
\fmffreeze
\fmf{phantom}{vl1,vm1}
\fmf{phantom}{vm1,vr1}
\fmf{phantom}{vl2,vm2}
\fmf{phantom}{vm2,vr2}
\fmffreeze
\fmf{phantom}{vl1,vlc}
\fmf{phantom}{vlc,vl2}
\fmf{phantom}{vm1,vmc}
\fmf{phantom}{vmc,vm2}
\fmf{phantom}{vr1,vrc}
\fmf{phantom}{vrc,vr2}
\fmffreeze
\fmfposition
{#1}
\fmf{double}{vin,vo}
\end{fmfchar*}}}}

\newgray{ogray}{0.85}
\newgray{hatchgray}{1}
\newgray{sgray}{0.8}

\newlength{\rad}
\newlength{\roff}
\newlength{\ri}
\psset{xunit=\unit,yunit=\unit,runit=\unit}
\newlength{\dlinewidth}
\newlength{\doublesep}
\psset{doublesep=\doublesep}
\psset{linewidth=\linew}
\newlength{\auxlen}
\newlength{\linearc}
\newlength{\xa}
\newlength{\ya}
\newlength{\xb}
\newlength{\yb}
\newlength{\xc}
\newlength{\yc}
\newlength{\xd}
\newlength{\yd}

\newlength{\ye}

\newlength{\yf}

\newlength{\armlen}
\newcounter{nnodenum}
\newcounter{mnodenum}
\newcommand{\vacpolp}[2][0.5]{%
\fmfcmd{
begingroup;
save t,v,tv,do,di,ppol,pstr,dia;
path ppol,pstr;
pair v[],tv[],do[],di[];
ppol=#2;
t3=arctime (0.5*arclength ppol) of ppol;
v3=point t3 of ppol;
dia=#1 arclength ppol; 
fill(fullcircle scaled dia shifted v3) withcolor 0.2black;
endgroup;
}
}
\newcommand{\swfone}[4][]{%
\settoheight{\eqoff}{$\times$}%
\setlength{\eqoff}{0.5\eqoff}%
\addtolength{\eqoff}{-6\unitlength}%
\raisebox{\eqoff}{%
\fmfframe(1,1)(1,1){%
\begin{fmfchar*}(14,10)
\fmfleft{v1}
\fmfright{v2}
\fmffixed{(0.5w,0)}{vc1,vc2}
\fmf{#2}{v1,vc1}
\fmf{#2}{vc2,v2}
\fmf{#3}{vc1,vc2}
\fmf{#4}{vc2,vc1}
\fmffreeze
\fmfposition
\fmfipath{p[]}
\fmfipair{vm[]}
\fmfiset{p1}{vpath(__v1,__vc1)}
\fmfiset{p2}{vpath(__vc1,__vc2)}
\fmfiset{p3}{reverse vpath(__vc2,__vc1) }
\fmfiset{p4}{vpath(__vc2,__v2)}
\svertex{vm1}{p2}
\svertex{vm2}{p3}
{#1}
\end{fmfchar*}}}}


%% file: Macro-Dhritiman.tex
\newcommand{\mK}{\mathcal{K}}

\newcommand{\mA}{\mathcal{A}}

\newcommand{\mO}{\mathcal{O}}
\newcommand{\mN}{\mathcal{N}}

\newcommand{\mF}{\mathcal{F}}
\newcommand{\mC}{\mathcal{C}}

\newcommand{\Fco}{\mF}
\newcommand{\Fb}{F}

\newcommand{\beq}{\begin{equation}}
\newcommand{\eeq}{\end{equation}}
\newcommand{\beqa}{\begin{eqnarray}}
\newcommand{\eeqa}{\end{eqnarray}}
\newcommand{\bea}{\begin{aligned}}
\newcommand{\eea}{\end{aligned}}
\newcommand{\bem}{\begin{multline}}
\newcommand{\eem}{\end{multline}}

\newcommand{\Gd}{\delta}

\newcommand{\Gl}{\lambda}

\newcommand{\twops}{\int\de \text{PS}_{2, \{l\}}}
\newcommand{\threeps}{\int \de \text{PS}_{3,\{l\}}}

\def\bra(#1,#2){[#1\,#2]}
\def\ket(#1,#2){\langle #1\,#2\rangle}
\def\mt(#1,#2,#3,#4){\langle #1\,#2\,#3\,#4\rangle}

\newcommand{\deta}{\de^8 \eta_{l_{1,2}}}

\newcommand{\be}{\begin{equation}}
\newcommand{\ee}{\end{equation}}


%% file: introduction.tex
\section{Introduction}

So far, the framework of quantum field theories (QFTs) is very successful in 
describing the high-energy processes measured at colliders such as the LHC.
However, theoretical predictions are usually restricted to the weak-coupling 
regime, which admits a perturbative expansion in terms of the small coupling constants.
The individual contributions to the perturbation series can be calculated via 
Feynman diagrams. 
Thereby, a large proliferation of diagrams is in general encountered when one proceeds to higher-order corrections, and hence concrete calculations are mainly restricted to the first few orders.

The investigation of alternative techniques that bypass this limitation is thus of high importance. It might not only allow to push perturbation theory to higher orders, but could also deepen our understanding of the fundamental principles and 
mechanisms encoded in QFTs. 
The so-called `on-shell' techniques are such an alternative.
They allow one to build amplitudes from simpler amplitudes with a lower number of external legs and loops via recursion relations \cite{Britto:2004ap,Britto:2005fq} and unitarity \cite{Bern:1994zx,Bern:1994cg}.
They have been successfully used in supersymmetric gauge theories as well as in QCD, see \cite{Bern:2007dw, Elvang:2013cua,Henn:2014yza} for pedagogical reviews and references therein.

In particular, the maximally supersymmetric Yang-Mills ($\mathcal{N}=4$ SYM) theory 
with gauge group $SU(N_{\text{c}})$
in four dimensions has played an important role in the aforementioned developments.
According to the
$\AdS/\text{CFT}$ correspondence \cite{Maldacena:1997re,Gubser:1998bc,Witten:1998qj},
 it has a dual description in terms of 
a string theory, allowing its study also at strong coupling. Moreover, in the planar limit \cite{'tHooft:1973jz}, it shows signs of integrability at weak as well as at 
strong coupling, which is
believed to be present even at any coupling. Based on the conjectured integrability, new predictions for the spectrum, i.e.\ for the anomalous scaling dimensions of gauge-invariant composite operators, were made; see \cite{Beisert:2010jr} for a review.
This rises the hope that the theory is exactly solvable, and it is hence sometimes even referred to as the ``harmonic oscillator of the $21^\text{st}$ century''.

Given the success of the aforementioned on-shell techniques for amplitudes, it is an intriguing question whether they can be applied for determining off-shell quantities such as correlation functions or the anomalous dimensions as well. A bridge between the purely on-shell amplitudes and the purely off-shell correlation functions 
is provided by form factors. In particular, they also contain the information necessary to determine 
the anomalous dimensions.
An $n$-point form factor describes the overlap of an off-shell initial state, described by a composite operator, into an on-shell final state consisting of $n$ elementary fields. 
It is given by
\begin{equation}
\label{formfactor}
{\cal F}_{\mathcal{O}}(1,\dots,n) = \int \de^D x \e^{-i q \cdot x} \langle 1\cdots n|{\cal O}(x) | 0\rangle= (2\pi)^D\delta^{(D)}\Big(q-\sum_{i=1}^n p_i\Big)\langle 1\cdots n|{\cal O}(0) | 0\rangle \col
\end{equation}
where the particles labeled by $i=1,\dots,n$ 
carry individual on-shell momenta $p_i$ and the operator $\mathcal{O}$ carries off-shell momentum $q$. If the number $n$ of the external fields exactly matches the
number of fields contained in $\mathcal{O}$, the form factor is called minimal.
Minimal form factors with $n=2$ points are denoted as Sudakov form factors.

In $\mathcal{N}=4$ SYM theory, the most intensively studied form factors are the ones of the half-BPS operator
\begin{equation}
\label{BPSinso6}
\mathcal{O}_\BPS=\tr(\phi_{(I}\phi_{J)})\col
\end{equation}
where the parentheses denote traceless-symmetrization of the indices $I,J=1,\dots,N_\phi$ of the $N_\phi$ scalar field flavors. This operator 
belongs to the stress-tensor supermultiplet.
Its Sudakov form factor
was first studied by van Neerven \cite{vanNeerven:1985ja} and analyzed up to four loops \cite{Gehrmann:2011xn, Boels:2012ew} in the recent past. 
The Sudakov form factor exhibits exponentiation \cite{Mueller:1979ih, Collins:1980ih, Sen:1981sd}, a feature which was seen to be the key for predicting the all-loop IR behavior of scattering amplitudes \cite{Bern:2005iz}.

The form factors of the stress-tensor multiplet with general $n$ external legs can be analyzed in 
analogy to scattering amplitudes with modern on-shell techniques.
The $n$-point form factor of the bosonic operator \eqref{BPSinso6} was first studied in \cite{Brandhuber:2010ad, Bork:2010wf}, and later generalized to the full stress-tensor multiplet in \cite{Brandhuber:2011tv, Bork:2011cj}. 
Up to one-loop order, compact expressions for general $n$-point maximally-helicity-violating (MHV) as well as some next-to-MHV (NMHV) form factors have been computed in
\cite{Brandhuber:2010ad,Brandhuber:2011tv, Bork:2011cj, Bork:2012tt, Engelund:2012re, Bork:2014eqa}. 
The two-loop three-point form factor was computed in \cite{Brandhuber:2012vm}.
The form factors of half-BPS operators with $k$ scalar fields, as well as the corresponding supermultiplets, 
have been studied in \cite{Bork:2010wf, Penante:2014sza, Brandhuber:2014ica}; $n$-point tree and one-loop MHV
results are presented in \cite{Penante:2014sza} and the mininal form factors (for $n=k$) were computed at two-loop \cite{Brandhuber:2014ica}. Form factors have also been studied at strong coupling via the AdS/CFT correspondence \cite{Alday:2007he}, and a Y-system formulation was given in \cite{Maldacena:2010kp} for $\AdS_3$ and in \cite{Gao:2013dza} for $\AdS_5$.

The aforementioned studies have shown that form factors share very similar recursive and analytic properties with scattering amplitudes, at least for the protected operators. 
Moreover, the robust set of on-shell techniques for computing on-shell objects is also applicable in this case. 
This rises the hope that also fully off-shell quantities can be studied using on-shell methods, and that such an enhancement of the toolkit allows to detect
new features of the theory.
Indeed, it was found that certain correlation functions can be constructed via generalized unitarity from amplitudes, form factors and their generalizations involving several operator insertions 
\cite{Engelund:2012re}.
In the recent parallel work \cite{Wilhelm:2014qua}, 
one of us has determined at tree level the minimal form factors
of a generic operator and at one-loop order their cut-constructible parts.
The one-loop results yield 
the complete one-loop dilatation operator of the theory.

Scattering amplitudes as well as form factors are themselves not physical observables, since they contain infrared (IR) divergences  from the integration of loop momenta. Adding the so-called bremsstrahlung contributions, their
IR divergences from the real emissions of soft and collinear particles cancel
the IR divergences coming from virtual loop corrections according to the Kinoshita-Lee-Nauenberg theorem \cite{Kinoshita:1962ur,Lee:1964is},  and one obtains an observable.
In particular, the cross sections are free of IR divergences and hence 
physical observables. They are, however, in general not well defined in a CFT such as $\mathcal{N}=4$ SYM theory, where asymptotic states are ill defined.
Some cross-section-type quantities have been defined by using coherent states as 
asymptotic states \cite{Bork:2009nc}. 
Alternatively, we can consider the decay of an initial off-shell state created by an operator $\mathcal{O}(q)$ with timelike momentum ($q^2>0$) into any final on-shell multi-particle state. 
The probability of this inclusive decay is the 
total decay rate of $\mathcal{O}(q)$. 
This decay process may occur as part of a total cross section of a 
scattering process in which $\mathcal{O}(q)$ is produced as an intermediate 
state.\footnote{The operator may be of different physical origin. For example, it can be part of a vertex that couples to a massive particle in an effective Lagrangian. Then, \eqref{crosssection} yields the decay rate of this particle.
A concrete example from the Standard Model is an effective Higgs-gluon vertex $H\tr(F_{\mu\nu}F^{\mu\nu})$ obtained by integrating out a heavy quark loop, see e.g.\ \cite{Schmidt:1997wr}.
The operator may also be a (conserved) current describing a two-particle scattering. Examples of this type are $e^+ e^-$ annihilation into a virtual photon or Drell-Yan scattering, where the two incoming particles are annihilated into a virtual photon or gluon, respectively, exciting the QCD vacuum and decaying into
quarks, gluons etc.}
The probability for the inclusive decay of $\mathcal{O}(q)$ into 
a final state $X$ with total momentum $q=p_X$ is defined by
\begin{equation}
\sigma_{\mathcal{O}}(q)=\sum_X \,\delta^{(D)}(q-p_X) \, | \langle X| {\cal O}(0) | 0 \rangle |^2 \col
\label{crosssection}
\end{equation}
where the sum ensures that the quantity is inclusive, i.e.\ 
all contributions, which are specified by the number and type of the particles
in the final states, are integrated over the respective phase space and are 
summed up.
This cross-section-type quantity depends on the matrix element $\langle X| {\cal O}(0) | 0 \rangle$, which is precisely the form factor of $\mathcal{O}$ with final 
state $X$.
Via the optical theorem, \eqref{crosssection} is related to the imaginary part of the (time-ordered) two-point correlation function $\langle 0|\bar{\cal O}(x){\cal O}(0) | 0 \rangle$ 
after transforming to momentum space.

Finally, although not considered in this paper, we would like to mention that
by modifying \eqref{crosssection}, 
`event shapes' such as energy  or charge correlation functions were studied
in ${\cal N}=4$ SYM theory \cite{Hofman:2008ar, Engelund:2012re, Belitsky:2013xxa, Belitsky:2013bja}. Also, Wilson coefficients for deep inelastic scattering were
considered  \cite{Bianchi:2013sta}.
For simplicity, we will follow the terminology of \cite{Belitsky:2013xxa} and denote the cross-section-type quantity defined in \eqref{crosssection} as total cross section, or simply cross section.

In this paper, we will study the form factor \eqref{formfactor} and the cross section
\eqref{crosssection} for the Konishi operator as a first example for an operator 
that is not protected by supersymmetry. Hence, UV divergences appear in 
addition to the aforementioned IR divergences that already emerge for protected 
operators.
The Konishi primary operator is given by
\begin{equation}\label{Kinso6}
\mK=\delta^{IJ}\tr(\phi_I\phi_J)\col
\end{equation}
where sums over all $I,J=1,\dots,N_\phi$ scalar field flavors are implicitly 
understood. In strictly $D=4$ dimensions, we have $N_\phi=6$. 
The Konishi scaling dimension 
$\Delta_{\mathcal{K}}=\Delta_{\mathcal{K}}^{(0)}+\gamma_{\mathcal{K}}$
consists of the bare dimension $\Delta_{\mathcal{K}}^{(0)}=2$ and an anomalous 
dimension 
$\gamma_{\mathcal{K}}$. It is a power series in the 
coupling constant
\begin{equation}\label{gdef}
g^2 = \frac{g_\YM^2N_{\text{c}}}{(4\pi)^2}(4\pi e^{- \gamma_{\text{E}}})^\peps
\col
\end{equation}
which depends on the Yang-Mills coupling constant $g_\YM$ as well as 
the number of colors 
$N_{\text{c}}$ and is the loop-counting parameter in the modified
dimensional reduction ($\overline{\text{DR}}$) 
scheme in $D=4-2\peps$ dimensions.\footnote{The $\overline{\text{DR}}$ scheme employs dimensional reduction of ten-dimensional $\mathcal{N}=1$ SYM theory to $D=4-2\peps$ as regularization  \cite{Siegel:1979wq,Capper:1979ns} and for the subtraction of the divergences a modified minimal subtraction which absorbs the same finite terms in addition to the UV divergences into the renormalization constant as the famous $\overline{\text{MS}}$ scheme \cite{Bardeen:1978yd}, leading to the factor $(4\pi e^{- \gamma_{\text{E}}})^\peps$.} 
In the planar limit, the Konishi anomalous dimension is given 
by\footnote{Note that there are no non-planar corrections to 
$\gamma_\mK$ at the first three
loop orders.}
\begin{equation}
\begin{aligned}
\label{gammaK}
\gamma_{\mK}&=6[2g^2-8g^4+56g^6-16(26-6\zeta_3+15\zeta_5)g^8\\
&\phantom{{}={}6[}
+16(158+72\zeta_3-54\zeta_3^2-90\zeta_5+315\zeta_7)g^{10}]
+\mathcal{O}(g^{12})\col
\end{aligned}
\end{equation}
where the one- and two-loop contributions, which we reproduce as a 
check in this paper, were 
obtained by explicit 
Feynman diagram calculations in \cite{Anselmi:1996mq,Anselmi:1996dd} and
\cite{Bianchi:1999ge,Bianchi:2000hn,Eden:2000mv}.\footnote{The Konishi anomalous dimension $\gamma_{\mK}$ is currently known up to five loops from field theory calculations
and up to nine loops from the conjectured integrability.
The three-loop result was conjectured in \cite{Kotikov:2004er} and confirmed in \cite{Eden:2004ua,Sieg:2010tz}. The four-loop result was determined by calculating the wrapping corrections to the integrability-based asymptotic dilatation operator in \cite{Fiamberti:2007rj,Fiamberti:2008sh} and by a computer-based direct calculation in \cite{Velizhanin:2008jd}. The integrability-based four-loop expression of \cite{Bajnok:2008bm} matches this result. The five-loop result was predicted from integrability in
\cite{Bajnok:2009vm,Arutyunov:2010gb,Balog:2010xa}, and confirmed in
\cite{Eden:2012fe} from an OPE analysis of the four-point correlation function of stress-tensor multiplets. The results at six \cite{Leurent:2012ab}, seven \cite{Bajnok:2012bz}, eight \cite{Leurent:2013mr} and nine loops \cite{Volin:IGST} 
are so far only based on the conjectured integrability.}

The operator \eqref{Kinso6} is the  primary operator
of the Konishi supermultiplet. Its anomalous dimension given in \eqref{gammaK}
was mainly obtained by considering certain descendent operators within the 
Konishi multiplet rather than the Konishi primary operator \eqref{Kinso6}.
This is possible  since all members of a supermultiplet have the same anomalous 
dimension.\footnote{Working with certain descendants which are 
non-singlet states of the $SU(4)$ R-symmetry instead of the primary 
operator \eqref{Kinso6}, which is an $SU(4)$ singlet, simplifies the calculations in both, the field theory and integrability-based approach.} In fact, we will see that the Konishi primary defined in \eqref{Kinso6} and involving a sum over the $N_\phi$ scalar field flavors depends on the dimension $D$, since $N_\phi=10-D$ is required 
to ensure supersymmetry. This becomes important when regulating the divergences by 
continuing the theory from $D=4$ to $D=4-2\peps$ dimensions.

We will apply four-dimensional unitarity in order to compute the form factors.
Within this framework, all on-shell component fields can be conveniently combined into Nair's $\mathcal{N}=4$ on-shell superfield \cite{Nair:1988bq}. 
The on-shell superfield reads
\begin{equation}\label{ossuperfield}
\Phi(p, \eta)= g_+(p) + \eta^A \, \psi_A(p) + \frac{ \eta^A \eta^B}{2!} \, \phi_{AB}(p) + \frac{\teps_{ABCD}\eta^A \eta^B \eta^C}{ 3!}  \tilde\psi^D(p) + \eta^1 \eta^2 \eta^3 \eta^4 g_-(p) \col
\end{equation}
where $\eta^A$ are Grassmann variables that encode the flavor and helicity of the component fields. Pairs of upper and lower $SU(4)$ R-symmetry indices $A,B,C,D=1,\dots,4$ are always understood to be summed.
In the above superfield, the six real on-shell scalars $\phi_I$ 
transforming in the fundamental representation of $SO(6)$
are represented via the anti-symmetric product representation 
of two fundamental $SU(4)$ representations,
$\phi_{AB}=\phi_I(\sigma_I)_{AB}$, employing the isomorphism of the Lie-algebras $\mathfrak{so}(6)$ and $\mathfrak{su}(4)$ induced by the $\sigma$-matrices
$(\sigma_I)_{AB}=-(\sigma_I)_{BA}$. 

Using \eqref{ossuperfield}, each $n$-point scattering amplitude with fixed total helicity can be efficiently packed into a single superamplitude.
In analogy, also the form factors for the BPS operator \eqref{BPSinso6}
can be packed into super form factors if the 
BPS operator is expressed in terms of the scalar fields $\phi_{AB}$ as
\begin{equation}
\label{BPSsu4-general}
\mathcal{O}_\BPS=\tr(\phi_{AB}\phi_{CD}) - \frac{1}{12} \teps_{ABCD} \, \tr(\phi^{EF} \phi_{EF})
\col
\end{equation}
where the last term subtracts the trace in the space of scalar 
flavors.\footnote{Note that
$(\sigma_I)^{AC}(\sigma_J)_{CB}+(\sigma_J)^{AC}(\sigma_I)_{CB}=-2\delta_{IJ}\delta^A{}_B$, where $(\sigma_I)^{AB}=\frac{1}{2}\teps^{ABCD}(\sigma_I)_{CD}$.}

Without loss of generality, we will focus in the rest of this paper on its particular component
\begin{equation}\label{BPSsu4}
\mathcal{O}_\BPS=\tr(\phi_{AB}\phi_{AB})
\col
\end{equation}
where doubled indices are not summed.
Expressing also the Konishi operator in terms of the scalar fields $\phi_{AB}$
yields
\begin{equation}\label{Kinsu4}
\mK_6=\frac{1}{8}\teps^{ABCD}\tr(\phi_{AB}\phi_{CD})
=\tr(\phi_{12}\phi_{34})-\tr(\phi_{13}\phi_{24})+\tr(\phi_{14}\phi_{23})
\col
\end{equation} 
where the subscript $6$ reminds us that the operator is identical to the Konishi
primary \eqref{Kinso6} only for $N_\phi=6$, i.e.\ only in strictly $D=4$ 
dimensions.

There is a subtlety originating from the fact that in $D\neq 4$ dimensions
the Konishi
operator $\mathcal{K}$ in \eqref{Kinso6} cannot be identified with 
$\mathcal{K}_6$ in \eqref{Kinsu4}.
The four-dimensional unitarity method directly
applies to the operator $\mathcal{K}_6$. In this formulation, the operator 
stays the same if the encountered IR and UV divergences are regularized
by changing the spacetime dimension from $D=4$ to $D=4-2\peps$. 
But in $D=4-2\peps$ dimensions the Konishi operator 
$\mathcal{K}$ is \emph{not} identical to the operator
$\mathcal{K}_6$. 
Hence, the unitarity-based results for $\mathcal{K}_6$ do not directly 
yield those for the Konishi operator $\mathcal{K}$.
Instead, modifications have to be made which take into account that
one should have used  
$\mathcal{K}$ and not $\mathcal{K}_6$ in order to obtain the results for the
Konishi operator regularized in $D=4-2\peps$ dimensions.

In the main part of the paper, we elaborate on the ideas mentioned above. 
In section \ref{section:CSNutshell}, we discuss two-point correlation functions of gauge-invariant local operators, their renormalization and the transformation to momentum space. We identify the imaginary part of such a correlation function with the cross section defined in \eqref{crosssection}. Finally, we  present the general strategy of computing the total cross section for a given operator using its form factors as building blocks.

In section \ref{sec:ffactor}, we present our computation of the form factors for $\mathcal{K}_6$ at the one- and two-loop orders, which are based on the unitarity method and on-shell superspace. 
Since the Konishi operator is not protected, several interesting features appear in the results which have not occurred for amplitudes or BPS form factors in $\mathcal{N}=4$ SYM theory, e.g.\ UV divergences and rational terms.

In section \ref{sec: subtleties in unitarity}, 
we discuss in detail the 
aforementioned subtleties arising from  
the fact that in $D=4-2\peps$ dimensions the Konishi operator $\mathcal{K}$ cannot be identified with $\mathcal{K}_6$.
We derive a rigorous prescription of how to implement the substitution
of $\mathcal{K}_6$ by $\mathcal{K}$ in the results of the previous section and 
give final results for $\mathcal{K}$.

In section \ref{sec:CS}, we present the computation of the cross section starting with the BPS operator up to one-loop order as a simple example to make the reader familiar with our strategy. We find the expected non-trivial cancelation of the IR divergences between real and virtual channels.
Then, we compute the cross section for the Konishi operator up to two loops. 
We extract the renormalization constant and hence the anomalous dimension from the UV divergence of the bare result. They match the known expressions.
We present the finite result for the renormalized cross section and discuss 
its dependence on the renormalization scheme. 

Finally, in section \ref{sec: conclusion and outlook} we summarize the main results of our paper and the interesting features associated with them. We also present some future directions and open questions.

In the appendices \ref{app:FTprop}, \ref{app:integrals} and \ref{app:PV}, we give 
some further conventions and explicit results for the occurring loop integrals
as well as Passarino-Veltman (PV) reduction formulae. Appendix \ref{app:check-3pt}
provides some cross checks for the one-loop three-point Konishi form factors.
In appendix \ref{App:phase-space}, we present some details on the phase space 
integrals occurring in section \ref{sec:CS}. A way to extract the 
anomalous dimension directly from the two-point Konishi form factor 
is given in appendix \ref{app:anomalous-from-2pt}.
In appendix \ref{app:scheme}, we discuss the renormalization-scheme dependence of 
the cross section.
In the final appendix \ref{app:feyndiag}, we summarize direct Feynman-diagrammatic calculations of the one- and two-loop form factors for the BPS and the Konishi operator, which serve as checks for our approach and
guided us to the modifications discussed in section \ref{sec: subtleties in unitarity}.

%% file: csnutshell.tex
\section{Cross sections for two-point correlation functions in a nutshell \label{section:CSNutshell} }

In this section, we review some facts about the form of the two-point 
correlation function of a renormalized composite operator in spacetime 
and in momentum space. Via the optical theorem, its imaginary part
yields a cross-section-type quantity.
We present our strategy of computing this quantity from the 
form factors of the respective operator.

\subsection{Renormalization of composite operators and their two-point functions}
\label{subsec:opreno}

Gauge-invariant local composite operators can be regarded as external states
of $\mathcal{N}=4$ SYM theory, 
and they can occur in correlation functions in the same way as the elementary fields. 
Such correlation functions in general contain UV divergences
which are associated with the presence of these operators, requiring 
their renormalization in analogy to that of the elementary fields and vertices of 
the theory.
In this paper, we only consider composite operators that are eigenstates under renormalization. Such a renormalized operator is given in terms of the bare operator
as 
\begin{equation}\label{opren}
\mathcal{O}_{\ren}
=\mathcal{Z}_{\mathcal{O}}(g,\peps)\mathcal{O}_{\bare}
\pnt
\end{equation}
The renormalization constant $\mathcal{Z}_{\mathcal{O}}$ depends
on the coupling constant $g$ and absorbs the UV divergences, which appear as 
poles in $\peps$ when the theory is regularized by changing the spacetime dimension from $D=4$ to $D=4-2\peps$. 
The renormalization constant determines the anomalous dimension
\begin{equation}
\gamma_{\mathcal{O}}=\sum_{\ell=1}^\infty g^{2\ell}\gamma_{\mathcal{O}}^{(\ell)}=\lim_{\peps\to0}\peps g\parderiv{g}\log\mathcal{Z}_{\mathcal{O}}\col
\end{equation}
which is added to the bare scaling dimension $\Delta_{\mathcal{O}}^{(0)}$ in order to obtain the conformal dimension $\Delta_{\mathcal{O}}$.
Since $\gamma_\mathcal{O}$ is finite when the limit $\peps\to0$ is taken in the 
above equation, the form of $\mathcal{Z}_{\mathcal{O}}$ as a power series in $g$ 
is fixed to
\begin{equation}
\mathcal{Z}_{\mathcal{O}}=\exp\bigg(\sum_{\ell=1}^\infty\frac{g^{2\ell}}{2\ell\peps}\gamma_{\mathcal{O}}^{(\ell)}\bigg)
=1+g^2\frac{\gamma_{\mathcal{O}}^{(1)}}{2\peps}
+g^4\bigg(\frac{(\gamma_{\mathcal{O}}^{(1)})^2}{8\peps^2}+\frac{\gamma_{\mathcal{O}}^{(2)}}{4\peps}\bigg)
+\mathcal{O}(g^6)
\pnt
\label{eq: relation of calZ and gamma}
\end{equation}

Conformal symmetry also completely fixes the form of the two-point function of 
the operator $\mathcal{O}_\ren$. In Minkowski spacetime, it reads
\begin{equation}\label{G2(x)}
G_{2\mathcal{O},\ren}(x)=\langle0|\bar{\mathcal{O}}_\ren(x)\mathcal{O}_\ren(0)|0\rangle 
=\frac{M}{(-x^2+i0)^{\Delta_{\mathcal{O}}}\mu^{2\gamma_{\mathcal{O}}}}\col\qquad \Delta_{\mathcal{O}}=\Delta_{\mathcal{O}}^{(0)}+\gamma_{\mathcal{O}}
\col
\end{equation}
where our conventions for 
the $i0$ description are given in appendix \ref{app:FTprop}.
The parameter $\mu$ has the dimension of mass and is introduced in order 
to fix the mass dimension of $G_{2\mathcal{O},\ren}$ to $2\Delta_{\mathcal{O}}^{(0)}$.
The coupling-dependent dimensionless factor $M$ 
has a perturbative expansion as
\begin{equation}\label{Mexpansion}
M=\sum_{\ell=0}^\infty g^{2\ell}M^{(\ell)}
\col
\end{equation}
and it can be absorbed into the normalization of $\mathcal{O}_\ren$.

We will work in momentum space, and hence need the Fourier transformation of 
\eqref{G2(x)}. According to appendix \ref{app:FTprop}, it is given by
\begin{equation}
\begin{aligned}\label{Pi(q^2)}
\tilde G_{2\mathcal{O}, \ren}(q^2)&=\int\de^Dx\e^{i q\cdot x}G_{2\mathcal{O},\ren}(x)
= (-i) 2^{D - 2\Delta_{\mathcal{O}}} \pi^{\frac{D}{2}} { \Gamma({D\over2} - \Delta_{\mathcal{O}}) \over \Gamma(\Delta_{\mathcal{O}})}  \frac{M}{(-q^2-i0)^{\frac{D}{2} - \Delta_{\mathcal{O}}}\mu^{2\gamma_{\mathcal{O}}}}\pnt  
\end{aligned}
\end{equation}

\begin{figure}[tp]
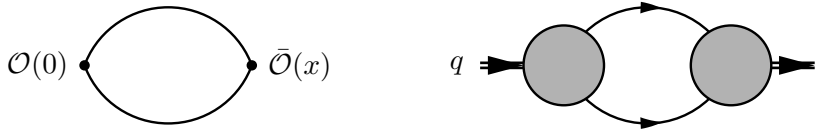

\begin{center}  
$
\settoheight{\eqoff}{$\times$}%
\setlength{\eqoff}{0.5\eqoff}%
\addtolength{\eqoff}{-12.0\unitlength}%
\raisebox{\eqoff}{%
\fmfframe(2,2)(2,2){
\qquad 
\begin{fmfchar*}(22,13)
\fmfleft{vq1}
\fmfright{vq2}
\fmf{plain,tension=0.5,left=0.7,label=$\phantom{l_2}$,l.s=left,l.d=8}{vq2,vq1}
\fmf{plain,tension=0.5,right=0.7,label=$\phantom{l_1}$,l.s=right,l.d=8}{vq2,vq1}
\fmfv{decor.shape=circle,decor.filled=100,decor.size=3}{vq1}
\fmfv{decor.shape=circle,decor.filled=100,decor.size=3}{vq2}
\fmffreeze
 \fmfcmd{pair vertqone, vertqtwo, vertone, verttwo; vertone = vloc(__v1); verttwo = vloc(__v2); vertqone = vloc(__vq1);  vertqtwo = vloc(__vq2); }
 \fmfiv{label=$\mathcal{O}(0)$}{vertqone}
 \fmfiv{label=$\bar{\mathcal{O}}(x)$}{vertqtwo}
\end{fmfchar*}%
\hskip 3cm
\begin{fmfchar*}(44,13)
\fmfleft{vq1}
\fmfright{vq2}
\fmf{dbl_plain_arrow,tension=2}{vq1,v1}
\fmf{plain_arrow,tension=0.5,left=0.7,label=$\phantom{l_2}$,l.s=left,l.d=8}{v1,v2}
\fmf{plain_arrow,tension=0.5,right=0.7,label=$\phantom{l_1}$,l.s=right,l.d=8}{v1,v2}
\fmf{dbl_plain_arrow,tension=2}{v2,vq2}
\fmfv{decor.shape=circle,decor.filled=30,decor.size=30,label=$\phantom{\mathcal{F}_{\mathcal{O}}}$,label.dist=0}{v1}
\fmfv{decor.shape=circle,decor.filled=30,decor.size=30,label=$\phantom{\mathcal{F}_{\mathcal{O}}}$,label.dist=0}{v2}
\fmffreeze
 \fmfcmd{pair vertqone, vertone, verttwo; vertone = vloc(__v1); verttwo = vloc(__v2); vertqone = vloc(__vq1); }
 \fmfiv{label=$q$}{vertqone}
\end{fmfchar*}%
}}%
$
\caption{The two-point function in position space and momentum space.}
\label{fig: two point function}
\end{center}
\end{figure}

When expanding the above expression first for small $g$ and then for small $\peps$, one obtains $\frac{1}{\peps^k}$-poles for any $k\ge1$, which for $k\ge2$ are proportional to powers of $\gamma_{\mathcal{O}}$ \cite{Penati:2001sv}. 
Since $G_{2\mathcal{O},\ren}(x)$ 
is the finite (renormalized) Green function, these poles cannot come from UV divergences.
In fact, they arise from integrating over the origin $x=0$ of spacetime, where $G_{2\mathcal{O},\ren}(x)$ is singular. 
This can be most easily seen 
for the half-BPS operator $\mathcal{O}_\BPS$
defined in \eqref{BPSinso6}. 
Since this operator is protected, 
$\gamma_{\BPS}=0$, and all poles of order $k\ge2$ disappear, but a simple $\frac{1}{\peps}$-pole remains.
In momentum space, this pole is associated with the one-loop bubble integral.
It is obtained when inserting Fourier expressions for the two 
scalar propagators\footnote{In $D$ dimensions, the scaling dimension of a scalar field is given by $\Delta_{\phi}^{(0)}=\frac{D}{2}-1=1-\peps$.} $\frac{1}{(-x^2+i0)^{1-\peps}}$ connecting the two operators as depicted in figure 
\ref{fig: two point function} and performing the integration over $x$
in \eqref{Pi(q^2)}, which yields a $\delta$-function of momentum conservation.
For the tree-level two-point function, the steps are as follows:
\begin{equation}
\frac{1}{(-x^2+i0)^{2-2\peps}}
\quad\underset{\scriptstyle\text{FT}}{\longrightarrow}\quad
\int\frac{\de^D l}{(2\pi)^D}\frac{1}{l^2(l-q)^2}\sim\frac{1}{(-q^2-i0)^\peps\peps}
\pnt
\label{eq: 2ptcorrelator-posi-mom}
\end{equation}
This simple pole (for the BPS operator) and
all the further $\frac{1}{\peps^k}$-poles, $k\ge 2$, (for non-protected operators) are absent when taking the imaginary part of the momentum-space Green function \eqref{Pi(q^2)}.

As we will see in the next subsection, via the 
optical theorem the imaginary part of \eqref{Pi(q^2)} yields a cross-section-type quantity: the probability of the inclusive decay of the renormalized operator $\mathcal{O}_\ren$. It has to be finite in the limit $\peps\to0$, since it is free of IR divergences and --- due to renormalization --- 
also of UV divergences.
 
\subsection{Two-point correlation functions and cross sections}

Via the optical theorem, the imaginary part of a two-point correlation function is related to the inclusive decay width of the renormalized operator $\mathcal{O}_\ren$ with off-shell momentum $q$, where $q^2>0$. As motivated in the introduction, we will simply denote this as cross section $\sigma_{\mathcal{O},\ren}$ in this paper. It is given by%
\footnote{Note that the factor of $(-i)$ appearing in \eqref{Pi(q^2)} due to the Wick rotation (see appendix \ref{app:FTprop}) must be removed before taking the imaginary part. 
Hence, we have to take the imaginary part of $i\,\tilde G_{2\mathcal{O}, {\ren}}$.}
\begin{equation}
\label{ImPi-sigma}
\sigma_{\mathcal{O},\ren}
=\im [2 i \, \tilde G_{2\mathcal{O}, {\ren}}(q^2)]
= \sum_X \,(2\pi)^D\delta^{(D)}(q-p_X) \, | \langle  X | {\cal O}_{\ren}(0) | 0 \rangle |^2
\col
\end{equation}
where one sums over all final on-shell states $X$, and the squared matrix element is given by the product of two form factors\footnote{Here and in the following, we understand that prefactors $(2\pi)^D\delta^{(D)}(q-p_X)$ ensuring momentum conservation have been stripped off from the form factors.}
\begin{equation}
\ncsusyF_{{\cal O},X} = \langle X| {\cal O}(0) | 0 \rangle \pnt
\end{equation}

The form factor has the perturbative expansion 
\begin{equation}
\ncsusyF_{{\cal O},X} =  \sum_{\ell=0}^\infty g^{2\ell} \ncsusyF_{{\cal O},X}^{(\ell)}\col
\end{equation}
where $g$ is the parameter of the loop expansion. 
Concretely, in $\mathcal{N}=4$ SYM theory in the modified dimensional reduction 
($\overline{\text{DR}}$) scheme, 
the coupling constant is given in \eqref{gdef}.
Moreover, the summation over all final states $X$ in \eqref{ImPi-sigma}
involves in particular a summation over the number $n$ of particles in the 
final state, i.e.\ of the $n$-point form factors $\ncsusyF_{{\cal O},n}^{(\ell)}$
over $n$. 
The number $n$ is directly related to powers of the Yang-Mills coupling constant $g_\YM$.
In analogy to amplitudes (see e.g. \cite{Mangano:1990by}), the $n$-point form factors
possess a decomposition in terms of the possible color structures as
\begin{equation}
\begin{aligned}
\ncsusyF_{{\cal O},n}^{(\ell)}(\{a_i, p_i, \eta_i\}) & = g_\YM^{n-2} \sum_{\sigma\in S_n/Z_n} \tr(\T^{a_{\sigma(1)}} \cdots \T^{a_{\sigma(n)}}) \susyF_{{\cal O},n}^{(\ell)}(\{p_{\sigma(i)}, \eta_{\sigma(i)}\})\\ & \,{}\phantom{=}{} + \text{multi-trace terms} \col
\label{eq: non-color-ordered-super-FF}
\end{aligned}
\end{equation}
where $\T^{a}$,  $a=1,\dots,N_{\text{c}}^2-1$, are the gauge-group generators 
of $SU(N_{\text{c}})$ normalized as
\begin{equation}
\tr(\T^a \T^b) = \delta^{ab}
\pnt
\end{equation}
In \eqref{eq: non-color-ordered-super-FF}, the $i^\text{th}$ particle, $i=1,\dots,n$, 
with momentum $p_i$ carries the adjoint gauge-group index $a_i$. 
Via Nair's superfield \eqref{ossuperfield},
its flavor and helicity are
encoded in terms of the Grassmann variables $\eta_i$, on which the 
color-ordered super form factors $\susyF_{{\cal O},n}^{(\ell)}$ on the rhs.\ 
also depend.

The imaginary part of  \eqref{Pi(q^2)} can be obtained by taking the 
discontinuity, which for timelike ($q^2>0$) momentum 
reads\footnote{Our conventions for the $i0$ description are given in appendix \ref{app:FTprop}.}
\begin{equation}
2 i \, \im {(-q^2-i0)^{x} }=(-q^2-i0)^{x}-(-q^2+i0)^{x}= \frac{2 \pi i}{\Gamma(x) \Gamma(1-x)}(q^2)^x \pnt  
\label{take-Im}
\end{equation}
Using this relation in order to determine the imaginary part of \eqref{Pi(q^2)}
and then inserting the result into \eqref{ImPi-sigma} yields
\begin{equation}\label{sigmaratio}
{\sigma_{\mathcal{O},\ren} \over \sigma^{(0)}_{\mathcal{O}} } = {M(g) \over M^{(0)}} {\Gamma(\Delta_{\mathcal{O}}^{(0)}) \Gamma({D\over2} - \Delta_{\mathcal{O}}) \over \Gamma(\Delta_{\mathcal{O}}) \Gamma({D\over2} - \Delta_{\mathcal{O}}^{(0)})} {\Gamma(\Delta_{\mathcal{O}}^{(0)}-{D\over2}) \Gamma(1+{D\over2} - \Delta_{\mathcal{O}}^{(0)}) \over \Gamma(\Delta_{\mathcal{O}}-{D\over2}) \Gamma(1+{D\over2} - \Delta_{\mathcal{O}}) } \Big(\frac{q^2}{4\mu^2}\Big)^{\gamma_{\mathcal{O}}}  \col
\end{equation}
where we have divided $\sigma_{\mathcal{O},\ren}$ by its classical part $\sigma_{\mathcal{O}}^{(0)} = \im[2i\,\tilde G_{2\mathcal{O},\ren}^{(0)}(q^2)]$.
Indeed, as mentioned at the end of the previous subsection, both $\sigma^{(0)}_{\mathcal{O}}$ and $\sigma_{\mathcal{O},\ren}$ are free of $\frac{1}{\peps}$-poles, since the poles are canceled by the extra $\Gamma$-functions introduced via \eqref{take-Im}. This can also directly be seen for the bubble integral in \eqref{eq: 2ptcorrelator-posi-mom}: its imaginary part is obtained by applying a double-cut, which just yields a finite constant.

By taking the logarithm of \eqref{sigmaratio}, we can expose the 
dependence on 
$q^2$ as follows:
\begin{equation}\label{logsigmarengeneral}
\log \bigg({\sigma_{\mathcal{O},\ren} \over \sigma^{(0)}_{\mathcal{O}} }\bigg) =  \gamma_{\mathcal{O}} \log\frac{q^2}{\mu^2} + C + {\cal O}(\peps) \col
\end{equation}
where the constant $C$ is scale-independent but depends on $\gamma_{\mathcal{O}}$
and the expansion coefficients of the normalization factor \eqref{Mexpansion} as
\begin{equation}
\label{eq:C-expansion}
\begin{aligned}
C={}&g^2\bigg(\frac{M^{(1)}}{M^{(0)}}-(1-2\gamma_\E)\gamma^{(1)}\bigg)\\&+g^4\bigg(\frac{M^{(2)}}{M^{(0)}}-\frac{1}{2}\bigg(\frac{M^{(1)}}{M^{(0)}}\bigg)^2+\frac{3-\pi^2}{6}\big(\gamma^{(1)}\big)^2-(1-2\gamma_\E)\gamma^{(2)}\bigg)+\mathcal{O}(g^6)
\pnt
\end{aligned}
\end{equation}  
It is also renormalization-scheme-dependent as discussed at the end of section
\ref{sec:CS}.
However, the $\log q^2$ term is universal and scheme-independent.  
The anomalous dimension is given by the coefficient of $\log\frac{q^2}{\mu^2}$.
In this paper, we will verify this structure for the Konishi operator  up to two loops.

\subsection*{Strategy of computing cross sections \label{sec:cs-general}}

The cross section is obtained from \eqref{ImPi-sigma} in more detail as follows:
\begin{equation}
\sigma =  \sum_n \int \de {\rm PS}_n  \underbrace{ \sum_{\rm colors}  \sum_{\substack{{\rm spins}\\{\rm helicities}}}  \left\{ \FDff[legs=5,scaled=1.1,fancy] ~ \FDff[legs=5,scaled=1.1,fancy,reflected] \right\} }_{{\cal M}_n}   \pnt
\label{eq:cs-picture}
\end{equation}
This relation holds for both, the bare and the renormalized cross section, 
if $\ncsusyF_n$ represents the bare and the renormalized form factors, respectively.
The evaluation of \eqref{eq:cs-picture} requires three main steps:
(1) determining the form factors $\ncsusyF_n$, (2) taking the absolute square of $\ncsusyF_n$, and (3) performing the $n$-particle phase-space integrals.
More concretely, \eqref{eq:cs-picture} is expanded in powers of $g$ as follows:
\begin{equation}
\label{eq:sigma-expansion-general}
\sigma  =  \sum_{\ell=0}^\infty g^{2 \ell} \sigma^{(\ell)} \col \qquad \sigma^{(\ell)} = \sum_{n=2}^{\ell+2}  g^{2(2-n)} \int \de {\rm PS}_n {\cal M}^{(\ell+2-n)}_{n} \col
\end{equation}
where the squared matrix elements are given by
\begin{equation}
{\cal M}_n^{(\ell)}  =  {1\over n!} \sum_{a_i} \int \prod_{i=1}^n \de^4 \eta_i  \sum_{k=0}^{m}  \sum_{l=0}^\ell  \ncsusyF_{{\cal O},n}^{ {\scriptscriptstyle{\rm N}}^k\scriptscriptstyle{\rm MHV},(l)}(\{a_i, p_i, \eta_i\}) \ncsusyF_{\bar {\cal O},n}^{*, \scriptscriptstyle{\rm N}^{m-k}\scriptscriptstyle{\rm MHV},(\ell-l)}(\{a_i, p_i, \eta_i\}) \col 
\label{calM_n^ell}
\end{equation}
in which $\ncsusyF_n^{(\ell)}(\{a_i, p_i, \eta_i\})$ is the $\ell$-loop $n$-point non-color-ordered super form factor defined in \eqref{eq: non-color-ordered-super-FF}, and 
$\ncsusyF^*_n(\{a_i, p_i, \eta_i\})$ is its complex conjugate.%
\footnote{Note that in \eqref{calM_n^ell} the complex conjugate of tree-level form factors is already encoded in replacing ${\cal O}$ by its conjugate $\bar{\cal O}$ and
changing the MHV degree from $k$ to $m-k$. Therefore, the `*' refers to taking the conjugate of the $\ell\ge1$ contributions only. This will be explained in explicit examples in section \ref{sec:CS}; see the discussion around \eqref{eq:taking-real-q^2}.}
Moreover, $k$ in $\text{N}^k\text{MHV}$ is called the MHV degree, which refers
to terms in $\ncsusyF_n^{(\ell)}$ with a specific degree in $\eta$. For the BPS and Konishi operator considered in this paper, the MHV form factors have degree 4 in $\eta$ and $m=n-2$ is fixed.
The squared matrix element involves sums over all numbers $n$ and types of external particles as well as their color degrees of freedom. The sum over the types of particles is given in terms of integrations over 
the fermionic variables $\eta_i^A$, $A=1,2,3,4$, and a sum over the MHV degree $k$.

Given the squared matrix elements, as a next step, the integration over 
the phase space of the $n$ particles in the final state has 
to be performed. The respective measure is given by
\begin{equation}
\de {\rm PS}_n = \bigg(\prod_{i=1}^n {\de^D p_i \over (2\pi)^D}  2\pi  \delta_+(p_i^2) \bigg) (2\pi)^D \delta^{(D)} \Big(q -\sum_{i=1}^n p_i \Big) \col 
\label{PS-def}
\end{equation}
where $\delta_+(p^2)= \delta(p^2) \theta(p_0)$ with $\theta(p_0)$ being the Heaviside step function which imposes the positivity condition on $p_0$. In appendix \ref{App:phase-space}, we give explicit parametrizations of the two-particle and three-particle phase-space integrals.

Finally, the sum over the different channels, i.e.\ over the different particle numbers $n$, has to be performed. This leads to a cancellation among the different soft and collinear IR divergences 
such that the final result is IR finite. If non-protected operators are involved, as in the Konishi case, their renormalization constants have to be taken into account.

%% file: FFactor.tex
\section{Form factors for \texorpdfstring{$\mathcal{K}_6$}{K6} via unitarity}
\label{sec:ffactor}

In the previous section, we have  defined the cross section for gauge-invariant operators $\mO$ in $\mN=4$ SYM theory in terms of its squared matrix elements. 
As discussed around \eqref{calM_n^ell}, the building blocks of these squared matrix elements are the non-color-ordered super form factors for the respective operator. 
In this section, we will present the building blocks necessary for computing the cross section of the Konishi operator \eqref{Kinso6} up to two loops, which are the two-point form factor up to two-loop order and the three-point form factor at one-loop order.%
\footnote{The tree-level four-point Konishi form factor essentially agrees with the BPS result, as we will discuss in subsection \ref{subsec: Konishi tree level}.}

We use the notation $\ncsusyF_{{\cal O},n}^{(\ell)}(\{a_i, p_i, \eta_i\})$ for the non-color-ordered super form factors  and $\susyF^{(\ell)}_{{\cal O},n}(\{p_i, \eta_i\})$ for the color-ordered super form factors, as introduced in  \eqref{eq: non-color-ordered-super-FF}. 
We denote the bosonic color-ordered form factors with fixed external states by $\bosoF^{(\ell)}_{{\cal O},n}(\{p_i\})$. If necessary, we
specify the external states by subscripts, e.g.\ in case of two scalars and one gluon as $\bosoF^{(\ell)}_{{\cal O}}(1_\phi, 2_\phi, 3_g)$ or simply $\bosoF^{(\ell)}_{{\cal O}, (\phi,\phi,g)}$.
These bosonic form factors can be obtained from $\susyF^{(\ell)}_{{\cal O},n}(\{p_i, \eta_i\})$ by taking a specific term in the $\eta_i$ expansion. 
We also introduce the normalized bosonic form factors $f_{{\cal O},n}^{(\ell)}$ as the ratio between the $\ell$-loop and tree-level color-ordered bosonic form factors:
\begin{equation}
f_{{\cal O},n}^{(\ell)}(\{p_i\}) \,=\, { F^{(\ell)}_{{\cal O},n}(\{p_i\}) \over F^{(0)}_{{\cal O},n}(\{p_i\}) } \pnt
\label{normFF}
\end{equation}
Our computation will focus on the colored-ordered form factors; via \eqref{eq: non-color-ordered-super-FF},
it is straightforward to obtain the full non-color-ordered super form factor from them.

The computation of form factors in this section are based on the on-shell superspace formulation \eqref{ossuperfield}. Therefore, the operator 
in the form factor is $\mK_6$ defined in \eqref{Kinsu4} and \emph{not} the 
Konishi operator $\mathcal{K}$ defined in \eqref{Kinso6}.
We denote the resulting form factors by $F_{\mK_6, n}^{(\ell)}$. 
In order to obtain the Konishi form factors $F_{\mK, n}^{(\ell)}$, we have to modify the results presented in this section, as
will be discussed in detail in section \ref{sec: subtleties in unitarity}.

\subsection{Some BPS form factor results}

We start by presenting some known results for BPS form factors, which are also useful building blocks for the Konishi form factors.
Unless otherwise specified, the BPS form factor in this paper will always refer to that of the half-BPS operator $\tr(\phi_{AB}^2)$ defined in \eqref{BPSsu4}, 
and we use the abbreviation $\Fco_{{\BPS,n}}^{(\ell)} = \Fco_{{\tr(\phi_{AB}^2)},n}^{(\ell)}$.

The $n$-point MHV tree-level BPS super form factor is given by \cite{Brandhuber:2011tv}
\beq
\Fco_{{\BPS,n}}^{(0), \text{MHV}}(1,2,\ldots ,n)=\frac{\Gd^{(4) AB}(\sum_{i=1}^n\Gl_{i}\eta_{i})}{\ket(1,2)\ket(2,3)\ldots\ket(n,1)} \col
\label{BPS-MHV-tree}
\eeq
where $\Gd^{(4)AB}(\sum_i\lambda_i \eta_i)$ is understood as taking $\eta$ in the delta function with only $A, B$ indices, or more explicitly
\beq
\Gd^{(4)AB}(\sum_i\lambda_i \eta_i) = \Big(\sum_{i<j} \ket(i,j)\eta_i^A \eta_j^A \Big) \Big(\sum_{k<l} \ket(k,l)\eta_k^B \eta_l^B \Big) \pnt
\eeq
Note that in this and all following expressions for form factors we do not explicitly write the momentum-conserving delta function $(2\pi)^4\Gd^{(4)}(q-\sum_{i=1}^n p_i)$, where $q$ is the four-momentum carried by the gauge-invariant operator.

We give the loop corrections to the BPS MHV form factor in terms of the normalized form factor defined in \eqref{normFF}. In this paper, we only need the following three results \cite{vanNeerven:1985ja, Brandhuber:2010ad, Bork:2010wf}: 
\begin{align}
f_{\BPS,2}^{(1)} &= -2 s_{12} \FDinline[triangle, twolabels, labelone=p_1, labeltwo=p_2] \col
\label{F2-BPS-1loop}\\
f_{\BPS,2}^{(2)} &= {s_{12}^2  \left( 4
\FDinline[rainbow, twolabels, labelone=p_1,labeltwo=p_2] + 
\FDinline[rainbownonplanar, twolabels, labelone=p_1,labeltwo=p_2] \right)} \col
\label{F2-BPS-2loop}\\
f_{\BPS,3}^{(1)}  &= -{s_{12}s_{23} \over2} 
\FDinline[box,threelabels,labelone=p_1,labeltwo=p_2,labelthree=p_3] - {s_{13}+s_{23} \over2} \FDinline[trianglethreetop,threelabels,labelone=p_1,labeltwo=p_2,labelthree=p_3] - {s_{12}+s_{31} \over2}
\FDinline[trianglethreebot,threelabels,labelone=p_1,labeltwo=p_2,labelthree=p_3] 
\nonumber\\
&\phantom{{}={}} + \text{cyclic perm.\ of }  \{p_1,p_2,p_3\} \col
\label{F3-BPS-1loop}
\end{align}
where $s_{ij\dots k}=(p_i+p_j+\dots+p_k)^2$.
Each graph corresponds to a Feynman integral which is defined in appendix \ref{app:integrals}. 
Throughout this paper, all external on-shell momenta $p_i$ are understood as outgoing.

For the two-point case only the MHV configuration exists, while at three points there are the MHV and the next-to-MHV (NMHV) configuration. The NMHV tree-level form factor can be obtained from \eqref{BPS-MHV-tree} by first taking the conjugation $\lambda \rightarrow \tilde\lambda$ and $\eta^A \rightarrow \tilde\eta_A$, and then applying a fermionic Fourier transformation as%
\footnote{Recall that the operator also becomes the conjugate one, $\tr(({\phi}^{AB})^2)$, where $\phi^{AB} = {1\over2}\teps^{ABCD}\phi_{CD}$.}
\beq
\Fco_{{{\overline\BPS},3}}^{(0), \text{NMHV}}(1,2,3)
=\bigg(\prod_{i=1}^3 \int \de^4 \tilde\eta_i  \e^{\eta_i^C \tilde\eta_{i,C}}\bigg) 
\frac{\Gd^{(4)}_{AB}(\sum_{j=1}^3\tilde\lambda_{j} \tilde\eta_{j})}{\bra(1,2) \bra(2,3) \bra(3,1)} \pnt
\label{BPS-3pt-NMHV-tree}
\eeq
The loop correction to both, the MHV and the NHMV three-point form factor, is given by \eqref{F3-BPS-1loop}.

\subsection{Tree-level two- and three-point form factors}
\label{subsec: Konishi tree level}

We now turn to the form factors of $\mK_6$. 
In this subsection, we consider its tree-level form factors. They are identical to those of the Konishi operator $\mK$. 
The expression for $\mK_6$ in \eqref{Kinsu4} 
contains the individual fields
$\phi_{AB}\phi_{CD}$ where $A,B,C,D$ assume distinct values instead of $\phi_{AB}^2$ as is the case for the BPS operator. For the tree-level bosonic form factor with specified external particles, however, the index structure of the external scalars and fermions do not play any role in the result, which is obvious from the Feynman diagram computation. Therefore, the tree-level bosonic form factors for the Konishi operator are identical to the corresponding BPS form factors.  

The super form factors, on the other hand, take different forms. Taking into account all the components, the two-point super form factor reads%
\footnote{The normalization factor is fixed to be consistent with the definition of the operator $\mK_6$ in \eqref{Kinsu4}.}
\beq
\bea
\Fco_{\mK_6}^{(0)}(1,2)&=-\frac{1}{4}\frac{\ket(1,2)^2}{\ket(1,2)\ket(2,1)} \teps_{ABCD}
(\eta_1^A \eta_1^B)(\eta_2^C \eta_2^D) \col
\label{Kon2pt}
\eea
\eeq
where $\teps_{1234}=1$.
The bosonic two-point form factor 
\beq
\bea
F_{\mK_6}^{(0)}(1_{\phi_{12}}, 2_{\phi_{34}})= - \frac{\ket(1,2)^2}{\ket(1,2)\ket(2,1)} = 1
\label{KF-boso-2pt-tree}
\eea
\eeq
can be obtained by taking the 
$(\eta_1^1 \eta_1^2)(\eta_2^3 \eta_2^4)$ component of the tree-level form factor $ \Fco_{\mK_6 , 2}^{(0)}$ in \eqref{Kon2pt}; 
it is identical to the BPS result as can be seen by taking the 
$(\eta_1^A \eta_1^B)(\eta_2^A \eta_2^B)$ component of \eqref{BPS-MHV-tree} at $n=2$.
There are two other possible scalar field configurations at the external legs, namely  $\{(\phi_{13},\phi_{24}),(\phi_{14},\phi_{23})\}$, and for both these cases we obtain the same bosonic form factor as above.

The three-point MHV super form factor is given by the following expression: 
 \beq
 \bea
\Fco_{\mK_6}^{(0)}(1,2,3)&=\frac{-1}{4\ket(1,2)\ket(2,3)\ket(3,1)}\Big(\ket(1,2)^2\teps_{ABCD} (\eta_1^A \eta_1^B)(\eta_2^C \eta_2^D) \\
&\hphantom{{}={}\frac{-1}{4\ket(1,2)\ket(2,3)\ket(3,1)}\Big(}
+2\ket(1,3)\ket(2,3)\teps_{ABCD} \eta_1^{A} \eta_2^B (\eta_3^C \eta_3^D) +\text{cyclic perm.}\Big) \pnt
\label{Kon3pt}
\eea
\eeq
It has two distinct configurations of the external states: scalar-scalar-gluon and fermion-fermion-scalar.
Taking the coefficients of 
$(\eta_1^1 \eta_1^2)(\eta_2^3 \eta_2^4)$ and $\eta_1^{1} \eta_2^2 (\eta_3^3 \eta_3^4)$, we find
\beq
\Fb_{\mK_6}^{(0)}(1_{\phi_{12}}, 2_{\phi_{34}}, 3_{g_+})= -\frac{\ket(1,2)^2}{\ket(1,2)\ket(2,3)\ket(3,1)} \col \qquad 
\Fb_{\mK_6}^{(0)}(1_{\psi_{1}}, 2_{\psi_{2}}, 3_{\phi_{34}})= -\frac{\ket(1,3) \ket(2,3)}{\ket(1,2)\ket(2,3)\ket(3,1)} \col
\label{KF-boso-3pt-tree}
\eeq
which are also identical to the corresponding BPS form factors.
The NMHV form factor can be obtained from the MHV result in a similar way as in the BPS case \eqref{BPS-3pt-NMHV-tree}.

\subsection{One-loop two-point form factor}

In this and the following subsection, we compute the form factor of $\mK_6$ at one- and two-loop level via four-dimensional unitarity 
\cite{Bern:1994zx,Bern:1994cg}. 

The general idea of unitarity in this context is to reconstruct loop corrections to the form factors at the integrand level from their discontinuities, i.e.\ by applying cuts. Here, a cut denotes setting a propagator on-shell according to 
\begin{equation}
 \frac{i}{l_i^2}\to 2\pi \delta_+(l_i^2)\col
\end{equation}
where $\delta_+(l_i^2)$ was defined after \eqref{PS-def}.
On the cut, the loop expression factorizes into a product of (known) tree-level or lower-loop form factors and amplitudes.
These have to be summed over all possible particles exchanged  in the cut channel, which can be achieved by integrating the super form factors as well as the super amplitudes over the Grassmannian degrees of freedom in the cut legs.
Then, one can apply the spinor algebra to write the result in a form that can be identified as a sum of cut integrals. 
In this way, an ansatz for the uncut integrals occurring in the loop correction is assembled. In general, not all integrals appear in a given cut, and additional cuts have to be taken to complement the ansatz. The complete ansatz has to be consistent with all possible cut. 
Finally, the cut integrals have to be lifted to the uncut integrals, as discussed in appendix \ref{app:integrals}.

In the following, we apply this technique to the form factor of $\mK_6$. We
start with 
the computation of the one-loop two-point form factor.
%
\begin{figure}[tbp]
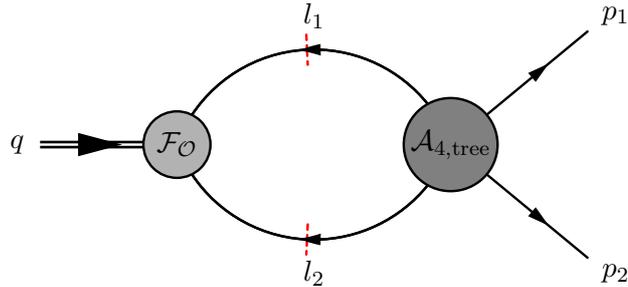

 \centering
$
\settoheight{\eqoff}{$\times$}%
\setlength{\eqoff}{0.5\eqoff}%
\addtolength{\eqoff}{-12.0\unitlength}%
\raisebox{\eqoff}{%
\fmfframe(2,2)(2,2){%
\begin{fmfchar*}(80,30)
\fmfleft{vq}
\fmfright{vp2,vp1}
\fmf{dbl_plain_arrow,tension=2}{vq,v1}
\fmf{plain_arrow,tension=0.5,left=0.7,label=$l_2$,l.s=left,l.d=8}{v2,v1}
\fmf{plain_arrow,tension=0.5,right=0.7,label=$l_1$,l.s=right,l.d=8}{v2,v1}
\fmf{phantom_smallcut,left=0.7,tension=0}{v1,v2}
\fmf{phantom_smallcut,right=0.7,tension=0}{v1,v2}
\fmf{plain_arrow}{v2,vp1}
\fmf{plain_arrow}{v2,vp2}
\fmfv{decor.shape=circle,decor.filled=30,decor.size=25,label=$\mathcal{F}_{\mathcal{O}}$,label.dist=0}{v1}
\fmfv{decor.shape=circle,decor.filled=50,decor.size=35,label=$\mathcal{A}_{4,,\text{tree}}$,label.dist=0}{v2}
\fmffreeze
 \fmfcmd{pair vertq, vertpone, vertptwo, vertone, verttwo; vertone = vloc(__v1); verttwo = vloc(__v2); vertq = vloc(__vq); vertpone = vloc(__vp1); vertptwo = vloc(__vp2); }
 \fmfiv{label=$q$}{vertq}
 \fmfiv{label=$p_1$}{vertpone}
 \fmfiv{label=$p_2$}{vertptwo}
\end{fmfchar*}%
}}%
$
\caption{The simple $(p_1+p_2)^2$ double cut.}
\label{fig: simple double cut}
\end{figure}
For the sake of explicitness,
we choose a fixed combination of external scalar states, namely $\{\phi_{12},\phi_{34}\}$.
As in the tree-level case, the other two choices of external scalars  $\{\phi_{13},\phi_{24}\}$ and $\{\phi_{14},\phi_{23}\}$ lead to the same result. 
We abbreviate $F_{{\cal K}_6}^{(\ell)}(1_{\phi_{12}}, 2_{\phi_{34}})$ as $F_{{\cal K}_6, (\phi, \phi)}^{(\ell)}$.

Only one cut needs to be considered: the two-particle cut in the channel $(p_1+p_2)^2=q^2$.\footnote{The other two two-particle cuts occur in the $p_1^2$ and $p_2^2$ channels. Since these legs have $p_1^2=p_2^2=0$, massless bubble integrals in these channels vanish identically when regularized in $D=4-2\peps$ dimensions. Hence, all integrals can be detected by the $q^2$-cut.}
It cuts the internal propagators carrying momenta $l_1$ and $l_2$ as shown in figure \ref{fig: simple double cut}. 
The building blocks on the two sides of the cut are the color-ordered two-point form factor \eqref{Kon2pt} and the color-ordered four-point MHV amplitude given in the standard MHV form \cite{Witten:2003nn} 
as\footnote{Recall that we are always suppressing the momentum-conserving delta function in the notation.}
\beq
\mA_{n}^{(0)}=i \frac{\Gd^{(8)}(\sum_{i=1}^{n}\lambda_i \eta_i)}{\ket(1,2)\ket(2,3)\ldots \ket(n,1)} \pnt
\label{Amhvtree}
\eeq
The sum over all possible particles exchanged along the cut is considered by integrating over the fermionic coordinates of the exchanged particles as $\int \deta =\int \de ^4\eta_{l_1}\de^4 \eta_{l_2}$ while keeping the external state fixed.

The $q^2$-cut integral reads\footnote{For reversed momenta $l \rightarrow - l$, occurring e.g.\ in $\Fco_{\mK_6 , 2}^{(0)}(-l_1,-l_2)$, we follow the convention $\lambda_l \rightarrow -\lambda_l$ and $\tilde\lambda_l \rightarrow \tilde\lambda_l$  in the spinor helicity formalism.}
\beq
\bea
F_{\mK_6 , (\phi, \phi)}^{(1)} \Big|_{q^2}&=\twops \deta\Fco_{\mK_6 , 2}^{(0)}(-l_1,-l_2)\times \mA_4^{(0)}(p_1,p_2,l_2,l_1)\Big|_{(\eta_1^1 \eta_1^2)(\eta_2^3 \eta_2^4)} \\
&=F_{\mK_6 , (\phi, \phi)}^{(0)} \, i \, \underbrace{ \twops \frac{\ket(l_1,2)^2\ket(l_2,1)^2+4\ket(l_1,1)\ket(l_1,2)\ket(l_2,1)\ket(l_2,2)+\ket(l_1,1)^2\ket(l_2,2)^2}{\ket(l_1,1)\ket(1,2)\ket(2,l_2)\ket(l_2,l_1)}}_{\mC_{ (\phi, \phi)}} \pnt
\label{eq: C_phiphi-def}
\eea
\eeq
Since the external states are fixed to be $\{\phi_{12},\phi_{34}\}$, we take the ${(\eta_1^1 \eta_1^2)(\eta_2^3 \eta_2^4)}$ component of the cut integrand.
The phase-space integration measure, $\de {\rm PS}_{2 , \{l\}}$, is defined according to \eqref{PS-def}, with the integration variables being the momenta of the cut propagators $\{ l_1,l_2\} $; hence the subscript in the notation for $\de {\rm PS}_{2, \{l\}}$.

The cut integral can be simplified at the integrand level as\footnote{The first line can be obtained via the Schouten identity $\ket(a,b)\ket(c,d) = \ket(a,c)\ket(b,d)+\ket(a,d)\ket(c,b)$ for $\ket(l_1,2)\ket(l_2,1)$ in \eqref{eq: C_phiphi-def}.}
\beq
\bea
\mC_{ (\phi, \phi)}&= \twops \left(\frac{\ket(l_1,l_2)\ket(1,2)}{\ket(l_1,1)\ket(l_2,2)}+6\frac{\ket(l_1,2)\ket(l_2,1)}{\ket(1,2)\ket(l_1,l_2)} \right)
= \twops\left(\frac{-s_{12}}{(l_1+p_1)^2} +6 \frac{(l_1+p_2)^2}{s_{12}}\right)\\
&=- \, s_{12}  \FDinline[triangle, doublecut,cutlabels, twolabels, labelone=p_1, labeltwo=p_2] + \, 6{}{}\frac{(l_1+p_2)^2}{s_{12}} \FDinline[bubble, doublecut,cutlabels, twolabels, labelone=p_1, labeltwo=p_2] \col
\label{Konishicut1loop}
\eea
\eeq
where the flow of the momenta is as specified in figure \ref{fig: simple double cut}. 
In the above equation, the integral over the two-particle phase space is shown by the dashed cut line of the triangle and bubble graph. For the triangle graph, the denominator in the integrand is the uncut propagator $\frac{1}{(l_1+p_1)^2}$ and the numerator coefficient is $-s_{12}$.
The shown bubble graph has no uncut propagator, but is has a loop-momentum-dependent numerator factor, which is written in front of the graph.

As described in appendix \ref{app:integrals}, the cut integrals \eqref{Konishicut1loop}
can be lifted to the full integrals. The full normalized form factor as defined in \eqref{normFF} then becomes\footnote{The coupling dependence can be recovered as shown in appendix \ref{app:integrals}.} 
\beq
\bea
f_{\mK_6, (\phi,\phi)}^{(1)} &=2 \left(-s_{12} 
\FDinline[triangle, twolabels, labelone=p_1, labeltwo=p_2] +6 \frac{s_{2 l}}{s_{12}} 
\FDinline[bubble, twolabels, labelone=p_1, labeltwo=p_2,momentum]
\right) \col
\label{Konishi1loopfull}
\eea
\eeq
where the factor of $2$ is due to the permutation of the two external legs, and we use the short notation $s_{i l} = (p_i+l)^2$.
Note that the prefactors that depend on the loop momentum are understood to appear in the integrand of the integral represented by the respective graph it multiplies.

The bubble integral with loop momentum in the numerator can be reduced to the scalar bubble integral via Passarino-Veltman (PV) reduction, see appendix \ref{app:PV} for details. Thus, we obtain the form factor\footnote{For convenience, we will from now on refer to the normalized form factor as form factor, too.} 
\beq
\bea
f_{\mK_6, (\phi,\phi)}^{(1)} &= \underbrace{-2s_{12} 
\FDinline[triangle, twolabels, labelone=p_1, labeltwo=p_2] }_{f_{\BPS,2}^{(1)}} 
\,-\, 6 
\FDinline[bubble, twolabels, labelone=p_1, labeltwo=p_2]
\pnt
\label{KF-2pt-1loop-final}
\eea
\eeq
The integrals corresponding to the graphs are given in appendix \ref{app:integrals}.
Note that the contribution to the form factor involving the triangle integral is the same as the BPS form factor $f_{\BPS,2}^{(1)}$ in \eqref{F2-BPS-1loop}.  An independent computation of this result via Feynman diagrams is shown in appendix \ref{app:feyndiag}.

From the above calculation at one-loop, we see that the IR-divergent part of the form factor of $\mK_6$ is the same as the one of the BPS operator. The extra contribution coming from the UV divergent bubble integral yields a non-vanishing anomalous dimension unlike in the BPS case. 
We will equally organize all subsequent results for the form factor in terms of a part that is identical to the BPS form factor and an additional contribution that is unique to the form factor of $\mK_6$.

\subsubsection*{Vanishing one-loop form factors}

Before proceeding to two-loop order, we briefly discuss two other possible form factors with gluon or fermion external states, namely the form factors $F^{(1)}_{\mK_6}(1_{g_-}, 2_{g_+})$ and  $F^{(1)}_{\mK_6}(1_{\psi_1}, 2_{\psi_{234}})$. 
At tree level, they are zero since no Feynman diagram for this configuration exists. 
At higher loops this is not obvious. Here, we use unitarity to show explicitly that they are zero at least at one-loop order.  
Consider the $q^2$-cut as in \eqref{eq: C_phiphi-def}, but to obtain $F^{(1)}_{\mK_6}(1_{g_-}, 2_{g_+})$ and  $F^{(1)}_{\mK_6}(1_{\psi_1}, 2_{\psi_{234}})$ take the components ${\eta_1^1\eta_1^2\eta_1^3 \eta_1^4}$ and ${\eta_1^1(\eta_2^2\eta_2^3 \eta_2^4)}$ of the cut integrand, respectively. 
This yields for the two cases
\begin{align}
F_{\mK_6,(g_-, g_+)}^{(1)} \Big|_{q^2}
&= i\twops \frac{6\ket(l_1,1)\ket(l_2,1)^2}{\ket(1,2)\ket(l_1,l_2)\ket(l_2,2)} 
= - i \, 6 {\langle 1 | l_1 | 2 ]^2 \over s_{12} }   \FDinline[triangle, doublecut,cutlabels, twolabels, labelone=p_1, labeltwo=p_2]  \col 
\label{Konishicut1loop-gg} \\
F_{\mK_6, (\psi_1, \psi_{234})}^{(1)} \Big|_{q^2}
&= i\twops \frac{3\ket(l_1,1)\ket(l_1,2)\ket(l_2,1)^2 + 3 \ket(l_1,1)^2\ket(l_2,1)\ket(l_2,2)}{\ket(1,2)\ket(l_1,l_2)\ket(l_1,1)\ket(l_2,2)} \nonumber\\
&= i \, 3 {\langle 1 | l_1 | 2 ] } \FDinline[triangle, doublecut,cutlabels, twolabels, labelone=p_1, labeltwo=p_2] + i \, 6 {\langle 1 | l_1 | 2 ] \over s_{12} } \FDinline[bubble, doublecut,cutlabels, twolabels, labelone=p_1, labeltwo=p_2] \pnt
\label{Konishicut1loop-psipsi}
\end{align}
When we lift these expressions to the full triangle and bubble integrals and perform the PV reduction, we obtain zero. Since we use four-dimensional unitarity, we also have to check that there is no contribution from potential rational terms. 
A similar (but simpler) study as in appendix \ref{app:check-3pt} shows that
rational terms are indeed absent. 

Finally, there is an easy way to see that $F_{\mK_6}(1_{g_-}, 2_{g_+}) =0$ to all loop orders. Using the gauge freedom, we can choose the polarization vectors of the outgoing gluons as  
$\teps_1^- = \teps_2^+ \propto \lambda_1 \tilde \lambda_2$. It is then obvious that the form factor must be zero, since it is proportional to $\teps_i \cdot p_j$ or $\teps_1 \cdot \teps_2$.

One can also compute $F_{\mK_6, (g, g)}^{(1)}$ directly by using Feynman diagrams. A simple computation gives 
\begin{equation}
\label{eq:Fgg-feynman}
F_{\mK_6, (g, g)}^{(1)} = \bigg[ 2 (\teps_1 \cdot \teps_2) - { (\teps_1 \cdot p_2) (\teps_2 \cdot p_1) \over s_{12} } \bigg]   I_3^D[ \ell_\peps^2 ] \col
\end{equation}
where the integral $I_3^D[ \ell_\peps^2] = {1\over2} + {\cal O}(\peps)$
is given in \eqref{rationaltermtriangle} for $p_2\to 0$ and the relabeling
$p_3\to p_2$. 
This result holds for the polarization vectors $\teps_{1,2}^\pm$ taken to be in general $D=4-2\peps$ dimensions. 
Since $I_3[ \ell_\peps^2]$ is finite and its prefactor is of order ${\cal O}(\peps)$ (as it vanishes when $D=4$), the form factor itself $F^{(1)}_{\mK_6,(g,g)}$ is of order ${\cal O}(\peps)$. This is consistent with the unitarity-based calculation.

\subsection{Two-loop two-point form factor}

Next, we compute the two-loop two-point form factor of $\mK_6$. As in the one-loop case, we specify the external states to be $\{\phi_{12},\phi_{34}\}$.

\subsubsection*{Two-particle cut}

We first study the two-particle cut in the $q^2$-channel. 
We follow a similar procedure as the one being used in computing the BPS form factor \cite{Brandhuber:2012vm}. We first quote the $q^2$-cut integral given in (2.6) of \cite{Brandhuber:2012vm}:\footnote{Note that \eqref{2loopKonecut} applies to any composite operator with two elementary fields, in particular to $\mK_6$.}
\beq
\bea
F_{{\cal O}, 2}^{(2)}\Big|_{q^2} 
&= \twops \deta\Fco_{{\cal O}, 2}^{(0)}(-l_1,-l_2)\Big( 4 \mA_4^{(1)}(p_1,p_2,l_2,l_1) +\mA_4^{(1)}(p_1,l_1,p_2,l_2) \Big)  \col
\label{2loopKonecut}
\eea
\eeq
where the building blocks are the two-point tree-level form factor \eqref{Kon2pt} and the one-loop color-ordered four-point amplitude \cite{Bern:1996je}
\beq
\bea
 \mA_4^{(1)}(p_1,p_2, p_3, p_4)&= \mA_4^{(0)}(p_1,p_2, p_3, p_4)(-\, s_{12} s_{23}) I_4^{(1)}(p_1,p_2,p_3, p_4) \pnt
 \label{MHVoneloop}
\eea
\eeq
The tree-level super amplitude $\mA_4^{(0)}(p_1,p_2, p_3, p_4)$ in \eqref{MHVoneloop} contains all the dependence on the fermionic coordinates, and the term multiplying it is a massless scalar box integral $I_4^{(1)}$ defined in \eqref{oneloopint:box}.\footnote{The minus sign in \eqref{MHVoneloop} is related to the convention of the box integral we use in \eqref{oneloopint:box}.}

\begin{figure}[tbp]
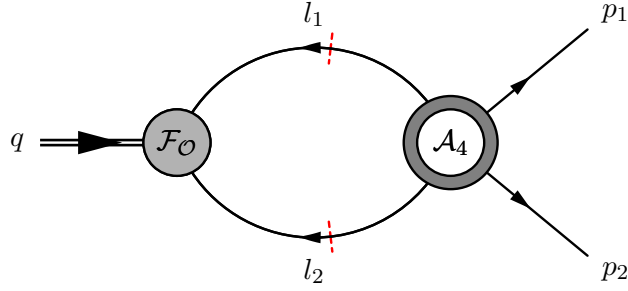

 \centering
$
\settoheight{\eqoff}{$\times$}%
\setlength{\eqoff}{0.5\eqoff}%
\addtolength{\eqoff}{-12.0\unitlength}%
\raisebox{\eqoff}{%
\fmfframe(2,2)(2,2){%
\begin{fmfchar*}(80,30)
\fmfleft{vq}
\fmfright{vp2,vp1}
\fmf{dbl_plain_arrow,tension=2}{vq,v1}
\fmf{plain_arrow,tension=0.5,left=0.7,label=$l_2$,l.s=left,l.d=8}{v2,v1}
\fmf{plain_arrow,tension=0.5,right=0.7,label=$l_1$,l.s=right,l.d=8}{v2,v1}
\fmf{phantom_smallcut,left=0.7,tension=0}{v1,v2}
\fmf{phantom_smallcut,right=0.7,tension=0}{v1,v2}
\fmf{plain_arrow}{v2,vp1}
\fmf{plain_arrow}{v2,vp2}
\fmfv{decor.shape=circle,decor.filled=30,decor.size=25,label=$\mathcal{F}_{\mathcal{O}}$,label.dist=0}{v1}
\fmffreeze
\fmfdraw
 \fmfcmd{pair vertq, vertpone, vertptwo, vertone, verttwo; vertone = vloc(__v1); verttwo = vloc(__v2); vertq = vloc(__vq); vertpone = vloc(__vp1); vertptwo = vloc(__vp2); }
 \fmfiv{label=$q$}{vertq}
 \fmfiv{label=$p_1$}{vertpone}
 \fmfiv{label=$p_2$}{vertptwo}
\fmfiv{decor.shape=circle,decor.filled=50,decor.size=35,label=$\mathcal{A}_{4}$,label.dist=0}{verttwo}
\fmfiv{decor.shape=circle,decor.filled=0,decor.size=25,label=$\mathcal{A}_{4}$,label.dist=0}{verttwo}
\end{fmfchar*}%
}}%
$
\caption{The $(p_1+p_2)^2$ double cut at two loops that contributes to the planar ladder integral. The building blocks are the color-ordered tree-level form factor and the color-ordered one-loop amplitude.}
\label{fig: planar double cut at two loop}
\end{figure}

\begin{figure}[tbp]
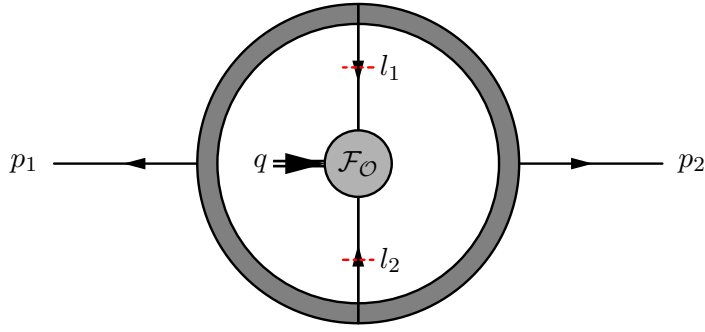

\centering
$
 \settoheight{\eqoff}{$\times$}%
 \setlength{\eqoff}{0.5\eqoff}%
 \addtolength{\eqoff}{-21.0\unitlength}%
 \raisebox{\eqoff}{%
 \fmfframe(2,6)(2,6){%
 \begin{fmfchar*}(80,30)
 \fmfleft{vp1}
 \fmfright{vp2}
 \fmftop{va}
 \fmf{plain_arrow,tension=2}{vl,vp1}
 \fmf{plain_arrow,tension=2}{vr,vp2}
 \fmf{plain,tension=0.5}{v1,vl}
 \fmf{plain,tension=0.5}{v1,vr}
 \fmf{phantom_smallcut,left=0,tension=0}{v1,vl}
 \fmf{phantom_smallcut,right=0,tension=0}{v1,vr}
 \fmf{phantom,tension=0.5,left=1}{vl,vr}
 \fmf{phantom,tension=0.5,right=1}{vl,vr}
 \fmffreeze
 \fmf{phantom,tension=3}{va,vq}
 \fmf{dbl_plain_arrow}{vq,v1}
 \fmffreeze
 \fmfdraw
   \fmfcmd{pair vertq, vertpone, vertptwo, vertone, vertl, vertr;
vertone = vloc(__v1); vertl = vloc(__vl); vertr = vloc(__vr); vertq =
vloc(__vq); vertpone = vloc(__vp1); vertptwo = vloc(__vp2);}
  \fmfiv{label=$q$,l.d=2,l.a=180}{vertone+0.352778(-90,0)}
   \fmfiv{label=$p_1$}{vertpone}
   \fmfiv{label=$p_2$}{vertptwo}
\fmfiv{decor.shape=circle,decor.filled=50,decor.size=120}{vertone}
\fmfiv{decor.shape=circle,decor.filled=0,decor.size=105}{vertone}
\fmfi{dbl_plain_arrow}{(vertone+0.352778(-90,0))--(vertone+0.352778(-35,0))}
 \fmfi{plain_arrow,label=$l_2$,l.s=right,l.d=8}{(vertone-0.352778(0,170))--(vertone-0.352778(0,35))}
 \fmfi{plain_arrow,label=$l_1$,l.s=left,l.d=8}{(vertone+0.352778(0,170))--(vertone+0.352778(0,35))}
 \fmfi{phantom_smallcut}{(vertone-0.352778(0,170))--(vertone-0.352778(0,35))}
 \fmfi{phantom_smallcut}{(vertone+0.352778(0,170))--(vertone+0.352778(0,35))}
    \fmfiv{decor.shape=circle,decor.filled=30,decor.size=25,label=$\mathcal{F}_{\mathcal{O}}$,label.dist=0}{vertone}
\end{fmfchar*}%
}}%
$
\caption{
The $q^2$-cut at two loops that contributes to the crossed ladder integral.  The building blocks are the color-ordered tree-level form factor and the color-ordered one-loop amplitude. }
\label{fig: nonplanar double cut two loop}
\end{figure}

Let us briefly explain \eqref{2loopKonecut}; see \cite{Brandhuber:2012vm} for a derivation in full details. The above cut integral is obtained by taking the product of the two-point form factor and the \emph{non-color-ordered} four-point amplitude. The one-loop four-point amplitude contains a single-trace contribution, as well as a double-trace contribution which is sub-leading in color. However, after the contraction of the color factors with the two-point form factor, both contribute to the cut integral with the single-trace color factor $\delta^{ab} = \tr(\T^a \T^b)$.\footnote{The enhancement of the power in $N_{\text{c}}$ of the apparently suppressed double-trace term in the amplitude is the wrapping effect analyzed earlier for the spectral problem \cite{Sieg:2005kd}.} The final building blocks in the cut integral are the \emph{color-ordered} form factor and amplitude as given in  \eqref{2loopKonecut}. The two contributions in the parentheses of \eqref{2loopKonecut} are depicted in figures \ref{fig: planar double cut at two loop} and \ref{fig: nonplanar double cut two loop} respectively.%
\footnote{The factor 4 in the first term comes from the different contributions of the color factor contraction; two of them come from the single-trace four-point amplitudes, the other two from the double-trace four-point amplitude, as explained in \cite{Brandhuber:2012vm}.
A different way to understand the factor 4 is to look at the two-particle cut with the one-loop form factor on the left hand side and the tree-level amplitude on the right hand side. It then arises from twice applying the reasoning that gave us the factor 2 at one loop.}
We consider them one by one below.

We first study the contribution from the first term in the parentheses of \eqref{2loopKonecut}, which is shown in figure \ref{fig: planar double cut at two loop}. 
Using the one-loop result \eqref{MHVoneloop} for the amplitude and taking the external states to be $\{\phi_{12}, \phi_{34}\}$, the corresponding cut integral can be written as 
\beq
\label{eq: first term in parentheses}
\bea
F_{\mK_6 , (\phi,\phi)}^{(2)}\Big|_{q^2}^{\text{I}}
&= \twops \deta\Fco_{\mK_6 , 2}^{(0)}(-l_1,-l_2)\mA_4^{(0)}(p_1,p_2,l_2,l_1) \Big|_{(\eta_1^1 \eta_1^2)(\eta_2^3 \eta_2^4)} \\
&\hphantom{{}={}\twops \deta}\times \, (-) s_{12} s_{1 l_1} I_4^{(1)}(p_1,p_2,l_2,l_1) \pnt
\eea
\eeq
The first line in \eqref{eq: first term in parentheses} given by the product of tree factors is the same as the cut integrand of the previously studied one-loop case in \eqref{eq: C_phiphi-def}. 
Therefore, we can perform exactly the same calculation as in \eqref{Konishicut1loop} and obtain
\beq
\bea
F_{\mK_6 , (\phi,\phi)}^{(2)}\Big|_{q^2}^{\text{I}}
& = - F_{\mK_6 , (\phi,\phi)}^{(0)} \,  \twops
\Big(\frac{-s_{12}}{s_{1 l_1}}+6\frac{s_{2 l_1}}{s_{12}}\Big)
s_{12} s_{1 l_1}  I_4^{(1)}(p_1,p_2,l_2,l_1)\\
& =F_{\mK_6 , (\phi,\phi)}^{(0)} \Big( s_{12}^2   - 6  s_{1 l_1} s_{2 l_1} \Big)
\FDinline[rainbow, doublecut,cutlabels, twolabels, labelone=p_1, labeltwo=p_2]  \pnt
\label{PL2loopcut}
\eea
\eeq

The above cut integral can be lifted to the two-loop planar ladder integral. 
This integral can be drawn in two different ways, namely 
\beq
\FDinline[rainbow,big,twolabels,labelone=p_1,labeltwo=p_2]\col
\hskip -0.8cm \FDinline[fliprainbow,big,twolabels,labelone=p_1,labeltwo=p_2] \pnt
\label{diagPL}
\eeq
Furthermore, there are two other planar graphs obtained by permuting the external legs  $p_1 \leftrightarrow p_2$.
So, altogether we have four diagrams which are drawn in different ways but 
all give equivalent planar ladder integrals. 
This provides a diagrammatic interpretation of the factor 4 in the first term of \eqref{2loopKonecut}. As we will see later in the triple cut, it is also important to separately draw the ladder graphs in different ways according to \eqref{diagPL} in order to compute the cut integrand correctly.

Next, we consider the second term inside the parentheses in \eqref{2loopKonecut}, which is depicted in figure \ref{fig: nonplanar double cut two loop}. The corresponding cut integral is given by
\beq
\bea
F_{\mK_6 , (\phi,\phi)}^{(2)}\Big|_{q^2}^{\text{II}}
&= -\twops \deta\Fco_{\mK_6 , 2}^{(0)}(-l_1,-l_2)\mA_4^{(0)}(p_1,l_1,p_2,l_2)\\
&\hphantom{{}={}\twops \deta}\times s_{12} s_{1 l_1} I_4^{(1)}(p_1,l_1,p_2,l_2) \Big|_{(\eta_1^1 \eta_1^2)(\eta_2^3 \eta_2^4)} \pnt
\label{F2loopCL}
\eea
\eeq
Following similar steps as in the previous case, the cut integral is expressed as a two-particle cut of the two-loop crossed ladder integral as shown below
\beq
\bea
F_{\mK_6 , (\phi,\phi)}^{(2)}\Big|_{q^2}^{\text{II}}=F_{\mK_6, (\phi,\phi)}^{(0)} \Big( s_{12}^2  - 6 s_{1 l_1} s_{2 l_1} \Big) \FDinline[fliprainbownonplanar, doublecut, big, cutlabels,twolabels,labelone=p_1,labeltwo=p_2, splitlabels]
\pnt
\label{2loop2case2}
\eea
\eeq
Now, lifting the cut integrals of the combined contributions \eqref{2loop2case2} and \eqref{PL2loopcut} as described in appendix \ref{app:integrals}, we find the following contribution to the two-loop form factor of $\mK_6$:
\beq
\bea
&4 f_{\mK_6, (\phi,\phi)}^{(2),{\text{I}}}+f_{\mK_6, (\phi,\phi)}^{(2),{\text{II}}} =  \Big( s^2_{12} - 6 s_{1 l} s_{2 l} \Big)
\left( 4 
\FDinline[rainbow,momentum,twolabels,labelone=p_1,labeltwo=p_2] +\FDinline[rainbownonplanar,momentum,twolabels,labelone=p_1,labeltwo=p_2] \right) \col
\label{doublecut2loopresult}
\eea
\eeq
where, as explained around \eqref{diagPL}, the factor $4$ is included for $f_{\mK_6, (\phi,\phi)}^{(2),{\text{I}}}$.

There is another $q^2$-cut which is similar to the one in figure \ref{fig: planar double cut at two loop}. It has the one-loop two-point form factor on the left hand side and the tree-level four-point amplitude on the right hand side.  This case is a bit subtle.  Naively, one would expect that only the one-loop two-point form factor with scalar external states $F_{\mK_6 , (\phi,\phi)}^{(1)}$ can occur on the left hand side, since the other possibly contributing 
form factors with gluon and fermion external states $F_{\mK_6,(g_-, g_+)}^{(1)}$
and $F_{\mK_6,(\psi_1, \psi_{234})}^{(1)}$, respectively, vanish as shown in the previous subsection.  However, this expectation turns out to be incorrect. There are non-vanishing contributions from these two cases: 
only the integrated one-loop form factors are zero, but the integrands are not, 
as we can see from \eqref{Konishicut1loop-gg} and \eqref{Konishicut1loop-psipsi}. 
Their integrands have to be taken into account in the unitarity cuts and 
then yield a result which is consistent with the one found from the 
$q^2$-cut of figure \ref{fig: planar double cut at two loop}.%
\footnote{Since the non-planar ladder does not contribute to this 
cut, one only obtains the contribution coming from the planar ladder integral in 
\eqref{doublecut2loopresult}.}

The result \eqref{doublecut2loopresult} obtained by using only the two-particle cuts is not guaranteed to give the full form factor.  
One problem is that the numerator coefficients of both integrals
are ambiguous w.r.t.\ terms that are proportional to $l^2$.
Due to the on-shell condition of the cut propagators, such terms are not detected by the double cuts. Moreover, there may be other basis integrals which cannot be detected by the double cuts.  Both these issues can be fixed by studying the three-particle cuts, which we do next.

\subsubsection*{Three-particle cut}

\begin{figure}[tbp]
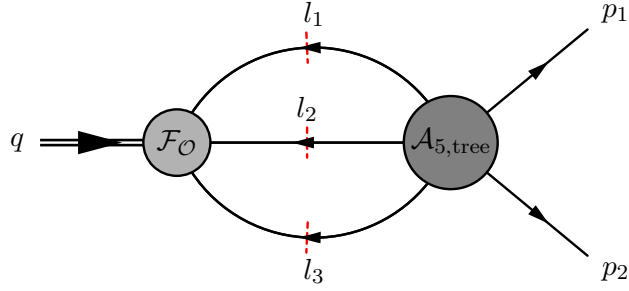

 \centering
$
\settoheight{\eqoff}{$\times$}%
\setlength{\eqoff}{0.5\eqoff}%
\addtolength{\eqoff}{-12.0\unitlength}%
\raisebox{\eqoff}{%
\fmfframe(2,2)(2,2){%
\begin{fmfchar*}(80,30)
\fmfleft{vq}
\fmfright{vp2,vp1}
\fmf{dbl_plain_arrow,tension=2}{vq,v1}
\fmf{plain_arrow,tension=0.333,left=0.7,label=$l_3$,l.s=left,l.d=8}{v2,v1}
\fmf{plain_arrow,tension=0.333,right=0.7,label=$l_1$,l.s=right,l.d=8}{v2,v1}
\fmf{plain_arrow,tension=0.333,right=0.0,label=$l_2$,l.s=right,l.d=8}{v2,v1}
\fmf{phantom_smallcut,left=0.7,tension=0}{v1,v2}
\fmf{phantom_smallcut,right=0.7,tension=0}{v1,v2}
\fmf{phantom_smallcut,right=0,tension=0}{v1,v2}
\fmf{plain_arrow}{v2,vp1}
\fmf{plain_arrow}{v2,vp2}
\fmfv{decor.shape=circle,decor.filled=30,decor.size=25,label=$\mathcal{F}_{\mathcal{O}}$,label.dist=0}{v1}
\fmfv{decor.shape=circle,decor.filled=50,decor.size=35,label=$\mathcal{A}_{5,,\text{tree}}$,label.dist=0}{v2}
\fmffreeze
 \fmfcmd{pair vertq, vertpone, vertptwo, vertone, verttwo; vertone = vloc(__v1); verttwo = vloc(__v2); vertq = vloc(__vq); vertpone = vloc(__vp1); vertptwo = vloc(__vp2); }
 \fmfiv{label=$q$}{vertq}
 \fmfiv{label=$p_1$}{vertpone}
 \fmfiv{label=$p_2$}{vertptwo}
\end{fmfchar*}%
}}%
$
\caption{The two loop $(p_1+p_2)^2=q^2$ triple cut.}
\label{fig: simple triple cut}
\end{figure}

The three-particle cut, or triple cut (TC), across the $q^2$-channel is shown in figure \ref{fig: simple triple cut}. 
Unlike for the BPS form factor, the triple cut will indeed give some new contribution to the form factor of $\mK_6$, which is not detectable by the previous double cut.

The cut integral is given as
\beq
\bea
F_{\mK_6, (\phi,\phi)}^{(2)}\Big|_{\text{TC}} & = \threeps \prod_{i=1}^3\de^4 \eta_{l_{i}} \Big(\Fco_{\mK_6,3}^{(0),{\rm MHV}}(-l_1, -l_2,-l_3) \mA_5^{(0),{\rm NMHV}}(p_1,p_2, l_3, l_2,l_1)\\
&\hphantom{\threeps  } 
+\Fco_{\mK_6,3}^{(0),{\rm NMHV}}(-l_1, -l_2,-l_3) \mA_5^{(0),{\rm MHV}}(p_1,p_2, l_3, l_2,l_1)\Big)\Big|_{(\eta_1^1 \eta_1^2)(\eta_2^3 \eta_2^4)} \\
&=F_{\mK_6 , (\phi,\phi)}^{(0)} \, i \,  \mathcal{C}_\text{TC} \pnt
\label{3ptcutpre}
\eea
\eeq
Note that besides the MHV form factors and amplitudes also the NMHV form factors and amplitudes appear as building blocks in \eqref{3ptcutpre}. The two terms in the above sum are in fact conjugate to each other.

After performing the fermionic integrations%
\footnote{To obtain the cut integrand in a compact form, it is convenient to take the product of the bosonic form factor and amplitude expressions and sum over all helicity configurations, since the NMHV result of both, the three-point form factors and the five-point amplitudes, take simple ${\overline{\rm MHV}}$ form. We have checked that the expression obtained in this way is equivalent to the expression by using super form factor and amplitudes and doing the fermionic integration directly.} 
and some spinor algebra, the cut integral can be simplified at the integrand level 
to obtain the following form:%
\footnote{In practice, this form can be obtained easily as follows. First, one can write down immediately the contribution of the first five terms by using the result \eqref{doublecut2loopresult} obtained from the double cuts, as explained below. Then, subtracting them from the cut integrand, the remaining terms take a very simple form which can be easily simplified into the last three terms.} 
\beq
\bea
\mathcal{C}_\text{TC}&= \threeps  \left(
\frac{s_{12}^2 - 6 s_{1l_1} s_{2l_1}}{s_{2l_3} s_{l_1l_2} s_{l_2l_3}}
   +\frac{s_{12}^2 - 6 s_{1l_3} s_{2l_3}}{s_{1l_1} s_{l_1l_2} s_{l_2l_3}}
   +\frac{s_{12}^2 - 6 s_{1l_2} s_{2l_2}}{s_{2l_3} s_{l_1l_2} s_{l_1l_3}}
 \right.
   \\
   & \hphantom{{}={} \threeps \bigg(} \left.    +\frac{s_{12}^2 - 6 s_{1l_2} s_{2l_2} }{s_{1l_1} s_{l_1l_3}
   s_{l_2l_3}} +\frac{s_{12}^2 - 6 s_{1l_2} s_{2l_2} }{s_{1l_1} s_{2l_3} s_{l_1l_3}}
   +\frac{18}{s_{12}}
   -\frac{18 s_{1l_3}}{s_{12} s_{l_1l_2}}-\frac{18 s_{2l_1}}{s_{12} s_{l_2l_3}} \right) \pnt
   \label{CstarTC}
   \eea
   \eeq

Note that the first five terms in \eqref{CstarTC} can be obtained directly from the result determined by the two-particle cut in the previous paragraph, namely the planar ladder contribution in \eqref{PL2loopcut} and the crossed ladder in \eqref{F2loopCL}. 
%
\begin{figure}[tbp]
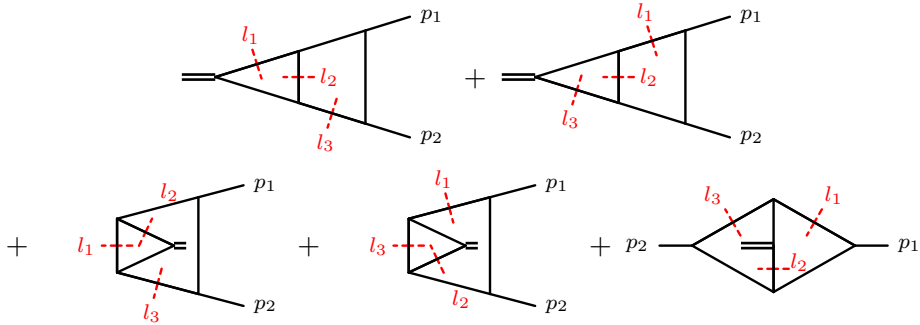

 \centering
\FDinline[rainbow,triplecut,big,cutlabels,twolabels,labelone=p_1,labeltwo=p_2]+\FDinline[rainbow,alttriplecut,big,cutlabels,twolabels,labelone=p_1,labeltwo=p_2]\\
+ \hskip -0.5cm \FDinline[fliprainbow,triplecut,big,cutlabels,twolabels,labelone=p_1,labeltwo=p_2]+
 \hskip -0.5cm \FDinline[fliprainbow,alttriplecut,big,cutlabels,twolabels,labelone=p_1,labeltwo=p_2]+\FDinline[splitlabels,fliprainbownonplanar,alttriplecut,big,cutlabels,twolabels,labelone=p_2,labeltwo=p_1]
\caption{Triple cut of the integrals that correspond to the first five terms in \eqref{CstarTC}. The flow of the momenta is as specified in figure \ref{fig: simple triple cut}. }
\label{ladder3cuts}
\end{figure}
Let us first look at the first term in \eqref{doublecut2loopresult}, the contribution from the planar ladder, which contains the numerical prefactor $4$. As mentioned earlier, this factor 4 stems from the four different ways of drawing the planar ladder graph. The two configurations shown in \eqref{diagPL} contribute to the above triple cut. In order to account for all possible triple cuts on these two diagrams, we cut each in two ways as shown in figure \ref{ladder3cuts}.  Thus, the first four terms in \eqref{CstarTC} correspond to the first four diagrams in figure \ref{ladder3cuts}, which are just the planar ladder integrals. The remaining fifth term in \eqref{CstarTC} correspond to the last diagram in figure \ref{ladder3cuts}, which is the crossed ladder integral with only one possible triple cut. Hence, the first five terms in \eqref{CstarTC} do not result in any new contribution but reproduce the double-cut result in  \eqref{doublecut2loopresult}.

 The remaining three terms in \eqref{CstarTC}, however,  are new contributions to the two-loop ansatz detected by the three-particle cut. They can be expressed as the three-particle cut of the following three integrals:
\beq
\bea
f_{\mK_6 , (\phi,\phi)}^{(2)} \Big|_{\text{TC}}^{\text{III}} = i\, 18 \left(
\frac{1}{s_{12}}\FDinline[sunrise,triplecut,cutlabels,twolabels,labelone=p_1,labeltwo=p_2] 
- \frac{s_{1 l_3} }{s_{12}}\FDinline[fishtop,triplecut,cutlabels,twolabels,labelone=p_1,labeltwo=p_2] 
- \frac{s_{2 l_1} }{s_{12}}\FDinline[fishbottom,triplecut,cutlabels,twolabels,labelone=p_1,labeltwo=p_2]
\right) \pnt
\label{tripcut1}
\eea
\eeq
These three cut integrals can be lifted to full integrals, which can be simplified further at the integral level to give a single scalar  integral:
\beq
f_{\mK_6 , (\phi,\phi)}^{(2),{\text{III}}} = 18\FDinline[fishtop,twolabels,labelone=p_1,labeltwo=p_2] \pnt
\label{tripcut2}
\eeq

\subsubsection*{Complete two-loop result}

Now, we combine the results from all the cuts, \eqref{doublecut2loopresult} and \eqref{tripcut2}, and obtain the two-loop two-point form factor,%
\footnote{The result  \eqref{F2looptot} matches the one in the unpublished notes of Boucher-Veronneau, Dixon and Pennington \cite{BDP-notes}.}
\beq
\bea
f_{\mK_6 , (\phi,\phi)}^{(2)} & =   4 f_{\mK_6, (\phi,\phi)}^{(2),{\text{I}}}+f_{\mK_6, (\phi,\phi)}^{(2),{\text{II}}} + 2 f_{\mK_6 , (\phi,\phi)}^{(2),{\text{III}}} \\
& =
-6 (l+p_1)^2(l+p_2)^2
\bigg(4\FDinline[rainbow,momentum,twolabels,labelone=p_1,labeltwo=p_2]+\FDinline[rainbownonplanar,momentum,twolabels,labelone=p_1,labeltwo=p_2]
\bigg)\\
&\phantom{{}={}}+ 36\FDinline[fishtop,twolabels,labelone=p_1,labeltwo=p_2] 
+ \underbrace{s_{12}^2\bigg(4
\FDinline[rainbow, twolabels, labelone=p_1,labeltwo=p_2] +
\FDinline[rainbownonplanar, twolabels, labelone=p_1,labeltwo=p_2]\bigg)}_{f_{\BPS,2}^{(2)}}\col
\label{F2looptot}
\eea
\eeq
where the integrals corresponding to the graphs are given in appendix \ref{app:integrals}.
Note that we have multiplied $f_{\mK_6, (\phi,\phi)}^{(2),{\text{III}}}$ by $2$ to include the contribution from the permutation of the external legs $p_1 \leftrightarrow p_2$.
As in the one-loop case, we have presented the result by separating a part that is identical to the BPS form factor ${f_{\BPS,2}^{(2)}}$ given in \eqref{F2-BPS-2loop}.

The double and triple cuts we have considered should be able to detect all possible basis integrals up to potential rational terms that might be missing when using four-dimensional unitarity. Comparing our result \eqref{F2looptot} with the 
one we obtained for $\mK_6$ from the Feynman diagrams of 
appendix \ref{app:feyndiag}, we have confirmed that such rational terms are absent.

As will be explained in section \ref{sec: subtleties in unitarity}, the result given by \eqref{F2looptot} is, however, only valid for the operator $\mK_6$ defined in \eqref{Kinsu4}, but not for the Konishi operator $\mK$ defined in \eqref{Kinso6}. 
This subtlety will be discussed in details in section \ref{sec: subtleties in unitarity}. We will see that by a rigorous prescription we can modify the above
result in order to obtain the Konishi form factor.

\subsection{One-loop three-point form factor}

\begin{figure}[tbp]
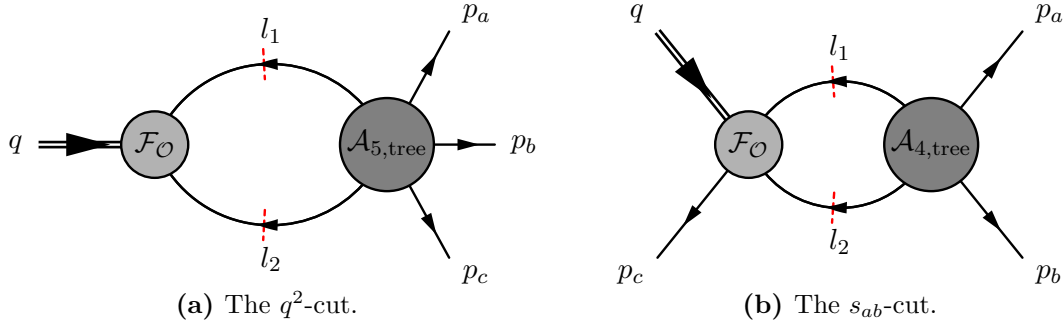

\begin{center}
\begin{subfigure}[c]{0.49\textwidth}
  \centering
$
\settoheight{\eqoff}{$\times$}%
\setlength{\eqoff}{0.5\eqoff}%
\addtolength{\eqoff}{-12.0\unitlength}%
\raisebox{\eqoff}{%
\fmfframe(2,2)(2,2){%
\begin{fmfchar*}(60,30)
\fmfleft{vq}
\fmfright{vp3,vp2,vp1}
\fmf{dbl_plain_arrow,tension=2}{vq,v1}
\fmf{plain_arrow,tension=0.5,left=0.7,label=$l_2$,l.s=left,l.d=8}{v2,v1}
\fmf{plain_arrow,tension=0.5,right=0.7,label=$l_1$,l.s=right,l.d=8}{v2,v1}
\fmf{phantom_smallcut,left=0.7,tension=0}{v1,v2}
\fmf{phantom_smallcut,right=0.7,tension=0}{v1,v2}
\fmf{plain_arrow}{v2,vp1}
\fmf{plain_arrow}{v2,vp2}
\fmf{plain_arrow}{v2,vp3}
\fmfv{decor.shape=circle,decor.filled=30,decor.size=25,label=$\mathcal{F}_{\mathcal{O}}$,label.dist=0}{v1}
\fmfv{decor.shape=circle,decor.filled=50,decor.size=35,label=$\mathcal{A}_{5,,\text{tree}}$,label.dist=0}{v2}
\fmffreeze
 \fmfcmd{pair vertq, vertpone, vertptwo, vertone, verttwo, vertpthree; vertone = vloc(__v1); verttwo = vloc(__v2); vertpthree = vloc(__vp3); vertq = vloc(__vq); vertpone = vloc(__vp1); vertptwo = vloc(__vp2); }
 \fmfiv{label=$q$}{vertq}
 \fmfiv{label=$p_a$}{vertpone}
 \fmfiv{label=$p_b$}{vertptwo}
 \fmfiv{label=$p_c$}{vertpthree}
\end{fmfchar*}%
}}%
$
\caption{The $q^2$-cut.}
\label{fig: simple double cut in three}
\end{subfigure}
\begin{subfigure}[c]{0.49\textwidth}
  \centering
$
\settoheight{\eqoff}{$\times$}%
\setlength{\eqoff}{0.5\eqoff}%
\addtolength{\eqoff}{-12.0\unitlength}%
\raisebox{\eqoff}{%
\fmfframe(2,2)(2,2){%
\begin{fmfchar*}(60,30)
\fmfleft{vp3,vq}
\fmfright{vp2,vp1}
\fmf{dbl_plain_arrow,tension=1}{vq,v1}
\fmf{plain_arrow}{v1,vp3}
\fmf{plain_arrow,tension=0.5,left=0.7,label=$l_2$,l.s=left,l.d=8}{v2,v1}
\fmf{plain_arrow,tension=0.5,right=0.7,label=$l_1$,l.s=right,l.d=8}{v2,v1}
\fmf{phantom_smallcut,left=0.7,tension=0}{v1,v2}
\fmf{phantom_smallcut,right=0.7,tension=0}{v1,v2}
\fmf{plain_arrow}{v2,vp1}
\fmf{plain_arrow}{v2,vp2}
\fmfv{decor.shape=circle,decor.filled=30,decor.size=25,label=$\mathcal{F}_{\mathcal{O}}$,label.dist=0}{v1}
\fmfv{decor.shape=circle,decor.filled=50,decor.size=35,label=$\mathcal{A}_{4,,\text{tree}}$,label.dist=0}{v2}
\fmffreeze
 \fmfcmd{pair vertq, vertpone, vertptwo, vertone, verttwo, vertpthree; vertone = vloc(__v1); verttwo = vloc(__v2); vertpthree = vloc(__vp3); vertq = vloc(__vq); vertpone = vloc(__vp1); vertptwo = vloc(__vp2); }
 \fmfiv{label=$q$}{vertq}
 \fmfiv{label=$p_a$}{vertpone}
 \fmfiv{label=$p_b$}{vertptwo}
 \fmfiv{label=$p_c$}{vertpthree}
\end{fmfchar*}%
}}%
$
\caption{The $s_{ab}$-cut.}
\label{fig: simple double cut in two one}
\end{subfigure}
\caption{The cuts needed to compute the one-loop three-point form factor of $\mK_6$.}
\label{fig: Konishi-3pt-cuts}

\end{center}
\end{figure}

In this subsection, we compute the one-loop three-point form factor of $\mK_6$. The computation is similar  to what we have done for the previous two-point case. We need to consider cuts in all possible kinematic channels, which, apart from the $q^2$-cuts employed earlier for the two-point form factors, contain also the $s_{ab}$-cuts, where $(a,b)=(1,2),(2,3),(3,1)$, as shown in figure \ref{fig: Konishi-3pt-cuts}. Combining the results from both types of cuts ensures that no contribution to the ansatz is missed.

Unlike for the BPS form factor, the loop corrections of the form factors of $\mK_6$ turn out to be different for different configurations of external particles. Therefore, we need to consider the form factors with specific configurations of the external states individually. 
We consider the scalar-scalar-gluon and fermion-fermion-scalar cases.%
\footnote{There could also be other external states composed of different fields, such as  $F^{(1)}_{\mK_6}(1_{g_-}, 2_{g_+}, 3_{g_+})$ and  $F^{(1)}_{\mK_6}(1_{\psi_1}, 2_{\psi_{234}},3_{g_+})$. Such form factors, however, do not contribute to the two-loop cross section studied in section \ref{sec:CS} as the corresponding tree-level results are zero, and we will not consider them in this paper.}
We will discuss the scalar-scalar-gluon case in some detail. The fermion-fermion-scalar result can be obtained in the same way and we only present the final result.

\subsection*{$F_{\mK_6}^{(1)}(1_{\phi_{AB}}, 2_{\phi_{CD}}, 3_{g_\pm})$}

We first consider the form factor of $\mK_6$ with scalar-scalar-gluon external states. For the sake of explicitness, we focus on $F_{\mK_6}^{(1)}(1_{\phi_{12}}, 2_{\phi_{34}}, 3_{g_+})$, which we abbreviate as $\Fb_{\mK_6, (\phi, \phi, g)}^{(1)}$.  
The result applies to all other non-vanishing cases, where $A,B,C,D$ are 
distinct and the $g$ may have positive or negative helicity.
As shown in figure \ref{fig: Konishi-3pt-cuts}, we need to consider both the $q^2$-cut and the $s_{ab}$-cut. Since the operator is a color singlet, we need to consider all possible cyclic permutations of external on-shell legs in the cuts, as they contribute to the same color-ordered form factor. Explicitly, we need to consider three cases for each channel in figure \ref{fig: Konishi-3pt-cuts}:
\begin{equation}
\{ a, b, c \} \ \rightarrow \  ({\rm I})\ \{1_{\phi_{12}}, 2_{\phi_{34}}, 3_{g_+}\}, \quad ({\rm II})\ \{2_{\phi_{34}}, 3_{g_+}, 1_{\phi_{12}} \} , \quad   ({\rm III}) \ \{3_{g_+}, 1_{\phi_{12}}, 2_{\phi_{34}} \} \pnt
\label{3pt1loopextstates}
\end{equation}
In total, there are six cut channels to consider: (a-I), (a-II), (a-III) and (b-I), (b-II), (b-III), where (a-I)--(a-III) are the $q^2$-cuts while (b-I)--(b-III) are the $s_{ab}$-cuts.
Note the (I) and (III) cases are actually related to each other by a flipping symmetry.

\subsubsection*{(a-I)-cut:}
This is the $q^2$-cut in  figure \ref{fig: Konishi-3pt-cuts} with the choice of external legs $\{ p_a, p_b, p_c \}$ corresponding to the particles $\{ 1_{\phi_{12}}, 2_{\phi_{34}}, 3_{g_+} \}$.
The cut integral is given by the following equation:
\beq
\bea
&\Fb_{\mK_6, (\phi, \phi, g)}^{(1)}\Big\vert_{\textrm{(a-I)}}=\twops \deta \Fco_{\mK_6 , 2}^{(0)}(-l_1,-l_2) \mA_5^{(0),\text{MHV}}(p_1,p_2,p_3,l_2,l_1) \Big|_{(\eta_1^1 \eta_1^2)(\eta_2^3 \eta_2^4)} \\
&=\Fb_{\mK_6, (\phi, \phi, g)}^{(0)} \, i \, \underbrace{\twops \frac{\ket(l_1,2)^2\ket(l_2,1)^2+4\ket(l_1,1)\ket(l_1,2)\ket(l_2,1)\ket(l_2,2)+\ket(l_1,1)^2\ket(l_2,2)^2}{\ket(l_1,1)\ket(3,l_2)\ket(l_2,l_1)}\frac{\ket(1,3)}{\ket(1,2)^2}}_{\mC_{\textrm{(a-I)}}} \col
\label{Kcutssg}
\eea
\eeq 
where the tree-level form factor $\Fb_{\mK_6, (\phi, \phi, g)}^{(0)}=\Fb_{\mK_6}^{(0)}(1_{\phi_{12}}, 2_{\phi_{34}}, 3_{g_+})$ is given in \eqref{KF-boso-3pt-tree}.

The above result can be reduced to an appropriate cut of integrals by using some spinor algebra. Without going through the detail, we present the result:
\beq
\bea
\mC_{\textrm{(a-I)}}&= -\frac{s_{12} s_{23}}{2}
\FDinline[box,doublecut,momentumarrow,cutlabels,threelabels,labelone=p_1,labeltwo=p_2,labelthree=p_3]
- \frac{s_{12}+s_{13}}{2}\FDinline[trianglethreebot,doublecut,cutlabels,momentumarrow,threelabels,labelone=p_1,labeltwo=p_2,labelthree=p_3]\\
   &\phantom{{}={}}
-\bigg[\frac{s_{13}+s_{23}}{2} - \frac{3  s_{2 l_1} \left( (s_{13}+s_{23}) s_{1 l_1} +s_{12}
   s_{23}\right)}{s_{12}^2} \bigg]
\FDinline[trianglethreetop,doublecut,cutlabels,momentumarrow,threelabels,labelone=p_1,labeltwo=p_2,labelthree=p_3] 
\\
&\phantom{{}={}}+ 3 \bigg[\frac{\left(q^2-2s_{13}\right) 
   (s_{13}+s_{1 l_2})}{s_{12}^2}   
   - \frac{s_{13} \left(s_{13} s_{3 l_2} -s_{12} s_{2 l_2} \right)}{s_{12}{}^2
   s_{123}}  \bigg] \FDinline[bubblethree,doublecut,momentumarrow,cutlabels,threelabels,labelone=p_1,labeltwo=p_2,labelthree=p_3]
   \pnt
\label{atilde}
\eea
\eeq

\subsubsection*{(a-II)-cut:}
This is the $q^2$-cut in  figure \ref{fig: Konishi-3pt-cuts} with the choice of external legs $\{ p_a, p_b, p_c\}$ corresponding to a different order of particles, namely $\{2_{\phi_{34}}, 3_{g_+}, 1_{\phi_{12}} \}$.
The cut integral can be computed as
\beq
\bea
\Fb_{\mK_6, (\phi, \phi, g)}^{(1)}&\Big\vert_{\textrm{(a-II)}}=\twops \deta\Fco_{\mK_6 , 2}^{(0)}(-l_1,-l_2) \mA_5^{(0),\text{MHV}}(p_2,p_3,p_1,l_2,l_1) \Big|_{(\eta_1^1 \eta_1^2)(\eta_2^3 \eta_2^4)} \\
&=\Fb_{\mK_6, (\phi, \phi, g)}^{(0)} \, i \,
\underbrace{ \twops \frac{\ket(l_1,1)^2\ket(l_2,2)^2+4\ket(l_1,1)\ket(l_1,2)\ket(l_2,1)\ket(l_2,2)+\ket(l_1,2)^2\ket(l_2,1)^2}{\ket(l_1,2)\ket(2,1)\ket(1,l_2)\ket(l_2,l_1)} }_{\mC_{\textrm{(a-II)}} } \pnt
\label{Kcutsgs}
\eea
\eeq
After some spinor algebra, the above result can be expressed as cut of the following integrals:
\beq
\bea
\label{eq:cut-a-II}
\mC_{\textrm{(a-II)}}&=-\frac{s_{23} s_{31}}{2}
\FDinline[box,doublecut,momentumarrow,cutlabels,threelabels,labelone=p_2,labeltwo=p_3,labelthree=p_1]
 - \frac{s_{12}+s_{13}}{2}
\FDinline[trianglethreetop,doublecut,momentumarrow,cutlabels,threelabels,labelone=p_2,labeltwo=p_3,labelthree=p_1]
- \frac{s_{12}+s_{23}}{2}\FDinline[trianglethreebot,doublecut,cutlabels,threelabels,labelone=p_2,labeltwo=p_3,labelthree=p_1]
\\&
\quad - 3\Big(1 + \frac{s_{1 l_1} s_{2 l_2}-s_{2 l_1} s_{1 l_2}}{s_{12}{}
   s_{123}}\Big)\FDinline[bubblethree,doublecut,cutlabels,threelabels,labelone=p_2,labeltwo=p_3,labelthree=p_1] \pnt
\eea
\eeq

\subsubsection*{(b-I) cut:}

This is the $s_{12}$-cut in  figure \ref{fig: Konishi-3pt-cuts} with the choice of external legs $\{p_a, p_b, p_c\}$ corresponding to the particles $\{ 1_{\phi_{12}}, 2_{\phi_{34}}, 3_{g_+} \}$.
In this case, one of the building blocks is the tree-level three-point form factor in \eqref{Kon3pt}.
The cut integral is given by,
\beq
\bea
&\Fb_{\mK_6, (\phi, \phi, g)}^{(1)}\big\vert_{\textrm{(b-I)}}=\twops \deta\Fco_{\mK_6 , 3}^{(0),\text{MHV}}(-l_1,-l_2,p_3)\mA_4^{(0)}(p_1,p_2,l_2,l_1) \Big|_{(\eta_1^1 \eta_1^2)(\eta_2^3 \eta_2^4)} \\
&=\Fb_{\mK_6, (\phi, \phi, g)}^{(0)} \, i \, \underbrace{\twops\frac{\ket(l_1,2)^2\ket(l_2,1)^2+4\ket(l_1,1)\ket(l_1,2)\ket(l_2,1)\ket(l_2,2)+\ket(l_1,1)^2\ket(l_2,2)^2}{\ket(l_1,1)\ket(2,l_2)\ket(3,l_2)\ket(l_1,3)}\frac{\ket(1,3)\ket(2,3)}{\ket(1,2)^2} }_{{\mC}_{\textrm{(b-I)}}} \pnt
\label{Kcutssgp3}
\eea
\eeq
It can be written as cut of integrals as
\beq
\bea
{\mC}_{\textrm{(b-I)}}&= -\frac{s_{12} s_{13}}{2} \FDinline[box,slashdoublecut,cutlabels,threelabels,labelone=p_3,labeltwo=p_1,labelthree=p_2] -\frac{s_{12} s_{23}}{2}
\FDinline[box,bslashdoublecut,cutlabels,threelabels,labelone=p_1,labeltwo=p_2,labelthree=p_3] \\
&\phantom{{}={}}+\bigg[ -\frac{s_{13}+s_{23}}{2} + \frac{3 s_{1 l_2} \left( (s_{13} +s_{23}) s_{2 l_2}+s_{12}
   s_{13}\right)}{s_{12}{}^2} \bigg] \FDinline[trianglethreebot,cutlabels,botdoublecut,threelabels,labelone=p_3,labeltwo=p_1,labelthree=p_2]
   \\
&\phantom{{}={}}+\bigg[ -\frac{s_{13}+s_{23}}{2} + \frac{3 s_{2 l_1} \left( (s_{13} +s_{23}) s_{1 l_1} +s_{12} s_{23}\right)}{s_{12}^2} \bigg]\FDinline[trianglethreetop,cutlabels,topdoublecut,threelabels,labelone=p_1,labeltwo=p_2,labelthree=p_3] \pnt
\eea
\eeq

\subsubsection*{(b-II) cut:}
This is the case of  the $s_{23}$-cut, the last of the independent cut channels, in  figure \ref{fig: Konishi-3pt-cuts} with the choice of external legs $\{p_a, p_b, p_c\}$ corresponding to the particles in the order of $\{ 2_{\phi_{34}}, 3_{g_+}, 1_{\phi_{12}} \}$. The cut integral is given by
\beq
\bea
\Fb_{\mK_6, (\phi, \phi, g)}^{(1)}\big\vert_{\textrm{(b-II)}}&=\twops \deta\Fco_{\mK_6 , 3}^{(0),\text{MHV}}(-l_1,-l_2,p_1)\mA_4^{(0)}(p_2,p_3,l_2,l_1) \Big|_{(\eta_1^1 \eta_1^2)(\eta_2^3 \eta_2^4)} \\
&=\Fb_{\mK_6, (\phi, \phi, g)}^{(0)} \, i \, \underbrace{\twops \frac{ \big(\ket(l_1,1)\ket(l_2,2)-\ket(l_1,2)\ket(l_2,1)\big)^2}{\ket(l_1,2)\ket(3,l_2)\ket(1,l_2)\ket(l_1,1)} \frac{\ket(3,1)}{\ket(2,1)} }_{{\mC}_{\textrm{(b-II)}}} \col
\label{Kcutsgsp3}
\eea
\eeq
which leads to
\beq
\bea
{\mC}_{\textrm{(b-II)}}&= -\frac{s_{12} s_{23}}{2} \FDinline[box,slashdoublecut,cutlabels,threelabels,labelone=p_1,labeltwo=p_2,labelthree=p_3]
-\frac{s_{23} s_{31}}{2} 
\FDinline[box,bslashdoublecut,cutlabels,threelabels,labelone=p_2,labeltwo=p_3,labelthree=p_1]\\
&\phantom{{}={}} - \frac{s_{12}+s_{13}}{2}\FDinline[trianglethreebot,cutlabels,botdoublecut,threelabels,labelone=p_1,labeltwo=p_2,labelthree=p_3] - \frac{s_{12}+s_{13}}{2}\FDinline[trianglethreetop,cutlabels,topdoublecut,threelabels,labelone=p_2,labeltwo=p_3,labelthree=p_1]
\pnt
\eea
\eeq

As previously mentioned, the cuts (a-III) and (b-III) give similar results to (a-I) and (b-I) and can be obtained by exchanging 
$p_1 \leftrightarrow p_2$ in the latter. Hence, we will not give them explicitly.

\subsubsection*{From the cuts to the full form factor}

We find that all the above cut results are consistent with each other, i.e.\ the prefactors of the graphs are identical when the same graph appears in different cut channels apart from terms that vanish due to the on-shell condition for the cut momenta
in the individual channels.
Given all these cut results, it is straightforward to lift the cut integrals, as described in appendix \ref{app:integrals}, to obtain the full form 
factor\footnote{Recall that the prefactors that depend on the loop momenta are understood to appear in the integrand of the integral represented by the respective graph each prefactor multiplies.}
\beq
\bea
f^{(1)}_{\mK_6,(\phi, \phi, g)} &=  \bigg\{ 3 \bigg[\frac{\left(s_{123} -2s_{13}\right) 
   (s_{13}+s_{1 l})}{s_{12}^2}
   - \frac{s_{13} \left(s_{13} s_{3 l}-s_{12} s_{2 l}\right)}{s_{12}^2
   s_{123}}  - {1\over2}
\bigg] \FDinline[bubblethree,momentum,threelabels,labelone=p_1,labeltwo=p_2,labelthree=p_3] 
\\
   &
\hphantom{{}={}\bigg\{} + \frac{3 s_{2 l} \left( s_{1 l} (s_{13}+s_{23}) +s_{12}
   s_{23}\right)}{s_{12}^2}
   \FDinline[trianglethreetop,momentum,threelabels,labelone=p_1,labeltwo=p_2,labelthree=p_3] 
+ \{p_1 \leftrightarrow p_2\} \bigg\}
+   f^{(1)}_{\BPS,3} \col
   \label{fulloneloop3ssg}
\eea
\eeq
where $f^{(1)}_{\BPS,3}$ denotes the BPS part that is given in \eqref{F2-BPS-2loop}.

\subsection*{$F_{\mK_6}^{(1)}(1_{\psi_A}, 2_{\psi_B}, 3_{\phi_{CD}})$}
 
For the form factor with fermion-fermion-scalar external states with distinct $A,B,C,D$, e.g.\ $F_{\mK_6}^{(1)}(1_{\psi_1}, 2_{\psi_2}, 3_{\phi_{34}})$, one can proceed along the above steps for computing the cut integrand in all possible channels and lifting the cut result to the full answer. 
Without giving details, we present the final result, denoted by $f_{\mK_6,(\psi,\psi,\phi)}^{(1)}$:
\beq
\bea
f^{(1)}_{\mK_6,(\psi, \psi, \phi)} &= \bigg\{
- \frac{3s_{23}}{2 }\FDinline[triangletwoonerev,threelabels,labelone=p_2,labeltwo=p_3,labelthree=p_1,splitlabels] 
-3\bigg( 1 - \frac{ s_{12} -s_{13} }{s_{23}} \bigg) \FDinline[bubbletwoonerev,threelabels,labelone=p_2,labeltwo=p_3,labelthree=p_1,splitlabels]  \\
&
\hphantom{{}={}\bigg\{}+ 3 \bigg( \frac{s_{12} s_{2 l}-s_{13} s_{3 l}}{s_{23} s_{123}}-\frac{ s_{2 l}-s_{3 l}}{s_{23}}\bigg)
\FDinline[bubblethree,momentum,threelabels,labelone=p_1,labeltwo=p_2,labelthree=p_3]\\
&
\hphantom{{}={}\bigg\{}+ 3\bigg( {s_{12}+s_{13}\over2} + \frac{ s_{12} s_{3 l} -s_{13} s_{2 l} }{s_{23}} \bigg)\FDinline[trianglethreebot,altmomentum,threelabels,labelone=p_1,labeltwo=p_2,labelthree=p_3]
+ \{ p_1 \leftrightarrow p_2 \} \bigg\} \\
&
\hphantom{{}={}} +\frac{3s_{23}s_{31}}{2 }
\FDinline[box,threelabels,labelone=p_2,labeltwo=p_3,labelthree=p_1] 
+f^{(1)}_{\BPS,3} \pnt
   \label{fulloneloop3ffs}
\eea
\eeq
Note that in the above result not all contributions from box graphs are incorporated in the corresponding BPS part given in \eqref{F2-BPS-2loop}, unlike in the expression for the scalar-scalar-gluon form factor \eqref{fulloneloop3ssg}.
Moreover, there is an additional one-mass triangle integral in the first line of 
\eqref{fulloneloop3ffs}, which does not appear in \eqref{fulloneloop3ssg}.

\subsubsection*{PV reduction and some interesting features of the results}

We have obtained the full integral expressions for the form factors  \eqref{fulloneloop3ssg} and \eqref{fulloneloop3ffs} of $\mK_6$. The results are obtained by using the unitarity method fully at the integrand level. As a result, the integrals still contain loop-momentum-dependent numerators. Such integrals can be reduced further via PV reduction, see appendix \ref{app:PV} for details.

After PV reduction, the results \eqref{fulloneloop3ssg} and \eqref{fulloneloop3ffs} are simplified to
\begin{align}
f_{\mK_6,(\phi, \phi, g)}^{(1)}
&=  -3 \left[ 1+ \frac{s_{13}^2+s_{23}^2}{(s_{13}+s_{23})^2} \right]
\FDinline[bubblethree,threelabels,labelone=p_1,labeltwo=p_2,labelthree=p_3] 
-\frac{ 6 s_{13}s_{23}}{(s_{13}+s_{23})^2}
\FDinline[bubbletwoone,threelabels,labelone=p_1,labeltwo=p_2,labelthree=p_3,splitlabels] \nonumber\\
&\quad + \frac{ 12 s_{13}s_{23}}{s_{12}(s_{13}+s_{23})} I_3^D[\ell_\peps^2]
+f_{\BPS,3}^{(1)} \col
\label{oneloop3Konscalar} \\
f_{\mK_6,(\psi, \psi, \phi)}^{(1)}&=
3\left( \frac{s_{12}-s_{13}}{s_{12}+s_{13}} + \frac{s_{12}-s_{23}}{s_{12}+s_{23}}\right)
\FDinline[bubblethree,threelabels,labelone=p_1,labeltwo=p_2,labelthree=p_3] \nonumber\\
&\quad -3 \left(1+ \frac{s_{12}-s_{13}}{s_{12}+s_{13}}\right)
\FDinline[bubbletwoone,threelabels,labelone=p_2,labeltwo=p_3,labelthree=p_1,splitlabels] -3 \left( 1+  \frac{s_{12}-s_{23}}{s_{12}+s_{23}} \right)
\FDinline[bubbletwoonerev,threelabels,labelone=p_3,labeltwo=p_1,labelthree=p_2,splitlabels] \nonumber\\
&\quad + \frac{3 s_{23}s_{31}}{2} \, \text{Fin}\bigg(
\FDinline[box,threelabels,labelone=p_2,labeltwo=p_3,labelthree=p_1]\bigg) +f_{\BPS,3}^{(1)} \col
\label{oneloop3Konfermion}
\end{align}
where all relevant integrals are given in appendix \ref{app:integrals}, and 
Fin extracts the finite part FB of the box integral defined
in \eqref{eq:finite-Box}.

There are several interesting features in the above results we would like to comment
on. 
\begin{itemize}
\item Going back to the expressions \eqref{fulloneloop3ssg} and \eqref{fulloneloop3ffs}, we notice that --- besides a part identical to the BPS form factor $f_{\BPS,3}^{(1)}$ --- there are still triangle or box integrals left, which separately contain IR divergences. This might cause a net IR divergences in addition to the one contained
in $f_{\BPS,3}^{(1)}$, i.e.\ it would spoil the universality of the IR divergence.
However, as evident from the results \eqref{oneloop3Konscalar} and \eqref{oneloop3Konfermion}, these additional IR divergences cancel after PV reduction and hence
only the universal IR divergence of the BPS part remains; see also 
\cite{Brandhuber:2010ad}.
\item
There are remaining divergences given by bubble integrals. These are the UV divergences which have to be canceled by renormalizing the composite operator $\mK_6$. 
See section \ref{sec:CS} for a further discussion.
\item Besides the common BPS part, the two results \eqref{oneloop3Konscalar} and \eqref{oneloop3Konfermion} are quite different from each other. 
This directly shows that the form factors of $\mK_6$ with different external legs (even with the same MHV degree) have very different structure and need to be studied case by case.
\item 
There is a term in \eqref{fulloneloop3ssg} involving the integral $I_3^D[\ell_\peps^2]$ given in \eqref{rationaltermtriangle}. It evaluates to a rational term. 
Interestingly, we have found this contribution by applying four-dimensional unitarity. This is possible, since we apply the four-dimensional unitarity to 
compute the integrand expressed in terms of a tensor-integral basis. The rational term only appears after the PV reduction when the basis is reduced to scalar integrals. 
In the usual one-loop (generalized) unitarity computation \cite{Britto:2004nc, Forde:2007mi}, one computes the coefficients of the scalar integrals directly
and hence one would miss this rational term.
In appendix \ref{app:check-3pt}, we have checked that the rational term in \eqref{fulloneloop3ssg} matches with an independent Feynman diagrammatic computation,
and hence the final result for the form factor is complete. 
\item The integral coefficients of the form factors in \eqref{oneloop3Konscalar} and \eqref{oneloop3Konfermion} appear to contain unphysical poles, such as  ${1\over s_{13}+s_{23}}={1\over s_{123}-s_{12}}$. These are, however, just spurious poles which cancel when all contributions are taken into account. We demonstrate this in detail in appendix \ref{app:check-3pt}. 
\end{itemize}
Finally, we would like to mention that most of the above features (except the IR divergence) do not occur for the one-loop scattering amplitudes and BPS form factors of the ${\cal N}=4$ SYM theory. In QCD, they are, however, common and appear e.g.\ for
one-loop amplitudes \cite{Bern:2007dw}.

Last but not least, recall that the above form factor results for $\mK_6$ still need to be modified to obtain the correct ones for the Konishi operator, as will be described in the next section. This does not affect any of the above listed properties.

%% file: subtleties.tex
\section{Konishi vs.\ \texorpdfstring{$\mathcal{K}_6$}{K6}}
\label{sec: subtleties in unitarity}

In this section, we discuss some important subtleties that arise when regulating 
the theory by continuing the spacetime dimension from $D=4$ to $D=4-2\peps$.
Our unitarity-based calculation made
use of the on-shell superfield \eqref{ossuperfield} that only captures 
all degrees of freedom in strictly $D=4$ dimensions. 
Hence, this approach does not
directly yield the correct form factors of 
dimension-dependent operators. 
We explain this problem and its resolution in details below, taking the Konishi form factor as a concrete example.

\subsection{A subtlety in choosing a regularization scheme}

When regulating the theory by continuing the spacetime dimension to $D=4-2\peps$,
one has to also specify how the various fields are continued. In conventional dimensional regularization (CDR) \cite{Collins:1984xc} and the 't Hooft Veltman (HV) scheme \cite{'tHooft:1972fi}, the number of fermion 
flavors $N_\psi$ and also the number of scalar flavors $N_\phi$ 
remain as in four dimensions and are hence kept as $N_\psi=4$ and $N_\phi=6$, respectively. 
This does, however, break supersymmetry, since the polarization vector $\epsilon^\mu$ is taken in $D=4-2\peps$ dimensions.

A scheme that preserves supersymmetry is dimensional reduction (DR) from ten dimensions \cite{Siegel:1979wq,Capper:1979ns}.
In this scheme, the number
of scalar fields is changed to $N_\phi=6+2\peps$, such that $D+N_\phi=10$ is 
independent of $\peps$. It exploits the fact that four-dimensional $\mathcal{N}=4$ SYM theory can be obtained by dimensional reduction of ten-dimensional $\mathcal{N}=1$ SYM theory. 
Performing the dimensional reduction 
to $D=4-2\peps$ rather than four dimensions, one
obtains a regulated theory that preserves $\mathcal{N}=4$ supersymmetry. 
The ten-dimensional gauge field $A_M$, $M=1,\dots,10$, then reduces to the $D$-dimensional gauge field $A_\mu$ and to $N_\phi=10-D=6+2\peps$ scalar fields $\phi_I$.
Similarly, the ten-dimensional metric $g^{MN}$ reduces to the $D$-dimensional metric $g^{\mu\nu}$ and $\delta^{IJ}$.

In a modified version of the DR scheme, known as four-dimensional-helicity (FDH) scheme \cite{Bern:1991aq,Bern:2002zk}, the $2\peps$ scalar  degrees of freedom are absorbed into the gluons. This apparently preserves supersymmetry in the sense that bosonic and fermionic degrees of freedom still match. In particular, it allows to use the $\mathcal{N}=4$ on-shell superfield \eqref{ossuperfield} and polarization vectors in $D=4$ dimensions.
So far, the FDH scheme has been successfully used in computing amplitudes and 
BPS form factors in $\mathcal{N}=4$ SYM theory.
However, as we will discuss below, the FDH scheme is incompatible with 
dimension-dependent operators, i.e.\ operators that are sensitive to the 
absorption of the $2\peps$ scalar degrees of freedom into the gluons.  
In particular, the Konishi operator \eqref{Kinso6} is dimension-dependent, 
since it contains a trace over $N_\phi=6+2\peps$ scalars in $D=4-2\peps$ dimensions.
The incompatibility arises in the FDH scheme since the $2\peps$ scalars
are absorbed into the gluons, giving direct results only for $\mK_6$ in \eqref{Kinsu4}.

In order to detect the differences between working with the FDH and the DR
scheme, we examine the underlying Feynman diagrams.
In Feynman diagrams, explicit factors of $D=g^\mu{}_\mu$ and $N_\phi=\delta^I{}_I$ arise whenever a gauge or scalar field runs in a loop in which the respective Lorentz or flavor index also forms a loop. We call such a loop an \emph{index loop}. 
Moreover, we call an index loop
\emph{externally closed} if fields of the composite operator are involved in the index loop and \emph{internally closed} if the operator is not involved in the index loop.\footnote{Note that the factors of $D=g^\mu{}_\mu$ and $N_\phi=\delta^I{}_I$ 
occur even if the index loop is apparently interrupted when the gauge or scalar field splits
into a pair of fermions that themselves build a loop.
This follows from the Clifford algebra for the
spacetime $(\sigma_\mu)_{\alpha\dot\alpha}$ matrices
and the flavor $(\sigma_I)_{AB}$ matrices into which the ten-dimensional 
$\sigma$ matrices split. 
}

\begin{figure}[htbp]
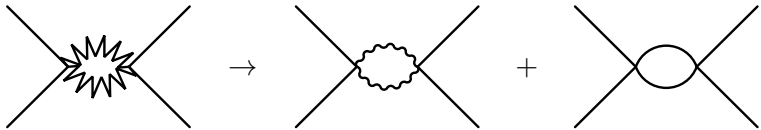

\centering
\scalebox{1}{$ \FDdimred[zigzagbubble] \rightarrow \FDdimred[gluonbubble]+\FDdimred[scalarbubble]$}
\caption{In dimensional reduction, a Feynman diagram with a closed ten-dimensional vector loop (zigzag) decomposes to one with a closed $4(-2\peps)$-dimensional vector loop (wiggly) and one with a $6(+2\peps)$-dimensional scalar loop (plain).
}
\label{fig: dimensional reduction of closed vector loops} 
\end{figure}

We look at internally closed index loops first.
From the dimensional reduction from ten dimensions, we know that an internally closed vector index loop always occurs together with an internally closed scalar index loop.
This is illustrated in figure \ref{fig: dimensional reduction of closed vector loops}. 
Hence, each factor of $D$ for a Lorentz index loop is accompanied by a factor of $N_\phi$ for a scalar flavor index loop.
This can also be seen in the concrete Feynman diagrams in appendix \ref{app:feyndiag}, e.g.\ by comparing the first two lines in table \ref{tab: twoloopdiagrams}.
The sum of both contributions is proportional to $D+N_\phi=10$, both in the DR scheme and in the FDH scheme.
Hence, as far as internally closed index loops are concerned, one is free to work in the FDH scheme.

The situation changes for externally closed index loops. Generically, the fields of a composite operator involved in such a loop are only a subset of the fields in the theory, e.g.\ only the scalar fields.
In this case, a diagram in which the externally closed scalar loop generates a factor $N_\phi$ is not paired with a diagram in which a vector field can circulate in the loop and generate a factor $D$. 
Hence, the result in the FDH scheme differs from the one in the DR scheme.
In scattering amplitudes and form factors of BPS operators such as $\tr(\phi_{12}^k)$, no externally closed index loops can occur. This is why the FDH scheme is directly applicable for calculating these quantities.

Let us consider the particular example of the Konishi primary operator \eqref{Kinso6};
it is defined as the trace over all scalars and can hence be part of an externally closed scalar index loop.
The Konishi primary is the highest-weight state of a so-called long supermultiplet of $\mathfrak{psu}(2,2|4)$. Supersymmetry guarantees that all members of this supermultiplet have the same anomalous dimension \eqref{gammaK} --- unless it is broken by the regularization scheme.
In the supersymmetry-preserving DR scheme, the Konishi primary is defined as the trace over all $N_\phi=10-D=6+2\peps$ scalars.
While the expression \eqref{Kinso6} can easily be modified to sum over this regularization-dependent number of scalars, the expression \eqref{Kinsu4} is only valid for $N_\phi=6$ and hence it is not the highest-weight state of the superconformal multiplet. In particular, its anomalous dimension is not given by \eqref{gammaK}, which is usually calculated using a descendent of the Konishi operator in the $SU(2)$ or $SL(2)$ sector.%
\footnote{The additional $2\peps$ scalar components in the $D$-dimensional continuation of the Konishi operator are an example of so-called evanescent operators, which also appear in the context of QCD \cite{Buras:1989xd}. See \cite{Collins:1984xc} for a textbook treatment. 
}  

A priori, this discrepancy requires one to abandon the unitarity techniques 
that employ the $\mathcal{N}=4$ on-shell superspace.
Fortunately, this is not the case.
In the following, we will show that --- at least for the cases at hand --- the strictly four-dimensional result can be lifted to the $D$-dimensional result with a simple prescription.

\subsection{\texorpdfstring{From $\mK_6$ to $\mK$}{From K6 to K}}

\begin{figure}[htbp]
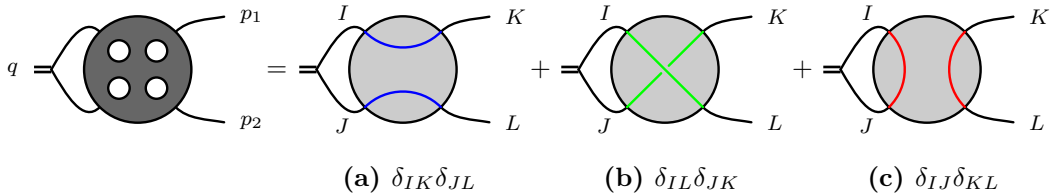

\centering
 $\quad\begin{subfigure}[c]{29\unitlength}
 \FDflow[black,spacetime]
 \end{subfigure}
   \,=\, 
   \begin{subfigure}[c]{29\unitlength}
   \FDflow[blue,flavour]\rule[0.585cm]{0cm}{1.2cm}
   \caption{$\delta_{IK}\delta_{JL}$}
   \label{subfig: a}
   \end{subfigure}
   \,+\,
   \begin{subfigure}[c]{29\unitlength}
   \FDflow[green,flavour]\rule[0.585cm]{0cm}{1.2cm}
   \caption{$\delta_{IL}\delta_{JK}$}
   \label{subfig: b}
   \end{subfigure}
   \,+\,
   \begin{subfigure}[c]{29\unitlength}
   \FDflow[red,flavour]\rule[0.585cm]{0.0cm}{1.2cm}
   \caption{$\delta_{IJ}\delta_{KL}$}
   \label{subfig: c}
   \end{subfigure}$ 
\caption{According to $R$-charge conservation, only three different contractions of the scalar flavors can exist in a generic multi-loop diagram with incoming operator $\tr(\phi_I\phi_J)$ and outgoing scalar fields $\phi_K$ and $\phi_L$: (a) $ \delta_{IK}\delta_{JL}$ (blue), (b) $\delta_{IL}\delta_{JK}$ (green) and (c) $\delta_{IJ}\delta_{KL}$ (red).}
 \label{fig: flavor flows}
\end{figure}

Consider a generic multi-loop diagram contributing to the two-point form factor 
of the operator $\tr(\phi_I\phi_J)$ with outgoing scalar fields $\phi_K$ and $\phi_L$.
It can only have one of the three types of $R$-charge flow depicted in figure 
\ref{fig: flavor flows} together with the respective tensor structures.
Only in the case (\subref{subfig: c}) an externally closed scalar index loop exists.
The BPS operator $\tr(\phi_{(I}\phi_{J)})$ defined in \eqref{BPSinso6} obtains contributions from the cases (\subref{subfig: a}) and (\subref{subfig: b}) but not from case (\subref{subfig: c}). The Konishi operator $\tr(\phi_{I}\phi_{I})$ defined in \eqref{Kinso6} obtains contributions from all three cases. The contributions it receives
from the cases (\subref{subfig: a}) and (\subref{subfig: b}) are identical to 
those of the BPS operator since the coefficients of the tensor structures in 
figure \ref{fig: flavor flows} do not depend on the $R$-charge. 
Thus, we can isolate the contribution from case (\subref{subfig: c}) by subtracting the result for the BPS operator from the result for the Konishi operator.  
In $D=4-2\peps$ dimensions, the single externally closed scalar index loop present in the case (\subref{subfig: c}) should not generate a factor $6$ as found in the FDH scheme but instead the factor $N_\phi=10-D=6+2\peps$ as prescribed in the DR scheme. 
Hence, in order to obtain the result for $\mK$ in the DR scheme from the one for 
$\mK_6$ in the FDH scheme, we simply have to multiply the contributions of case (\subref{subfig: c}), i.e.\ the difference of the Konishi and BPS case, by the ratio 
\begin{equation}\label{rphi}
r_\phi=\frac{N_\phi}{6}=\frac{6+2\peps}{6}
\pnt
\end{equation}

\begin{figure}[htbp]
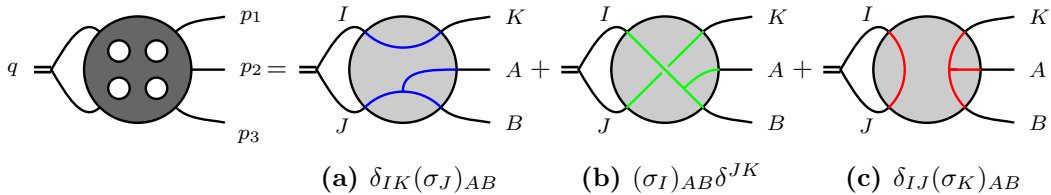

\centering
 $\quad\begin{subfigure}[c]{29\unitlength}
 \FDflow[black,spacetime,fermion]
 \end{subfigure}
   \,=\, 
   \begin{subfigure}[c]{29\unitlength}
   \FDflow[blue,flavour,fermion]\rule[0.585cm]{0cm}{1.2cm}
   \caption{$\delta_{IK}(\sigma_{J})_{AB}$}
   \label{subfig: a fermion}
   \end{subfigure}
   \,+\,
   \begin{subfigure}[c]{29\unitlength}
   \FDflow[green,flavour,fermion]\rule[0.585cm]{0cm}{1.2cm}
   \caption{$(\sigma_{I})_{AB}\delta^{JK}$}
   \label{subfig: b fermion}
   \end{subfigure}
   \,+\,
   \begin{subfigure}[c]{29\unitlength}
   \FDflow[red,flavour,fermion]\rule[0.585cm]{0cm}{1.2cm}
   \caption{$\delta_{IJ}(\sigma_{K})_{AB}$}
   \label{subfig: c fermion}
   \end{subfigure}$ 
\caption{According to $R$-charge conservation, only three different contractions of the scalar flavors can exist in a generic multi-loop diagram with incoming operator $\tr(\phi_I\phi_J)$, outgoing scalar fields $\phi_K$ and outgoing fermion fields $\psi_A$ and $\psi_B$: a) $\delta_{IK}(\sigma_{J})_{AB}$ (blue), b) $(\sigma_{I})_{AB}\delta_{JK}$ (green) and c) $\delta_{IJ}(\sigma_{K})_{AB}$ (red).}
 \label{fig: flavor flows fermion}
\end{figure}

Similar arguments are valid for the three-point form factor of the Konishi operator. 
In our calculation, only its components with either two scalar legs and one gluon leg or two fermion legs and one scalar leg appear.
While in the former case the previous arguments directly apply, 
in the latter case a slight modification is necessary as shown in the following.
A generic multi-loop diagram of the latter type, which has incoming operator $\tr(\phi_I\phi_J)$ and outgoing fields $\phi_K$,  $\psi_A$ and $\psi_B$,
can only have one of the three possible $R$-charge flows show
in figure \ref{fig: flavor flows fermion}.
In addition to the Kronecker $\delta$, the tensor structures contain the flavor
$(\sigma_{I})_{AB}$ matrices obeying the Clifford algebra.
An externally closed scalar index loop exists only in the case (\subref{subfig: c fermion}).
In analogy to the case of the two-point form factor, we can isolate this 
case by subtracting the result for the BPS operator from the result for the Konishi operator. Then, we modify the number of scalars by multiplying this difference by $r_\phi$.

The above arguments can be generalized to any number $n$ of points. Moreover, 
they also hold when reexpressing the real scalars $\phi_I$ in terms 
of the complex scalars $\phi_{AB}$.
This allows us to perform the calculations in the FDH scheme, using the superfields \eqref{ossuperfield} of $\mathcal{N}=4$ on-shell super space, as done in the previous section. 
We formulate the modification necessary to obtain the correct results of the DR scheme explicitly below.

In the previous section, we have split the form factor ratios of $\mK_6$ 
as
\begin{equation}
 f_{\mK_6,n}^{(\ell)}
=f_{{\BPS},n}^{(\ell)}+{\tilde f}_{\mK_6,n}^{(\ell)}
\col
\label{eq: K6decomposition}
\end{equation}
where $f_{\BPS,n}$ coming from (\subref{subfig: a}) and (\subref{subfig: b}) in figures \ref{fig: flavor flows} and \ref{fig: flavor flows fermion}
is the part identical to the BPS form factor 
and ${\tilde f}_{\mK_6,n}$ coming from (\subref{subfig: c}) is the part unique 
for the operator \eqref{Kinsu4}. 
To obtain the form factor ratio for $\mK$
\begin{equation}
 f_{\mK,n}^{(\ell)}
=f_{{\BPS},n}^{(\ell)}+{\tilde f}_{\mK,n}^{(\ell)}
\col
\label{eq: Kdecomposition}
\end{equation}
we apply the replacement rule
\begin{equation}
{\tilde f}_{\mK_6,n}^{(\ell)}
\quad \xrightarrow{r_\phi} \quad 
r_\phi{\tilde f}_{\mK_6,n}^{(\ell)}={\tilde f}_{\mK,n}^{(\ell)}
\col
\label{eq: ftilde-correction-rule}
\end{equation}
where $r_\phi$ is defined in \eqref{rphi}. According to our discussion, it should be valid to all loop orders.\footnote{This statement relies on the validity of the DR scheme, which, however, is known to have inconsistencies at 
higher loop orders \cite{Siegel:1980qs,Avdeev:1981vf,Avdeev:1982np,Avdeev:1982xy}.}

We have focused on the form factor of the Konishi operator. A similar discussion should also be applicable to other operators containing a contraction of flavor or vector indices. As in the Konishi case, it is essential to be able to formulate the results in terms of two parts, one that contains an externally closed index loop and the other that does not. The part without externally closed index loop should be independently computable, such as the BPS part in the Konishi form factor. Given such a decomposition, one can then use the efficient on-shell techniques, together with a simple modification rule as \eqref{eq: ftilde-correction-rule}. 
Another example of an operator with contracted flavor indices is $\tr(\phi_I\phi_I\phi_K)$, which has one-loop anomalous dimension $8$. 
An example with contracted vector indices is $\tr(D^\mu\phi_{12}D_\mu\phi_{12})$, which has one-loop anomalous dimension $12$. In the latter case, the differences in the one-loop two-point form factor between intermediate states in $D=4$ and $D=4-2\peps$ dimensions are precisely given by the rational terms in the PV reduction formula \eqref{eq: PV reduction of triangles}. 
Similarly, $f_{\mK,n}^{(1)}$ and $f_{\mK_6,n}^{(1)}$ differ by rational terms that are introduced by the replacement \eqref{eq: ftilde-correction-rule};  these rational terms arise when multiplying the $\frac{1}{\peps}$-pole from the bubble integral with the term in $r_\phi$ that is linear in $\peps$.

\subsection{Final Konishi form factors \label{sec:final-konishi-ff} }

Finally, we list the non-vanishing results for the Konishi form factor, which will be used as the input in the next section to calculate the cross section. They can be obtained by using the form factors computed in section \ref{sec:ffactor} and integral results in appendix \ref{app:integrals}. 
Note that the ${\tilde f}_{\mK}$ parts have been modified by the $r_\phi$ factor according to the prescription \eqref{eq: ftilde-correction-rule}.  The full form factors can be obtained as ${f}_{\mK,n}^{(\ell)} = f_{{\BPS},n}^{(\ell)}+{\tilde f}_{\mK,n}^{(\ell)}$. In the following equations, $(\frac{\mu^2}{-q^2})^{\ell\peps}$ is always understood
as $(\frac{\mu^2}{-q^2-i0})^{\ell\peps}$ and analogously for $s_{ab}$.

\subsubsection*{Two-point one-loop}

\begin{equation}
\label{eq:final-2pt-1loop}
\begin{aligned}
f_{{\BPS},2}^{(1)} &= \Big({\mu^2\over - q^2}\Big)^{\peps}  \left[- {2\over \peps^2} +  {\pi^2 \over6} + {14\over 3} \zeta_3 \peps + {47\over 720} \pi^4 \, \peps^2 \right] + {\cal O} (\peps^3) \col 
\\
{\tilde f}_{\mK,(\phi,\phi)}^{(1)} & = \Big({\mu^2\over - q^2}\Big)^{\peps} \left[- {6\over \peps} - 14 - \Big(28 - {\pi^2 \over2} \Big) \peps - \Big( 56 - {7\pi^2 \over6} - 14 \zeta_3 \Big) \peps^2 \right] + {\cal O} (\peps^3) \pnt
\end{aligned}
\end{equation}

\subsubsection*{Two-point two-loop}

\begin{equation}
\label{eq:final-2pt-2loop}
\begin{aligned}
f_{{\BPS},2}^{(2)} &= \Big({\mu^2\over - q^2}\Big)^{2\peps} \left[ {2\over \peps^4} -  {\pi^2 \over6 \peps^2} - {25  \zeta_3\over 3 \peps} - {7 \pi^4 \over 60} \right] + {\cal O} (\peps) \col
\\
{\tilde f}_{\mK,(\phi,\phi)}^{(2)} &= \Big({\mu^2\over - q^2}\Big)^{2\peps}  \left[ {12\over \peps^3} +  {46 \over \peps^2} + {152 - 2\pi^2 \over \peps} + \Big( 484 - {35 \pi^2 \over 3} - 56 \zeta_3\Big) \right] + {\cal O} (\peps) \pnt
\end{aligned}
\end{equation}

\subsubsection*{Three-point one-loop}

\begin{equation}
\label{eq:final-3pt-1loop}
\begin{aligned}
f_{{\BPS},3}^{(1)}  = & -\frac{c_\Gamma}{\peps^2} \Big[ \Big({\mu^2\over - s_{12}}\Big)^{\peps} +
\Big({\mu^2\over - s_{23}}\Big)^{\peps}  + \Big({\mu^2\over - s_{31}}\Big)^{\peps}  \Big] \\
 &  + \text{FB}(p_1,p_2,p_3,-q) + \text{FB}(p_2,p_3,p_1,-q) + \text{FB}(p_3,p_1,p_2,-q) \, ,
\\
{\tilde f}_{\mK,(\phi,\phi,g)}^{(1)} 
= & - {r_\phi c_\Gamma \over \peps(1-2\peps)} \bigg\{ 3 \left[ 1+ \frac{s_{13}^2+s_{23}^2}{(s_{13}+s_{23})^2} \right]
 \Big({\mu^2\over - q^2}\Big)^{\peps}  
+ \frac{ 6 s_{13}s_{23}}{(s_{13}+s_{23})^2} \Big({\mu^2\over - s_{12}}\Big)^{\peps}  \\
&+ \frac{ 12 s_{13}s_{23}}{s_{12}(s_{13}+s_{23})} {\peps \over (2-2\peps) } {1\over s_{13} + s_{23}} \Big[s_{12} \Big({\mu^2\over - s_{12}}\Big)^{\peps}  - q^2 \Big({\mu^2\over - q^2}\Big)^{\peps}  \Big] \bigg\}
\col
\\
{\tilde f}_{\mK,(\psi, \psi, \phi)}^{(1)} = & 
- {r_\phi c_\Gamma \over \peps(1-2\peps)} \bigg\{ 3 \left(1+ \frac{s_{12}-s_{13}}{s_{12}+s_{13}}\right)
 \Big({\mu^2\over - s_{23}}\Big)^{\peps} 
+ 3 \left( 1+  \frac{s_{12}-s_{23}}{s_{12}+s_{23}} \right)
 \Big({\mu^2\over - s_{13}}\Big)^{\peps}  \\
& - 3\left( \frac{s_{12}-s_{13}}{s_{12}+s_{13}} + \frac{s_{12}-s_{23}}{s_{12}+s_{23}}\right)
 \Big({\mu^2\over - q^2}\Big)^{\peps}  \bigg\} - 3 r_\phi  \, \text{FB}(p_2,p_3,p_1,-q) 
\col
\end{aligned}
\end{equation}
where $c_\Gamma$ is given in \eqref{famous-c_Gamma}, $q^2 = s_{12}+s_{23}+s_{31}$
and we have defined the (rescaled) finite part of the one-mass box integral as
\begin{equation}
\label{eq:finite-Box}
\begin{aligned}
&\text{FB}(p_1,p_2,p_3,-q) \\
&\qquad=  -\frac{c_\Gamma}{\peps^2} \bigg[ \Big({\mu^2\over - s_{12}}\Big)^{\peps}  h\Big(-\frac{s_{31}}{s_{23}}\Big) + \Big({\mu^2\over - s_{23}}\Big)^{\peps}  h\Big(-\frac{s_{31}}{s_{12}}\Big) 
- \Big({\mu^2\over - q^2 }\Big)^{\peps}  h\Big(-\frac{s_{31} q^2}{s_{12} s_{23}}\Big)  \bigg]
\end{aligned}
\end{equation}
with $h(x) = \mbox{}_2F_1(1,-\peps,1-\peps,x)-1$.

%% file: crosssection.tex
\section{BPS and Konishi cross sections}
\label{sec:CS} 

In this section, we compute the cross section discussed in section \ref{section:CSNutshell}. We first discuss in detail the case of the BPS operator \eqref{BPSinso6} as a warm-up example.
Then, we compute one of our main results: the Konishi cross section to two-loop order. We will use the Konishi form factors $f_{\mK,n}^{(\ell)}$ and ${\tilde f}_{\mK,n}^{(\ell)}$ given in subsection \ref{sec:final-konishi-ff}
that were obtained from the form factors of $\mK_6$ computed in section \ref{sec:ffactor} by applying the prescription of section \ref{sec: subtleties in unitarity}.

\subsection{BPS cross section up to one-loop order\label{sec:BPS-CS} }

As a warm-up, we first consider in detail the cross section  corresponding to the imaginary part of the two-point correlation function of the BPS operator $\tr(\phi_{12}^2)$ and its conjugate $\tr(\phi_{34}^2)$, $\langle 0 |\tr(\phi_{12}^2)(x) \tr(\phi_{34}^2)(0) | 0 \rangle$.
Since the operators are protected, the cross section has no UV-divergent loop corrections. Moreover, finite corrections do not occur either \cite{Gubser:1997se,Anselmi:1997am}, i.e.\
\begin{equation}
\sigma_{\BPS} = \sigma_{\BPS}^{(0)} + {\cal O}(\peps) \pnt 
\label{sigmaBPS}
\end{equation}
We check this explicitly up to one-loop level.
\begin{figure}[tbp]
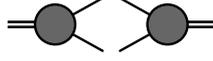

\begin{center}
\FDff[legs=2,scaled=0.5] \FDff[legs=2,reflected,scaled=0.5]
\caption{Tree-level squared matrix element.}
\label{fig:cs-tree}
\end{center}
\end{figure}

\begin{figure}[tbp]
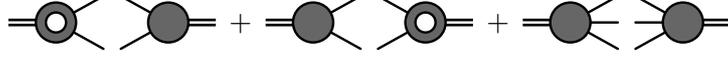

\begin{center}
 \FDff[oneloop, legs=2,scaled=0.5] \FDff[legs=2,reflected,scaled=0.5]
$+$ \FDff[legs=2,scaled=0.5] \FDff[oneloop, legs=2,reflected,scaled=0.5] 
$+$ \FDff[legs=3,scaled=0.5] \FDff[legs=3,reflected,scaled=0.5]
\caption{One-loop squared matrix elements.}
\label{fig:cs-oneloop}
\end{center}
\end{figure}

\subsection*{Tree level}

Let us start with the tree-level cross section. The squared matrix element, as shown in figure \ref{fig:cs-tree}, is the product of two two-point tree-level BPS form factors, one for $\tr(\phi_{12}^2)$ and one for its conjugate $\tr(\phi_{34}^2)$.
The tree-level non-color-ordered BPS super form factor can be obtained from \eqref{eq: non-color-ordered-super-FF} and \eqref{BPS-MHV-tree}.
It is easy to perform the color factor summation and the fermionic integration.
This yields the squared matrix element
\begin{equation}\label{MBPS2(0)}
{\cal M}_{{\BPS},2}^{(0)} = {1\over2!} \sum_{a_1, a_2} \int \de^4 \eta_1 \de^4 \eta_2 \,{\ncsusyF}_{{\BPS}}^{(0)}(1,2) \, {\ncsusyF}_{{\BPS}}^{*(0)}(1,2) =  {N_{\text{c}} ^2 - 1 \over2} \pnt
\end{equation}
The tree-level cross section is given by the integral \eqref{PS-2pt} of ${\cal M}_{{\BPS},2}^{(0)}$ over the two-particle phase space in $D=4-2\epsilon$ dimensions. This yields
\begin{equation}
\sigma_{\BPS}^{(0)} = \int \de {\rm PS}_2 \, {\cal M}^{(0)}_{{\BPS},2}  =  
\Big(\frac{\mu^2}{q^2}\Big)^\peps
{1 \over 4 (16\pi)^{{1\over2}-\peps} \, \Gamma({3\over2}-\peps)} \, { N_{\text{c}} ^2 -1 \over2}  \pnt
\end{equation}

\subsection*{One loop}

The one-loop cross section is given by the sum of a two-particle and a three-particle channel, as shown in figure \ref{fig:cs-oneloop}:
\begin{equation}\label{sigmaBPS(1)}
\sigma_{\BPS}^{(1)} = \int \de {\rm PS}_2 \, {\cal M}^{(1)}_{{\BPS},2} + \frac{1}{g^2}\int \de {\rm PS}_3 \, {\cal M}^{(0)}_{{\BPS},3} \pnt 
\end{equation}

\subsubsection*{Two-particle channel}

The squared matrix element of the two-particle channel corresponds to the first two graphs of figure \ref{fig:cs-oneloop}. As an equation, it reads
\begin{equation}
\begin{aligned}\label{MBPS2(1)}
{\cal M}^{(1)}_{{\BPS},2} & = {1\over2!}\sum_{a_1, a_2} \int \de^4 \eta_1 \de^4 \eta_2 \, \Big[ \ncsusyF_{{\BPS},2}^{(1)} \, {\ncsusyF}_{{\BPS},2}^{*(0)} + \ncsusyF_{{\BPS},2}^{(0)} \, {\ncsusyF}_{{\BPS},2}^{*(1)} \Big]
=2{\cal M}^{(0)}_{{\BPS},2}\, \Re\big(\ratioF_{{\BPS},2}^{(1)}\big)
\col
\end{aligned}
\end{equation}
where $\Re$ denotes the real part, and 
$\ratioF_{\mathcal{O},n}^{(\ell)}$ is the ratio between the $\ell$-loop and tree-level $n$-point form factor of the operator $\mathcal{O}$ as defined in \eqref{normFF}. The tree-level form factor is absorbed into ${\cal M}^{(0)}_{{\BPS},2}$.
For short notation, we denote $\ncsusyF_{{\cal O}}(1,\dots,n)$ as $\ncsusyF_{{\cal O},n}$.

There is an important point related to the $i0$-prescription to be explained here. The two-point form factors acquire a factor of $(-q^2- i0)^{-\peps}$ for each loop. 
The function ${\ratioF}^{*(1)}_{{\BPS},2}$ is the complex conjugate of ${\ratioF}^{(1)}_{{\BPS},2}$ and can be obtained from the latter by replacing $(-q^2 - i0)^{-\peps}$ with $(-q^2 + i0)^{-\peps}$.
The sum of both terms amounts to taking the real part of ${\ratioF}^{(1)}_{{\BPS},2}$. 
Hence, we need the real part of $(-q^2 \pm i0)^{-\peps}$, which for $q^2>0$ is given by (see e.g. \cite{vanNeerven:1985xr})
\begin{equation}
\Re (-q^2 \pm i0)^{x} = {\Gamma(1+x) \Gamma(1-x) \over \Gamma(1+2x)\Gamma(1-2x)} (q^2)^{x} \pnt
\label{eq:taking-real-q^2}
\end{equation}

Using this result to determine the real part of the form factor \eqref{eq:final-2pt-1loop} and then inserting it into \eqref{MBPS2(1)} together with the tree-level
result \eqref{MBPS2(0)} and performing 
the two-particle phase space integral \eqref{PS-2pt}, we obtain for the first
term in \eqref{sigmaBPS(1)}:
\begin{equation}
\sigma_{\BPS,2}^{(1)} = \int \de {\rm PS}_2 \, {\cal M}^{(1)}_{{\BPS},2} = \sigma_{\BPS}^{(0)} \, \Big(\frac{\mu^2}{q^2}\Big)^\peps \left( -{4\over \peps^2} + {7 \pi^2 \over 3} \right) + {\cal O}(\peps) \pnt
\label{sigma-bps-1loop2pt}
\end{equation}

\subsubsection*{Three-particle channel}

The squared matrix element of the three-particle channel is given by the last graph of figure \ref{fig:cs-oneloop}. 
The MHV and NMHV tree-level three-point non-color-ordered  form factor \eqref{eq: non-color-ordered-super-FF} can be obtained using \eqref{BPS-MHV-tree} and \eqref{BPS-3pt-NMHV-tree}.
Performing the color summation and fermionic integration, we find the squared matrix element
\begin{equation}
\begin{aligned}
 {\cal M}^{(0)}_{{\BPS},3} & = {1\over 3!} \sum_{a_1, a_2, a_3} \int \de^4 \eta_1 \de^4 \eta_2 \de^4 \eta_3 \, \Big[ \ncsusyF_{{\BPS},3}^{ {\rm MHV}, (0)}\, \ncsusyF_{{\BPS},3}^{*{\rm NMHV}, (0)} + \ncsusyF_{{\BPS},3}^{ {\rm NMHV}, (0)} \, \ncsusyF_{{\BPS},3}^{*{\rm MHV}, (0)} \Big]  \\ 
& = {2\over3} g_{\YM}^2 N_{\text{c}} \, (N_{\text{c}} ^2 -1) \, {(q^2)^2 \over s_{12} s_{23} s_{31} } 
\pnt
\end{aligned}
\end{equation}
Performing the three-particle phase space integral via 
\eqref{PS-3pt}, we obtain for the second term in \eqref{sigmaBPS(1)}:
\begin{equation}
\sigma_{\BPS,3}^{(1)} = \frac{1}{g^2}\int \de {\rm PS}_3 \, {\cal M}^{(0)}_{{\BPS},3} = \sigma_{\BPS}^{(0)}  \, \Big(\frac{\mu^2}{q^2}\Big)^\peps \left( {4\over \peps^2} - {7 \pi^2 \over 3} \right) + {\cal O}(\peps) \pnt
\label{sigma-bps-1loop3pt}
\end{equation}

Summing \eqref{sigma-bps-1loop2pt} and \eqref{sigma-bps-1loop3pt} together as prescribed by \eqref{sigmaBPS(1)}, we see that both contributions cancel and hence that \eqref{sigmaBPS} holds at one-loop level.

\subsection{Konishi cross section up to two-loop order \label{sec:konishi-cs} }

Next, we compute the Konishi cross section. We start with the discussion of an important simplification for the computation, which exploits the fact that a part of the Konishi cross section is identical to the BPS cross section \eqref{sigmaBPS}, which is protected.

At tree level, the squared matrix elements of the Konishi and the BPS cross section satisfy the following simple relation\footnote{Note that the tree-level  Konishi form factors with specified external legs are identical to the corresponding BPS form factors.}
\begin{equation}
\begin{aligned}\label{MKMBPStreerel}
{\cal M}^{(0)}_{\mK,n} = \sum_{\rm colors} \, \sum_{\substack{{\rm spins}\\{\rm helicities}}} \, \ncsusyF^{(0)}_{\mK,n} \ncsusyF^{*(0)}_{\mK,n}  \,= \, 
6 \sum_{\rm colors} \, \sum_{\substack{{\rm spins}\\{\rm helicities}}} \, \ncsusyF^{(0)}_{{\BPS},n} \ncsusyF^{*(0)}_{{\BPS},n} = 
6 \, {\cal M}^{(0)}_{{\BPS},n} \col
\end{aligned}
\end{equation}
where the factor $6$ originates from the contribution of all scalar flavor 
degrees of freedom in the two-point function of the Konishi operator \eqref{Kinso6}, which does not occur for the BPS 
operator.\footnote{The additional prefactor compared to the BPS result comes from the trace over the scalar degrees of freedom, which here (in the FDH scheme) is $6$. 
In the DR scheme, one would have to replace $6$ by $N_\phi = 6 + 2\peps$. 
In any case, this factor cancels out when the cross section is divided by the
tree-level cross section as e.g.\ in \eqref{sigmaratio}.}

Furthermore, the loop correction to the Konishi form factor can be written as linear combination of two contributions as defined in \eqref{eq: Kdecomposition}: one that is identical to the BPS form factor and the other that is unique for the Konishi operator. 
We can introduce a corresponding squared matrix element that includes a subtraction of the BPS part as
\begin{equation}
\tilde{\cal M}^{(\ell)}_{\mK,n}= {\cal M}^{(\ell)}_{\mK,n} - 
6 \, {\cal M}^{(\ell)}_{{\BPS},n} \col
\label{calMtilde}
\end{equation}
where the factor $6$
takes into account that at any loop order $\ell$ the contribution ${\cal M}^{(\ell)}_{{\BPS},n}$ built from two BPS-type components of the Konishi form factor receives a factor $6$
as in \eqref{MKMBPStreerel}.

Since the BPS cross section \eqref{sigmaBPS} receives no loop corrections,
\begin{equation}
\label{eq:sigma-Mtilde}
\sigma_\mK^{(\ell)} 
= \sum_{n=2}^{\ell+2}  g^{2(2-n)} \int \de {\rm PS}_n{\cal M}^{(\ell+2-n)}_{\mK, n}
= \sum_{n=2}^{\ell+1}  g^{2(2-n)} \int \de {\rm PS}_n \tilde{\cal M}^{(\ell+2-n)}_{\mK, n} \col \quad \ell \geq1 \pnt
\end{equation}
Note in particular that $\tilde{\cal M}^{(0)}_{\mK,n}=0$. Hence, the sum over $n$
can be terminated already at $\ell+1$. As we will see, 
this simplifies the computation dramatically.

From \eqref{MKMBPStreerel}, it immediately follows that the tree-level
cross section for the Konishi operator also contains an extra factor $6$
compared to the one of the BPS-operator, i.e.\ 
\begin{equation}
\sigma^{(0)}_\mK =  \int \de {\rm PS}_2 \, {\cal M}^{(0)}_{\mK,2} = 
6 \, \sigma^{(0)}_\BPS \pnt
\end{equation}
At loop-level, it is convenient to 
factor out $\sigma^{(0)}_\mK$.

\subsubsection{One-loop result}

The bare one-loop Konishi cross section receives contributions from 
products of tree-level and one-loop two-point form factors and of tree-level
three-point form factors as shown in figure \ref{fig:cs-oneloop}.
The squared matrix element of the two-particle channel is given by
\begin{equation}
\begin{aligned}
{\cal M}^{(1)}_{\mK,2} & = {1\over2!}   \sum_{a_1, a_2} \int \de^4 \eta_1 \de^4 \eta_2 \, \Big( \ncsusyF_{\mK,2}^{(1)}  \, \ncsusyF_{\mK,2}^{*(0)} + \ncsusyF_{\mK,2}^{(0)}  \, \ncsusyF_{\mK,2}^{*(1)} \Big) 
= 2{\cal M}^{(0)}_{\mK,2} \, \Re\big({f}_{\mK,(\phi,\phi)}^{(1)}\big)
 \col
\label{M2-Konishi-1loop}
\end{aligned}
\end{equation}
where we use the abbreviation $f^{(1)}_{\mK,(\phi,\phi)}=f^{(1)}_{\mK,2}(1_{\phi_{12}}, 2_{\phi_{34}})$.\footnote{The tree-level two-point form factor must contain two external scalar legs to be non-vanishing. Therefore, it is not necessary to consider one-loop form factors with other external states such as $f_{\mK,2}^{(1)}(1_{g_+}, 2_{g_-})$, which are moreover zero as shown in section \ref{sec:ffactor}. A similar argument applies also to the following two-loop computation.}

As discussed above, the result for the three-particle channel cancels with the BPS part in the two-particle channel. Therefore, we can subtract the BPS part from \eqref{M2-Konishi-1loop}, as in \eqref{calMtilde}.
This yields
\begin{equation}
\tilde{\cal M}^{(1)}_{\mK,2} = 2{\cal M}^{(0)}_{\mK,2}  \, \Re\big( \tilde{f}^{(1)}_{\mK,(\phi,\phi)}\big)\col
\label{tilde-calM2-Konishi-1loop}
\end{equation}
where $\tilde{f}^{(1)}_{\mK,(\phi,\phi)}$ is given in \eqref{eq:final-2pt-1loop}.
Performing the two-particle phase space integral \eqref{PS-2pt}, the one-loop bare Konishi cross section reads
\begin{equation}
\sigma_{\mK}^{(1)} =  \int \de {\rm PS}_2 \, \tilde{\cal M}^{(1)}_{\mK,2} 
= \sigma^{(0)}_\mK\, \Big(\frac{\mu^2}{q^2}\Big)^\peps \left( - {12\over \peps} - 28 \right) + {\cal O}(\peps) \pnt
\label{sigma(1)-B}
\end{equation}

The divergence in \eqref{sigma(1)-B} has to be canceled by the 
one-loop correction of the Konishi operator obtained from the one-loop
term $\mathcal{Z}^{(1)}_\mK$ in the operator renormalization 
constant $\mathcal{Z}_\mK$. As  shown in figure \ref{fig:cs-Konishi-calZ1}, 
$\mathcal{Z}^{(1)}_\mK$ contributes as
\begin{equation}\label{sigmaZ11}
\sigma_{{\cal Z}^{(1)}\mK}^{(1)} 
= 2 {\cal Z}^{(1)} \sigma^{(0)}_\mK \pnt
\end{equation}
Requiring the sum of \eqref{sigma(1)-B} and \eqref{sigmaZ11}
to be finite, we immediately find
\begin{equation}
\label{eq:calZ-1loop}
 {\cal Z}^{(1)}_\mK = 
{6 \over \peps} \pnt
\end{equation}
Comparing this result with the one-loop term of 
the expansion \eqref{eq: relation of calZ and gamma} reproduces 
the known one-loop Konishi anomalous dimension $\gamma^{(1)}_\mK = 12$, which was
first obtained in \cite{Anselmi:1996mq,Anselmi:1996dd}.

\begin{figure}[t]
\begin{center}
$\Big({\cal Z}^{(1)}_\mK$\FDff[legs=2,scaled=0.5]\Big)\FDff[legs=2,reflected,scaled=0.5] + \FDff[legs=2,scaled=0.5]\Big(\FDff[legs=2,reflected,scaled=0.5]${\cal Z}^{(1)}_\mK$\Big)
\caption{One-loop correction  from one-loop renormalization constant.}
\label{fig:cs-Konishi-calZ1}
\end{center}
\end{figure}

The renormalized one-loop cross section is hence given by
\begin{equation}\label{sigmaren1}
\sigma_{\cal K,\ren}^{(1)} =  \sigma_{\mK}^{(1)}  + \sigma_{{\cal Z}^{(1)}\mK}^{(1)}  = \sigma^{(0)}_\mK\Big( {12 \log\frac{q^2}{\mu^2}} 
- 28 \Big) + {\cal O}(\peps)  \pnt
\end{equation}
As predicted in \eqref{logsigmarengeneral}, the coefficient of $\log\frac{q^2}{\mu^2}$ also reproduces the correct one-loop anomalous dimension.

\subsubsection{Two-loop result}

\begin{figure}[h]
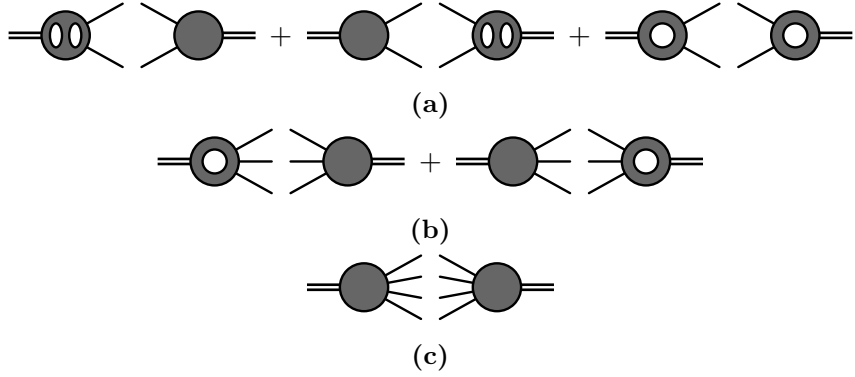

\begin{center}
\begin{subfigure}[c]{\textwidth}
\centering
\FDff[twoloop, legs=2,scaled=0.6] \FDff[legs=2,reflected,scaled=0.6]  
$+$ \FDff[legs=2,scaled=0.6] \FDff[twoloop, legs=2,reflected,scaled=0.6]  
$+$ \FDff[oneloop, legs=2,scaled=0.6] \FDff[oneloop, legs=2,reflected,scaled=0.6] 
\caption{}
\label{fig:cs-2loop-a}
\end{subfigure}
\\ 
\begin{subfigure}[c]{\textwidth}
\centering
\FDff[oneloop, legs=3,scaled=0.6] \FDff[legs=3,reflected,scaled=0.6]  
$+$ \FDff[legs=3,scaled=0.6] \FDff[oneloop, legs=3,reflected,scaled=0.6] 
\caption{}
\label{fig:cs-2loop-b}
\end{subfigure}
\\ 
\begin{subfigure}[c]{\textwidth}
\centering
\FDff[legs=4,scaled=0.6] \FDff[legs=4,reflected,scaled=0.6]  
\caption{}
\label{fig:cs-2loop-c}
\end{subfigure}
\caption{The two-loop bare squared matrix element.}
\label{fig:cs-2loop}
\end{center}
\end{figure}

The two-loop cross section is obtained from the contributions to the squared matrix elements depicted in figure \ref{fig:cs-2loop}. 
As discussed at the beginning of this subsection, we can neglect the contribution that is proportional to the BPS cross section. In particular, it is not necessary to consider the contribution from the four-particle channel in figure \ref{fig:cs-2loop-c}, which involves the complicated four-particle 
phase space integral.
This simplifies the computation significantly. In the following, we separately compute the contributions from the two-particle and three-particle channel.

\renewcommand{\thesubfigure}{\alph{subfigure}}
\setcounter{subfigure}{0}

\subsubsection*{Two-particle channel}

The full contribution of the two-particle channel consists of the three terms 
\begin{equation}\label{twoparticlefull}
\int \de {\rm PS}_2 \, \tilde{\cal M}^{(2)}_{\mK,2} + \int \de {\rm PS}_2 \, {\cal M}^{(2)}_{{\cal Z}^{(1)}\mK,2} + \int \de {\rm PS}_2 \, {\cal M}^{(2)}_{{\cal Z}^{(2)}\mK,2}  \col
\end{equation}
where the first term is the bare contribution, and the second and the third term 
involve the one- and two-loop contributions of the renormalization constant 
$\mathcal{Z}_\mK$, respectively.
We compute the first two terms. After having considered also the three-particle channel, we will determine ${\cal Z}^{(2)}_\mK$ from the 
condition that all divergences are canceled.

The squared matrix element obtained from the bare form factors is shown in figure \ref{fig:cs-2loop-a}. In analogy to \eqref{tilde-calM2-Konishi-1loop}, the first two graphs 
yield
\begin{equation}
\tilde{\cal M}^{(2),\textrm{I}}_{\mK,2} = 
2{\cal M}^{(0)}_{\mK,2} \, \Re\big(\tilde {f}_{\mK,(\phi,\phi)}^{(2)}\big)\col
\label{tilde-calM-2pt-2loop-a}
\end{equation}
where $\tilde {f}_{\mK,(\phi,\phi)}^{(2)}$ is given in \eqref{eq:final-2pt-2loop}.

The third graph of figure \ref{fig:cs-2loop-a} has no lower-loop counterpart and needs to be discussed in detail. It is the product of two one-loop Konishi form factors, and each of them is a linear combination of the BPS part ${f}^{*(1)}_{{\BPS},2}$ and the additional $\tilde{f}_{\mK,2}^{(1)}$ part. 
After subtracting the product of two BPS parts, we obtain%
\begin{equation}
\tilde{\cal M}^{(2),\textrm{II}}_{\mK,2} = {\cal M}^{(0)}_{\mK,2} \, \Big[ 2\,\Re\Big( \tilde{f}_{\mK,(\phi,\phi)}^{(1)} {f}^{*(1)}_{{\BPS},2} \Big) +  \tilde{f}_{\mK,(\phi,\phi)}^{(1)} \tilde{f}_{\mK,(\phi,\phi)}^{*(1)} \Big] \col
\label{M2-simp-1}
\end{equation}
where the form factors are given in \eqref{eq:final-2pt-1loop}.\footnote{Recall that for the Konishi form factors we have to specify the external legs while for the universal BPS ones this is not necessary.}

Integrating the sum of the two previous contributions 
over the two-particle phase space \eqref{PS-2pt} yields the
bare cross section of the two-particle channel with its BPS part subtracted. 
It explicitly reads
\begin{equation}
\begin{aligned}
\tilde\sigma_{\mK,2}^{(2)} 
&= \int \de {\rm PS}_2 \, \big( \tilde{\cal M}^{(2),\textrm{I}}_{\mK,2} + \tilde{\cal M}^{(2),\textrm{II}}_{\mK,2} \big) \\
&= \sigma^{(0)}_\mK  \, \Big(\frac{\mu^2}{q^2}\Big)^{2\peps} \bigg[ {48 \over \peps^3} + {184 \over \peps^2} + {584 - 56\pi^2 \over \peps} + {1724 - {668 \over3}\pi^2 - 224 \zeta_3}  \bigg] + {\cal O}(\peps) 
\pnt
\label{eq: sigma-2loop-bare-2pt}
\end{aligned}
\end{equation}
%

\begin{figure}[t]
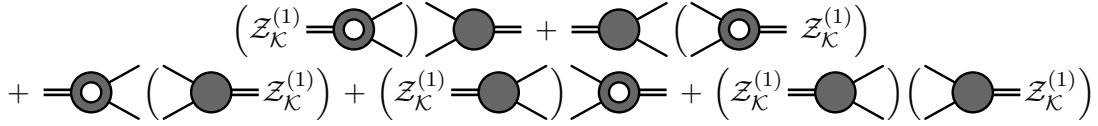

\begin{center}
$\Big({\cal Z}^{(1)}_\mK$\FDff[oneloop,legs=2,scaled=0.5]\Big)\FDff[legs=2,reflected,scaled=0.5] + \FDff[legs=2,scaled=0.5]\Big(\FDff[oneloop,legs=2,reflected,scaled=0.5] ${\cal Z}^{(1)}_\mK$\Big) \\ + \FDff[oneloop,legs=2,scaled=0.5]\Big(\FDff[legs=2,reflected,scaled=0.5]${\cal Z}^{(1)}_\mK$\Big)  + $\Big({\cal Z}^{(1)}_\mK$\FDff[legs=2,scaled=0.5]\Big)\FDff[oneloop,legs=2,reflected,scaled=0.5] + $\Big({\cal Z}^{(1)}_\mK$\FDff[legs=2,scaled=0.5]\Big)\Big(\FDff[legs=2,reflected,scaled=0.5]${\cal Z}^{(1)}_\mK$\Big)
\caption{The contribution from the one-loop renormalization constant to the two-particle channel.}
\label{fig:cs-Konishi-2pt-calZ1-2loop}
\end{center}
\end{figure}
Next, we consider the contribution involving the one-loop renormalization constant. %
It is shown in figure \ref{fig:cs-Konishi-2pt-calZ1-2loop} and leads to the squared matrix element
\begin{equation}
\begin{aligned}\label{calMZ1K2}
{\cal M}^{(2)}_{{\cal Z}^{(1)} \mK,2} 
= 
{\cal M}^{(0)}_{\mK,2} 
\Big[ 4\Re\big({f}_{\mK,(\phi,\phi)}^{(1)}\big){\cal Z}^{(1)}_\mK  
+ \big( {\cal Z}^{(1)}_\mK \big)^2 \Big]
\pnt
\end{aligned}
\end{equation}
Inserting the explicit expressions \eqref{eq:final-2pt-1loop} into \eqref{calMZ1K2} and
performing the two-particle phase space integration \eqref{PS-2pt}, we obtain 
\begin{equation}
\begin{aligned}
\sigma_{{\cal Z}^{(1)} \mK, 2}^{(2)} 
& = \int \de {\rm PS}_2 \, {\cal M}^{(2)}_{{\cal Z}^{(1)} \mK,2} \\ 
& 
=  \sigma^{(0)}_\mK \bigg[ \Big(\frac{\mu^2}{q^2}\Big)^\peps 
\bigg( -{48 \over \peps^3} - {144 \over \peps^2} - { 336 - 28 \pi^2 \over \peps} - 672 + 84 \pi^2 + 112 \zeta_3  \bigg) \\ 
& \phantom{{}={}\sigma^{(0)}_\mK \bigg[} 
+  
{36 \over \peps^2} - { 12 \over \peps} +4\bigg] + {\cal O}(\peps) \pnt
\label{eq: sigma-2loop-calZ1-2pt}
\end{aligned}
\end{equation}
%

\subsubsection*{Three-particle channel}

There are two contributions to the two-loop cross section in the three-particle channel:
\begin{equation}\label{threeparticlestwoloop}
\frac{1}{g^2}\bigg[\int \de {\rm PS}_3 \, \tilde{\cal M}^{(1)}_{\mK,3} + \int \de {\rm PS}_3 \, {\cal M}^{(1)}_{{\cal Z}^{(1)} \mK,3}\bigg]\pnt
\end{equation}

The contribution involving the bare form factor is determined from the diagrams 
of figure \ref{fig:cs-2loop-b}. The resulting expression reads
\begin{equation}
\begin{aligned}
{\cal M}^{(1)}_{\mK,3} 
& = {1\over 3!}\sum_{a_i} \int \prod_{i=1}^3 \de^4 \eta_i \, \sum_{\ell=0}^1 \Big[ \ncsusyF_{\mK,3}^{(\ell), {\rm MHV}} \ncsusyF_{\mK,3}^{*(1-\ell), {\rm NMHV}}  +  \ncsusyF_{\mK,3}^{(\ell),{\rm NMHV}} \ncsusyF_{\mK,3}^{*(1-\ell), {\rm MHV}} \Big] \\  
& = 6 {\cal M}^{(0)}_{\mK,3} \,  \bigg[ 2\Re\Big( {f}_{\mK,(\phi,\phi,g)}^{(1)} \Big) \, {s_{12}^2 \over (q^2)^2} + 2\Re\Big({f}_{\mK,(\psi,\psi,\phi)}^{(1)} \Big) \, {s_{13} s_{23} \over (q^2)^2} \bigg] \pnt
\label{eq: calM-bare-3pt}
\end{aligned}
\end{equation}
In the second line, we have not indicated the MHV degree, since the loop correction is the same for the MHV and the NMHV form factor. This allows us to use the abbreviation ${f}_{\mK,3}^{(1)}={f}_{\mK,3}^{(1),{\rm MHV}}={f}_{\mK,3}^{(1),{\rm NMHV}}$
for any fixed three-particle final state. Moreover, we have abbreviated 
the form factors of the two different final states as ${f}_{\mK,(\phi,\phi,g)}^{(1)}={f}_{\mK}^{(1)}(1_{\phi_{12}}, 2_{\phi_{34}}, 3_{g_+})$ and ${f}_{\mK,(\psi,\psi,\phi)}^{(1)} = {f}_{\mK}^{(1)}(1_{\psi_1}, 2_{\psi_2}, 3_{\phi_{34}})$. 
Since the one-loop corrections for different external states differ from each other, as given in \eqref{eq:final-3pt-1loop},
we need to treat the contribution of these two form factors separately.
The factors ${s_{12}^2 \over (q^2)^2}$ and 
${s_{13} s_{23} \over (q^2)^2}$ stem from the squares of the corresponding tree-level form factors divided by the tree-level matrix element  ${\cal M}^{(0)}_{\mK,3}$.

After subtracting the BPS part, we find
\begin{equation}
\tilde{\cal M}^{(1)}_{\mK,3}  = 6 {\cal M}^{(0)}_{\mK,3} \, \bigg[ 2\,\Re\Big( \tilde{f}_{\mK,(\phi,\phi,g)}^{(1)} \Big) \, {s_{12}^2 \over (q^2)^2} + 2\,\Re\Big(\tilde{f}_{\mK,(\psi,\psi,\phi)}^{(1)}  \Big) \, {s_{13} s_{23} \over (q^2)^2} \bigg]  \pnt \label{M2-simp-2}
\end{equation}
Inserting the explicit results \eqref{eq:final-3pt-1loop} and performing the three-particle phase space integration \eqref{PS-3pt}, 
we find that the 
contribution to the cross section is given by
\begin{equation}
\begin{aligned}
\tilde\sigma_{\mK, 3}^{(2)} 
& = \sigma^{(0)}_\mK \, \Big(\frac{\mu^2}{q^2}\Big)^{2\peps} \bigg[ -{48 \over \peps^3} - {112 \over \peps^2} - {224 - 56 \pi^2 \over \peps} - 544 + {632\over3} \pi^2 + 848 \zeta_3  \bigg] + {\cal O}(\peps) \pnt
\label{eq: sigma-2loop-bare-3pt}
\end{aligned}
\end{equation}
%

%
\begin{figure}[t]
\begin{center}
$\Big({\cal Z}^{(1)}_\mK$\FDff[legs=3,scaled=0.5]\Big)\FDff[legs=3,reflected,scaled=0.5] + \FDff[legs=3,scaled=0.5]\Big(\FDff[legs=3,reflected,scaled=0.5]${\cal Z}^{(1)}_\mK$\Big)
\caption{The contribution from the one-loop renormalization constant to the three-particle cross section.}
\label{fig:cs-Konishi-calZ1-3pt}
\end{center}
\end{figure}
The one-loop renormalization constant \eqref{eq:calZ-1loop} contributes as shown in figure \ref{fig:cs-Konishi-calZ1-3pt}. 
The squared matrix element reads
\begin{equation}
\begin{aligned}
{\cal M}^{(1)}_{{\cal Z}^{(1)}\mK,3} & = 2{\cal M}^{(0)}_{\mK,3} \,{\cal Z}^{(1)}_\mK \pnt
\end{aligned}
\end{equation}
The corresponding contribution to the cross section can be computed as in \eqref{sigma-bps-1loop3pt}, and it is given by
\begin{equation}
\begin{aligned}
\sigma_{{\cal Z}^{(1)} \mK, 3}^{(2)} & = \frac{1}{g^2}\int \de {\rm PS}_3 \, {\cal M}^{(1)}_{{\cal Z}^{(1)}\mK,3}  = \sigma^{(0)}_\mK \, \Big(\frac{\mu^2}{q^2}\Big)^\peps 
\bigg[ {48 \over \peps^3} - { 28\pi^2 \over \peps} - 400 \zeta_3  \bigg] + {\cal O}(\peps) \pnt
\label{eq: sigma-2loop-calZ1-3pt}
\end{aligned}
\end{equation}

Summing \eqref{eq: sigma-2loop-bare-2pt}, \eqref{eq: sigma-2loop-calZ1-2pt}, \eqref{eq: sigma-2loop-bare-3pt} and \eqref{eq: sigma-2loop-calZ1-3pt}, we find
\begin{equation}
\begin{aligned}
& \tilde\sigma_{\mK, 2}^{(2)} + \sigma_{{\cal Z}^{(1)}\mK, 2}^{(2)} +\tilde\sigma_{\mK, 3}^{(2)} + \sigma_{{\cal Z}^{(1)}\mK, 3}^{(2)} \\ 
&\qquad= \sigma^{(0)}_\mK \, \bigg[ -{36\over \peps^2} + {24 
\over\peps} 
+ 72 \, \log^2\frac{q^2}{\mu^2} -384 \, \log\frac{q^2}{\mu^2} +{508 + 72 \pi^2 + 336 \zeta_3}  \bigg] +  {\cal O}(\peps) \pnt
\label{eq:sigma without Z2}
\end{aligned}
\end{equation}
Since the one-loop UV subdivergences in the bare contributions are canceled by 
the second and fourth terms, all divergences in the above result originate from the
two-loop overall UV divergence.\footnote{All infrared divergences should be already canceled between the different channels.}

\subsubsection*{Two-loop renormalization constant}

The two-loop overall UV divergence has to be canceled
by the third contribution of \eqref{twoparticlefull}, which
involves the two-loop renormalization constant. In analogy to \eqref{sigmaZ11},
this contribution reads
\begin{equation}\label{sigmaZ22}
\sigma_{{\cal Z}^{(2)}\mK}^{(2)} = 2\sigma^{(0)}_\mK{\cal Z}^{(2)}_\mK\pnt
\end{equation}
Requiring the sum of \eqref{eq:sigma without Z2} and \eqref{sigmaZ22}
to be finite, 
we immediately find
\begin{equation}\label{gammaK12}
 {\cal Z}^{(2)}_\mK = 
 {18\over \peps^2} - {12 \over \peps}  \pnt
\end{equation}
Comparing this with the expansion of \eqref{eq: relation of calZ and gamma} 
in terms of the anomalous dimension to two-loop order yields
the known one- and two-loop Konishi anomalous dimension first obtained
in \cite{Anselmi:1996mq,Anselmi:1996dd}:
\begin{equation}
\gamma^{(1)}_\mK = 12 \col \qquad \gamma^{(2)}_\mK = -48\pnt
\end{equation}

\subsubsection*{Final result}

Summing \eqref{sigmaZ22} and \eqref{eq:sigma without Z2} yields the renormalized two-loop cross section
\begin{equation}
\begin{aligned}
\sigma_{\cal K,\ren}^{(2)} & = \sigma^{(0)}_\mK \,\bigg[ 72 \log^2\frac{q^2}{\mu^2} - 384 \, \log\frac{q^2}{\mu^2}+ {{508 + 72 \pi^2 + 336 \zeta_3} } \bigg] + {\cal O}(\peps) \pnt
\end{aligned}
\end{equation}
Finally, we compute the second order term in the expansion of \eqref{logsigmarengeneral}
and obtain
\begin{equation}
\left[ \log \bigg( {\sigma_{\cal K,\ren} \over \sigma_\mK^{(0)} } \bigg)\right]^{(2)} = {\sigma_{\cal K,\ren}^{(2)} \over \sigma_\mK^{(0)} } - {1\over2} \bigg({\sigma_{\cal K,\ren}^{(1)} \over \sigma_\mK^{(0)} }\bigg)^2 =  {-48 \log\frac{q^2}{\mu^2}} + {116 + 72 \pi^2 + 336 \zeta_3} + {\cal O}(\peps) 
\pnt
\end{equation}
We find that the coefficient of $\log\frac{q^2}{\mu^2}$ gives  the correct two-loop anomalous dimension, as expected from \eqref{logsigmarengeneral}.

Including also the one-loop result \eqref{sigmaren1}, the logarithm of the 
normalized Konishi cross section is given by
\begin{equation}\label{sigmaR}
\begin{aligned}
\log \bigg( {\sigma_{\cal K,\ren} \over \sigma_\mK^{(0)} } \bigg)
&=  g^2 \, \bigg( {12 \,\log\frac{q^2}{\mu^2}} - 28 \bigg) + g^4 \,  \bigg( { -48 \, \log\frac{q^2}{\mu^2}} + {116 + 72 \pi^2 + 336 \zeta_3}  \bigg) + {\cal O}(g^6, \peps) \pnt
\end{aligned}
\end{equation}
The finite terms that are independent of $\log\frac{q^2}{\mu^2}$
yield the constant $C$, and by a comparison with \eqref{eq:C-expansion} 
they determine the one- and two-loop terms of the constant $M$ in \eqref{G2(x)}. 
In particular, this yields the full two-point function \eqref{G2(x)} up to two-loop order.

\subsection*{Some discussion}

The renormalized cross section can be computed in different ways. In the above presentation, we have treated the bare contribution and the terms involving the renormalization constant separately at the cross section level. One can also first perform the renormalization of the operators via form factors, as described in appendix \ref{app:anomalous-from-2pt}, and then compute the renormalized cross section directly from them. Furthermore, the terms involving the renormalization constant can be obtained directly by expanding relation \eqref{sigmaRinsigmaB}, which gives the renormalized cross section in terms of the bare one and the renormalization constant. For example, the sum of \eqref{eq: sigma-2loop-calZ1-2pt} and \eqref{eq: sigma-2loop-calZ1-3pt} that involve the one-loop renormalization constant can be obtained as\footnote{In this case, one needs the result of the one-loop cross section up to ${\cal O}(\peps)$ order. The result \eqref{sigma(1)-B} based on \eqref{eq:sigma-Mtilde} is not enough, since $\sigma_\BPS$ is not zero at  ${\cal O}(\peps)$ order.} 
\begin{equation}
\begin{aligned}
\sigma_{{\cal Z}^{(1)} \mK, 2}^{(2)} + \sigma_{{\cal Z}^{(1)} \mK, 3}^{(2)}
= 2 {\cal Z}_{\mK}^{(1)} \sigma_{\mK}^{(1)} + ({\cal Z}_{\mK}^{(1)})^2 \sigma_{\mK}^{(0)} \pnt
\label{eq:simple-Z-computation}
\end{aligned}
\end{equation}
We have checked that these different ways  give the same result.

As discussed in appendix \ref{app:scheme}, the above result depends on the renormalization scheme. 
One can define the new coupling at which the subtraction is performed 
as $g_\varrho=g\e^{\varrho\peps}$ and then expand the expressions in terms of the original coupling $g$. 
This scheme change can be implemented by replacing ${\cal Z}^{(\ell)}_\mK\to {\cal Z}^{(\ell)}_\mK\e^{2\varrho\ell\peps}$
in all the above equations.
With such a modification in the above computation, one finds that
the renormalized cross section \eqref{sigmaR} acquires a finite additive contribution $2\gamma_\mK\varrho$, demonstrating that $M$ in \eqref{G2(x)} is scheme-dependent.
This agrees with the expectation from \eqref{eq:s-scheme-change}, since the scheme 
change can be understood as a change of the 't Hooft mass $\mu\to\mu\e^{-\varrho}$.

Finally, let us briefly comment on the FDH scheme we have chosen in the computation. In the FDH scheme, we set the number of external scalars to $6$ and use polarization vectors in $D=4$ dimensions for the form factors.  As discussed in section \ref{section:CSNutshell}, this corresponds to the prescription given in \eqref{calM_n^ell}, where the sum of the degrees of freedom for the external legs is performed by the $\eta$-integration based on the $SU(4)$ representation.  One can also perform a detailed analysis at the diagrammatic level, as for the form factors in section \ref{sec: subtleties in unitarity}, which leads to an alternative prescription for obtaining the cross section in $D=4-2\peps$ dimensions. We will not present the details in the paper, but we have checked that both prescriptions give identical results at least up to the two-loop order.

%% file: conclusion.tex
\section{Conclusion and outlook}
\label{sec: conclusion and outlook}

In this paper, we have studied form factors of non-protected operators in $\mN=4$ SYM theory, specifically of the Konishi operator, using on-shell unitarity techniques. 
Importantly, we have found that this requires an extension of these techniques.
We have obtained explicit new results of the three-point form factor at one-loop and two-point form factors up to two-loop order, given in 
\eqref{eq:final-2pt-1loop}--\eqref{eq:final-3pt-1loop}.  The application of on-shell methods to determine such form factors, which are partial off-shell quantities involving both, composite operators and on-shell states, provides a step to
deepen our understanding of the connection between modern on-shell techniques and 
the off-shell world of correlation functions.

Another important aspect of this paper is to provide a physical observable within $\mN=4$ SYM theory, given by a cross-section-type quantity: the inclusive
decay rate of a state, described by a composite operator carrying timelike off-shell 
momentum $q$, into any final on-shell multi-particle state. 
We gave a formulation of how to compute this observable. 
Using the Konishi form factor results mentioned above, we performed an explicit computation of the total cross section up to two-loop order, given in \eqref{sigmaR}.  
Via the optical theorem, this also yields the two-point function up to this order.

The UV divergences appearing in the Konishi form factors together with the IR divergences require the renormalization of the operator. This is carried out explicitly in the computation of the total cross section in which the IR divergences cancel. 
We have reproduced the known Konishi anomalous dimension up to two-loop order from 
the renormalization constant and also identified it as the coefficient of the $\log\frac{q^2}{\mu^2}$ term in the renormalized cross section \eqref{sigmaR}.

Since the Konishi operator is not protected by supersymmetry, interesting subtleties and new features appeared, which we now summarize.

First, an important subtlety occurs in the unitarity-based computation of the Konishi form factors. The Konishi primary is a trace of all scalars. In order to preserve supersymmetry, the Konishi operator has to be continued to $D = 4 - 2\peps$ dimensions, i.e. the number of scalar field flavors that is summed over has to be $N_\phi = 10 - D$.  However, working with four-dimensional unitarity based on Nair's on-shell superspace, one can only directly compute the form factor for the different operator $\mK_6$, 
which in $D=4-2\peps$ dimensions has $N_\phi = 6$ scalars rather 
than $N_\phi = 6 + 2\peps$.
In order to find the Konishi form factor, the results based on four-dimensional 
unitarity have to be modified when they are lifted to $D=4-2\peps$ dimensions where the occurring divergences are regularized.
We provide a rigorous prescription \eqref{eq: ftilde-correction-rule} for this modification 
that yields the form factors of the Konishi operator.

Second, the Konishi form factors contain some interesting features that are not present in other on-shell quantities studied in ${\cal N}=4$ SYM theory studied so far, such as scattering amplitudes or the BPS form factors as partial off-shell quantities. The bare Konishi form factor contains bubble integrals and bubble subintegrals, which are UV divergent.
Moreover, the one-loop three-point result contains a rational term.%
\footnote{A similar rational term also occurs for the minimal form factor of operators in the $SL(2)$ subsector \cite{Wilhelm:2014qua}.} 
In addition, the coefficients of the individual integrals occurring in the form factor results involve spurious poles, which only disappear after multiplication with the integrals and summation over all contributions.
Last but not least, the loop corrections of the Konishi form factors with different external states  turn out to have quite different structures, even if they are in the 
same MHV sector.
The emergence of these features that are familiar from QCD  
can be traced back to the fact that a non-protected operator has been inserted into 
the action, formally breaking its supersymmetry.

Finally, let us briefly mention some further directions one can pursue following this work.

First, it should be straightforward to generalize the computation of the one-loop Konishi form factors to the higher-point cases. 
It is also interesting to proceed to higher loop orders.
In particular, using the known IR exponentiation property of the Sudakov form factor, the knowledge of the two-point Konishi form factor alone is sufficient to extract the Konishi anomalous dimension. We explain this in appendix \ref{app:anomalous-from-2pt}.
Turning the logic around, we also give 
a prediction for the three-loop two-point Konishi form factor apart from finite terms there, only using the known three-loop anomalous dimension and the IR exponentiation in addition to our form factor results.

Second, our detailed example of how to apply four-dimensional unitarity to compute the Konishi form factor by understanding an encountered subtlety and providing 
a solution 
is a solid stepping stone for further studies of other non-protected operators, based on generalizing the prescription we give in section \ref{sec: subtleties in unitarity}. Combining our insights with those from the 
recent one-loop calculation in \cite{Wilhelm:2014qua}, it should be possible to  
compute  the minimal form factors for general operators at two-loop order
via on-shell methods. This would allow us to determine the complete two-loop dilatation operator of $\mathcal{N}=4$ SYM theory which yields all two-loop anomalous dimensions as eigenvalues. 
Besides the anomalous dimensions, the other important CFT data is given by the structure constants, which can be computed from three-point functions. It would be very interesting to use similar unitarity-based techniques to compute them.

Furthermore,
as given in \cite{Hofman:2008ar, Engelund:2012re, Belitsky:2013xxa, Belitsky:2013bja}, the so-called energy energy correlation function can be computed as a weighted cross section which is very similar to the total cross section studied in this paper.  In \cite{Belitsky:2013xxa, Belitsky:2013bja}, different techniques not relying on cross sections have been used to evaluate them up to two-loop order. 
It would be interesting to obtain them also from direct cross section computations.
The interpretation of the cross-section-type quantities at strong 
coupling via the AdS/CFT correspondence is also an open problem. In particular, it would be interesting to consider the phase-space integration with strong coupling form factors in the framework of string theory.

Finally, as a cousin of $\mN=4$ SYM, the so-called ABJM theory \cite{Aharony:2008ug} has been intensively studied in recent years. In particular, the form factors for half-BPS operators have been determined in this theory as well \cite{Brandhuber:2013gda, Young:2013hda, Bianchi:2013iha, Bianchi:2013pfa}. It would be interesting to pursue  a similar study as in this paper for the ABJM theory, especially for the form factors of non-protected operators.

%% file: acknowledgements.tex
\acknowledgments

It is a great pleasure to thank Lance Dixon and Gregory Korchemsky for enlightening discussions and suggestions. 
We would also like to thank Zvi Bern, Andreas Brandhuber, Burkhard Eden, Valentina Forini, Sergey Frolov, Laura Koster, Brenda Penante, Jan Plefka, Radu Roiban, Matthias Staudacher, Gabriele Travaglini, Peter Uwer and Vitaly Velizhanin for helpful discussions.
%
We are grateful to Camille Boucher-Veronneau, Lance Dixon and Jeffrey Pennington for sharing their unpublished notes.
GY would also like to thank Andreas Brandhuber and Gabriele Travaglini for initial collaboration on the Konishi form factor. 
We thank the Marie Curie network GATIS (gatis.desy.eu)
of the European Unions Seventh Framework Programme FP7/2007-2013/ under REA
Grant Agreement No 317089 for support.
DN's research is supported by the SFB 647 ``Raum-Zeit-Materie. Ana\-ly\-tische
und Geometrische Strukturen".
MW dankt der Studienstiftung des deutschen Volkes f\"ur ein Promotionsf\"orderstipendium. GY is supported by a DFG grant in the framework of the SFB 647 ``Space-Time-Matter", and he also thanks the KITPC/ITP-CAS in Beijing for hospitality during the workshop ``Quantum Gravity, Black Holes and Strings", where part of this work was done.
The work of DN and MW  was supported by the Marie Curie International Research Staff Exchange Network UNIFY (FP7-People-2010-IRSES under grant
agreement number 269217), which allowed them to visit Stony Brook University. Furthermore, DN and MW thank the Simons Center for Geometry and Physics and the C.N. Yang Institute for Theoretical Physics, Stony Brook, for warm hospitality during the final stage of this project.

%% file: appendices.tex
\section{Fourier transformation of the two-point function}
\label{app:FTprop}

In this appendix, we give our conventions for the Fourier transformation and 
for the transition from Euclidean to Minkowski signature.

In Euclidean signature, the coordinate dependence of the two-point function 
has the following representation in terms of the momentum space integral:
\begin{equation}\label{xfouriertrafoE}
{1 \over (x_\E^2)^\Delta} \, = \, 2^{D - 2 \Delta} \pi^{\frac{D}{2}} { \Gamma({D\over2} - \Delta) \over \Gamma(\Delta)} \int {\de^D q_\E \over (2\pi)^D} \, {\e^{i q_\E \cdot x_\E} \over (q_\E^2)^{\frac{D}{2} - \Delta}}  \col
\end{equation}
where $q_\E \cdot x_\E = q_{\E,0} x_{\E,0} + \sum_{i=1}^{D-1} q_i x_i$.

In Minkowski signature with a mostly-minus metric, the exponent in the momentum space integral is given by $-iq\cdot x$, and
the integrand has poles at $q_0=\pm|\vec q|$. We want
positive energies $q_0>0$ to propagate into the future $x_0>0$.
Hence, the pole at $q_0=|\vec q|$ has to be picked
when for $x_0>0$ the integral over $q_0$ is closed in the negative imaginary half plane such that the exponential factor vanishes
 for $q_0\to -i\infty$. This is guaranteed if we replace $q_\E^2\to -q^2-i0$
in the denominator of the above expression. 
The position of the poles fixes the Wick-rotation
to be counterclockwise in momentum space, i.e.\ $q_0=iq_{\E,0}$, and for
$x_0=-ix_{\E,0}$ to be clockwise in configuration space. This leaves the exponential invariant, and 
it can be transformed to Minkowski signature by flipping the sign of the spatial momenta $q_i$.
We hence find
\begin{equation}\label{xfouriertrafo}
{1 \over (-x^2+i0)^\Delta} = (-i) 2^{D - 2\Delta} \pi^{\frac{D}{2}} { \Gamma({D\over2} - \Delta) \over \Gamma(\Delta)} \int {\de^D q \over (2\pi)^D} \,  {e^{-i q \cdot x} \over (-q^2-i0)^{\frac{D}{2}- \Delta}} \col
\end{equation}
where $q \cdot x = q_0 x_0 - \sum_{i=1}^{D-1} q_i x_i$.

\section{Feynman integrals}
 \label{app:integrals} 
 
In this appendix, we present all integrals that enter the form factor results in section \ref{sec:ffactor}, as well as our conventions. Moreover, we show how the cut integrals are lifted to full integrals.

As a regularization procedure, the four-dimensional $\mathcal{N}=4$ SYM 
theory can be continued to $D=4-2\peps$ dimensions. Both IR and UV divergences are then 
captured as poles in $\peps$. Moreover, the Yang-Mills coupling constant 
$g_\YM$ has to be replaced by $g_\YM\mu^\peps$, where $\mu$ is the 't Hooft mass
which is introduced in order to keep $g_\YM$ dimensionless \cite{'tHooft:1973mm}.
Hence, the combination $g\mu^\peps$ with $g$ given in \eqref{gdef} occurs as the effective loop expansion parameter of the perturbation series in the 
large-$N_{\text{c}}$ expansion.

From Feynman diagrams, the following combination of the coupling constant, the 't Hooft mass $\mu$ and the loop integral occurs at $\ell$-loop order
\begin{equation}
(g_\YM\mu^\peps)^{2\ell}N_{\text{c}}^{\ell} (-i)^\ell \int\frac{\de^Dl_1}{(2\pi)^D}\dots\frac{\de^Dl_\ell}{(2\pi)^D}
\frac{f(l_1,\dots,l_\ell)}{\prod_j D_j}
=g^{2\ell} I^{(\ell)}[f(l_1,\dots,l_\ell)]\col
\label{integral-measure-relation}
\end{equation}
where the integral $I^{(\ell)}$ is of the following form:
\begin{equation}
\label{Ielldef}
I^{(\ell)}[f(l_1,\dots,l_\ell)]=(\e^{\gamma_\E}\mu^2)^{\ell\peps}\int\frac{\de^Dl_1}{i\pi^{\frac{D}{2}}}\dots\frac{\de^Dl_\ell}{i\pi^{\frac{D}{2}}}
\frac{f(l_1,\dots,l_\ell)}{\prod_j D_j}
\pnt
\end{equation}
In these formulae, the $D_j$'s are the propagators, i.e.\ $D_j = k_j^2 + i0$ for $k_j$ being the combination of external momenta and loop momenta that flows through the propagator.

\subsection*{Lifting the cut integral}

Let us explain our procedure and conventions for lifting the cut integrals to the full integrals.

Consider the triangle term in \eqref{Konishicut1loop} as an explicit example. We have
\beq
\bea
g_{\YM}^2 N_{\text{c}} \twops \frac{ s_{12}}{(l_1+p_1)^2} 
= g_{\YM}^2 N_{\text{c}} s_{12} \FDinline[triangle, doublecut,cutlabels, twolabels, labelone=p_1, labeltwo=p_2]  
\ \underset{\scriptstyle\text{lifting}}{\longrightarrow} \ - i g^2 s_{12} \FDinline[triangle, twolabels, labelone=p_1, labeltwo=p_2]  \col
\eea
\eeq
where the phase-space integration measure $\de {\rm PS}_{2 , \{l\}}$ is defined according to \eqref{PS-def} with measure factor $\frac{\de^Dl}{(2\pi)^D}$. To lift the cut integral to the full integral, two steps have to be performed.

One is to replace the cut propagator as 
\beq
\bea
2\pi \delta_+(l_j^2) \ \longrightarrow \ {i \over l_j^2}  \pnt
\eea
\eeq

The other is to change the measure factor and coupling constant as in \eqref{integral-measure-relation}, such that the integrals of  \emph{uncut} graphs are defined in terms of \eqref{Ielldef}. This prescription is employed throughout section \ref{sec:ffactor}.

\subsection*{List of integrals}

In order to present the expressions for the required integrals in a compact form, 
we define $q=\sum_i p_i$
and introduce the factor
\begin{equation}
\begin{aligned}
c_\Gamma = \e^{\gamma_\E \peps} \, \frac{\Gamma(1-\peps)^2\Gamma(1+\peps)}{\Gamma(1-2\peps)} \pnt
\label{famous-c_Gamma}
\end{aligned}
\end{equation}
Moreover, in the results for the integrals, all $(-q^2)^{\ell \peps}$ should be understood as $(-q^2 - i 0)^{\ell \peps}$ and similarly for  $(-s_{ij})^{\ell \peps}$.

In the conventions introduced in \eqref{Ielldef}, the one-loop integrals that are required to compute the one-loop form factors read 
\begin{align}
\label{oneloopint:bubble}
\FDinline[bubble, twolabels, labelone=p_1, labeltwo=p_2]
&={(\e^{\gamma_\E}\mu^2)^{\peps}} \int\frac{\de^Dl}{i\pi^{\frac{D}{2}}}\frac{1}{ l^2 (l+q)^2}
= {c_\Gamma \over \peps (1-2\peps)} \Big({-q^2 \over \mu^2} \Big)^{-\peps}
\col\\
\label{oneloopint:1masstriangle}
\FDinline[triangle, twolabels, labelone=p_1, labeltwo=p_2] 
&={(\e^{\gamma_\E}\mu^2)^{\peps}} \int\frac{\de^Dl}{i\pi^{\frac{D}{2}}}
\frac{1}{(l+p_1)^2 l^2 (l-p_2)^2}
= - {c_\Gamma \over \peps^2} {1\over (-q^2)} \Big({-q^2 \over \mu^2} \Big)^{-\peps} 
\col\\
\label{oneloopint:2masstriangle}
\FDinline[trianglethreetop,threelabels,labelone=p_1,labeltwo=p_2,labelthree=p_3] 
&={(\e^{\gamma_\E}\mu^2)^{\peps}} \int\frac{\de^Dl}{i\pi^{\frac{D}{2}}}
\frac{1}{(l+p_1+p_2)^2 l^2 (l-p_3)^2}\nonumber \\*
&= - {c_\Gamma \over \peps^2} {1\over s_{13} + s_{23}} \Big[\Big({-s_{12} \over \mu^2} \Big)^{-\peps} - \Big({-q^2 \over \mu^2} \Big)^{-\peps}\Big]
\col
\\
\label{oneloopint:box}
\FDinline[box,threelabels,labelone=p_1,labeltwo=p_2,labelthree=p_3] 
&
={(\e^{\gamma_\E}\mu^2)^{\peps}} \int\frac{\de^D l}{i \pi^{\frac{D}{2}}} \frac{1}{l^2 (l^2+p_1)^2(l+p_1+p_2)^2(l-q)^2}
\\*
&=\frac{c_\Gamma}{\peps^2} 
\frac{2}{s_{12}s_{23}} \Big[-\Big({-q^2 \over \mu^2} \Big)^{-\peps} \,_2 F_1\big(1,-\peps, 1-\peps, -\tfrac{q^2s_{13}}{s_{12}s_{23}} \big) \nonumber\\*
& \hphantom{{}={}\frac{c_\Gamma}{\peps^2}\frac{2}{s_{12}s_{23}}\Big[}
+\Big({-s_{12} \over \mu^2} \Big)^{-\peps} \,_2 F_1\big(1,-\peps, 1-\peps, -\tfrac{s_{13}}{s_{23}} \big) \nonumber \\*
& \hphantom{{}={}\frac{c_\Gamma}{\peps^2}\frac{2}{s_{12}s_{23}}\Big[}+\Big({-s_{23} \over \mu^2} \Big)^{-\peps} \,_2 F_1\big(1,-\peps, 1-\peps, -\tfrac{s_{13}}{s_{12}} \big)
 \Big] \nonumber
\col
\end{align}
where ${}_2F_1$ denote the Gaussian hypergeometric function. The results \eqref{oneloopint:bubble}--\eqref{oneloopint:box} can be found for example in \cite{Bern:1995ix,Anastasiou:1999cx,Smirnov:2004ym}.

Furthermore, we need the following integral, which evaluates to a rational term \cite{Bern:1995ix}:
\begin{equation}
\label{rationaltermtriangle}
\begin{aligned}
I_3^D[l_\peps^2]  &={(\e^{\gamma_\E}\mu^2)^{\peps}} \int\frac{\de^Dl}{i\pi^{\frac{D}{2}}}
\frac{l_\peps^2}{(l+p_1+p_2)^2 l^2 (l-p_3)^2}\\
&=  - {c_\Gamma  \over (1-2\peps)(2-2\peps) } {1\over s_{13} + s_{23}} \Big[s_{12}\Big({-s_{12} \over \mu^2} \Big)^{-\peps} - q^2 \Big({-q^2 \over \mu^2} \Big)^{-\peps}\Big] \col
\end{aligned}
\end{equation}
where the momentum $l_\peps$ in the $2\peps$ dimensions is introduced in \eqref{eq:l_peps}.

To compute the two-loop two-point form factor, we need the following two-loop integrals.  Using IBP identities as e.g.\ implemented in {\texttt LiteRed} \cite{Lee:2012cn}, they can be reduced to master integrals as\footnote{Recall that the loop-momentum-dependent prefactors are understood to appear in the numerators of the depicted integrals.}
%
\begin{align}
\FDinline[fishtop,twolabels,labelone=p_1,labeltwo=p_2] 
&= {2 - 3 \peps \over \peps (-q^2)} \FDinline[sunrise,twolabels,labelone=p_1,labeltwo=p_2]\col
\\
(q^2)^2\FDinline[rainbow, twolabels, labelone=p_1,labeltwo=p_2] 
&=  -{3(1-2 \peps) (1-3\peps) (2 - 3 \peps) \over \peps^3 (-q^2)} \FDinline[sunrise,twolabels,labelone=p_1,labeltwo=p_2] \\* \nonumber
& \phantom{{}={}} +  {3(1-2 \peps) (1-3\peps) \over 2\peps^2} \FDinline[itwo,twolabels,labelone=p_1,labeltwo=p_2] 
+  {(1 - 2 \peps)^2 \over \peps^2} \bigg( \FDinline[bubble,twolabels,labelone=p_1,labeltwo=p_2] \bigg)^2\col \\
s_{1l}s_{2l}
\FDinline[rainbow,momentum,twolabels,labelone=p_1,labeltwo=p_2]
&= {(2 - 3 \peps)(2-9\peps+10\peps^2 - 4\peps^3) \over (1-\peps) (1-2\peps) \peps^2 (-q^2)} \FDinline[sunrise,twolabels,labelone=p_1,labeltwo=p_2] \\* \nonumber
& \phantom{{}={}} -  {1-4 \peps + 2\peps^2 \over (1-\peps)\peps} \FDinline[itwo,twolabels,labelone=p_1,labeltwo=p_2] 
-  {2- 3 \peps +2 \peps^2 \over 2 (1-\peps) \peps} \bigg(\FDinline[bubble,twolabels,labelone=p_1,labeltwo=p_2] \bigg)^2\col
\\
s_{1l}s_{2l}
\FDinline[rainbownonplanar,momentum,twolabels,labelone=p_1,labeltwo=p_2]
&= {(1-2 \peps) (2-3\peps) (3 - 5\peps) \over \peps^2(1-4\peps) (-q^2) } \FDinline[sunrise,twolabels,labelone=p_1,labeltwo=p_2] \\* \nonumber
& \phantom{{}={}} - {(1+ \peps) (1-2\peps) \over \peps (1-4\peps)} \FDinline[itwo,twolabels,labelone=p_1,labeltwo=p_2] -  {\peps \, (-q^2)^2 \over (1-4\peps)} \FDinline[rainbownonplanar, twolabels, labelone=p_1,labeltwo=p_2]
\col
\end{align}
where the master integrals are \cite{Gehrmann:2005pd}
\begin{equation}
\begin{aligned}
\FDinline[sunrise,twolabels,labelone=p_1,labeltwo=p_2] &= \e^{2\gamma_\E\peps} \frac{\Gamma(1-\peps)^3\Gamma(1+2\peps)}{2\peps (1-2\peps)\Gamma(3-3\peps)}  (-q^2) \left(-{q^2 \over \mu^2} \right)^{-2\peps}  \col
\\
\FDinline[itwo,twolabels,labelone=p_1,labeltwo=p_2] &= \e^{2\gamma_\E\peps} \frac{\Gamma(1-\peps)^2\Gamma(1+\peps)\Gamma(1+2\peps)\Gamma(1-2\peps)}{2\peps^2 (1-2\peps)\Gamma(2-3\peps)}  \left(-{q^2 \over \mu^2} \right)^{-2\peps} \col
\\
\FDinline[rainbownonplanar, twolabels, labelone=p_1,labeltwo=p_2] &= 
\e^{2\gamma_\E\peps} {1\over (-q^2)^2} \left(-{q^2 \over \mu^2} \right)^{-2\peps} \bigg[ \frac{\Gamma(1-2\peps)^4\Gamma(1+2\peps)^3\Gamma(1-\peps)\Gamma(1+\peps)}{\peps^4 (1-4\peps)^2\Gamma(1+4\peps)} \\
& \phantom{{}={}}+ \frac{4\Gamma(1-\peps)^2\Gamma(1-2\peps)\Gamma(1+2\peps)}{\peps^2 (1+\peps)(1+2\peps)\Gamma(1-4\peps)} \,_3 F_2\big(1,1,1+2\peps; 2+\peps, 2+2\peps; 1 \big)  \\
& \phantom{{}={}}+ \frac{\Gamma(1-\peps)^2\Gamma(1+\peps)\Gamma(1-2\peps)\Gamma(1+2\peps)}{2\peps^4 \Gamma(1-3\peps)} \,_3 F_2\big(1,-4\peps,-2\peps; 1-3\peps, 1-2\peps; 1 \big)  \\
& \phantom{{}={}}+ \frac{\Gamma(1-\peps)^3\Gamma(1+2\peps)}{2\peps^4 \Gamma(1-3\peps)} \,_4 F_3\big(1,1-\peps, -4\peps, -2\peps; 1-3\peps, 1-2\peps, 1-2\peps; 1 \big) \bigg] \col
\end{aligned}
\end{equation}
and the one-loop bubble integral is given in \eqref{oneloopint:bubble}.

\section{Passarino-Veltman reductions }
\label{app:PV}

In this appendix, we summarize some results on Passarino-Veltman (PV) reduction \cite{Passarino:1978jh}, which we need in section \ref{sec:ffactor}.

We use the four-dimensional-helicity (FDH) scheme of \cite{Bern:1991aq, Bern:2002zk}, and decompose the $D$-dimensional loop momentum $l$ into a four-dimensional part $l_{(4)}$ and a $(D-4)=-2\peps$ dimensional part $l_\peps$, where we assume that $\peps<0$.
This yields the following decomposition of the scalar product:
\begin{equation} 
\label{eq:l_peps}
\eta_{\mu \nu} l^\mu l^\nu = l_{(4)}^2 = l^2 + l_\peps^2 \col
\end{equation}
where $\eta_{\mu \nu}$ is the four-dimensional metric.\footnote{Note that, although $D=4-2\peps$, there is a plus sign in front of $l_\peps^2$  since the metric has  mostly-minus signature.}
Arbitrary four-dimensional external reference momenta are denoted as $k_i$.

\noindent {\bf Bubble}. 
The $D$-dimensional bubble integral with external momentum $q$ 
defined in \eqref{oneloopint:bubble} may include a non-trivial polynomial $f(l)$ of 
the loop momentum $l$ and the reference momenta $k_i$ in its numerator.
Denoting this integral as $I_2^D[f(l)](q^2)$, we find the following relations
for the reduction of tensor integrals to scalar integrals:
\begin{eqnarray}
I_2^D[(l \cdot k_1)](q^2) &=& - {(q\cdot k_1) \over 2} \, I_2^D(q^2) \col \\
I_2^D[(l \cdot k_1)(l\cdot k_2)](q^2) &=& \left( {(q \cdot k_1) \, (q \cdot k_2) \over 3} - { q^2 \,(k_1 \cdot k_2) \over 12} \right) I_2^D(q^2) \nonumber\\ &&- \left( {(q \cdot k_1) \, (q \cdot k_2) \over 3 q^2}  - {(k_1 \cdot k_2) \over 3} \right) \, I_2^D[l_\peps^2](q^2) \pnt 
\end{eqnarray}

\noindent {\bf Triangle}.
Next, we consider the $D$-dimensional triangle integral with numerator $f(l)$, which depends on 
two arbitrary momenta $q_1$ and $q_2$. It is defined as 
\begin{equation}
I_3^D[f(l)](q_1,q_2)= {(\e^{\gamma_\E}\mu^2)^{\peps}} \int\frac{\de^Dl}{i \pi^{\frac{D}{2}}}\frac{f(l)}{l^2(l +q_1)^2(l +q_2)^2} \pnt 
\end{equation}
We find
\begin{equation}
\label{eq: PV reduction of triangles}
\begin{aligned}
I_3^D[(l \cdot k_1)](q_1,q_2) &= \sum_{i=0}^2 {a_i \over2} \, I_2^{D,(i)}  -  \sum_{i=1}^2 {a_i \, q_i^2 \over2} I_3^D \col \\ 
I_3^D[(l \cdot k_1)(l \cdot k_2)](q_1,q_2) &= \sum_{i=0}^2 C_2^{(i)} \, I_2^{D,(i)} + C_{3,0}I_3^D + C_{3,\peps} I_3^D[l_\peps^2](q_1,q_2)  \col
\end{aligned}
\end{equation}
where $I_2^{D,(0)}=I_2^D((q_1-q_2)^2)$, $I_2^{D,(1)}=I_2^D(q_2^2)$, $I_2^{D,(2)}=I_2^D(q_1^2)$,
\begin{equation}
\begin{aligned}
C_2^{(1)} &= - {1\over4} \Big( \sum_{i=1}^2 a_{1i} \, q_i^2 \Big) \, a_{21} - {1\over4}a_{11} (q_2 \cdot k_2) + {1\over8}  [(k_1 \cdot k_2) - b] \, q_2^2 \sum_{i=1}^2 (A^{-1})_{2i} \col \\ 
C_2^{(2)} &= - {1\over4} \Big( \sum_{i=1}^2 a_{1i} \, q_i^2 \Big) \, a_{22} - {1\over4}a_{12} (q_1 \cdot k_2) + {1\over8} [(k_1 \cdot k_2) -b] \, q_1^2 \sum_{i=1}^2 (A^{-1})_{1i} \col \\ 
C_2^{(0)} &= - \sum_{i=1}^2 C_2^{(i)} + {1\over4}(k_1 \cdot k_2) \col \\ 
C_{3,0} &= {1\over4}\Big( \sum_{i=1}^2 a_{1i} \, q_i^2 \Big) \, \Big( \sum_{j=1}^2 a_{2j} \, q_j^2 \Big) + {1\over8} \big[ b - (k_1 \cdot k_2)  \big] \, q_1^2 \, q_2^2\, \tilde A  \col\\
C_{3,\peps} &= {(k_1 \cdot k_2) - b \over2} 
\end{aligned}
\end{equation}
with
\begin{equation}
\begin{aligned}
a_{ij} &= \sum_{m=1}^2 (k_i \cdot q_m) (A^{-1})_{mj} \col \qquad A_{ij} = q_i \cdot q_j \col\qquad  i,j=1,2 \col\\
a_0 &= - \sum_{i=1}^2 a_i \col \qquad  a_i = a_{1i} \col  \\
b&= \sum_{i,j=1}^2 (k_1 \cdot q_i) (A^{-1})_{ij} (q_j \cdot k_2) \col \qquad \tilde A = \sum_{i,j=1}^2 (A^{-1})_{ij}\pnt
\end{aligned}
\end{equation}

The integrals involving  $l_\peps^2$ give rational terms  \cite{Bern:1995ix}. The rational term for the rank-two tensor triangle integral is given in \eqref{rationaltermtriangle}.

\section{Checks of the three-point one-loop Konishi form factor}
 \label{app:check-3pt}
\subsubsection*{Rational term in $F_{{\cal K}}^{(1)}(1_{\phi_{12}}, 2_{\phi_{34}}, 3_{g^+})$}

An interesting feature of the Konishi form factor is the occurrence of rational terms at one loop. 

For  the form factor $F_{{\cal K}}^{(1)}(1_{\phi_{12}}, 2_{\phi_{34}}, 3_{g^+})$, denoted as $F_{\mK, (\phi,\phi,g)}^{(1)}$, this corresponds to the triangle integral containing the $l_\peps^2$-term; see \eqref{oneloop3Konscalar}.
Using \eqref{rationaltermtriangle},
the rational term, denoted by ${\cal R}[\bullet]$, can be computed as
\begin{equation}
{\cal R}[F_{\mK, (\phi,\phi,g)}^{(1)}] = F_{\mK, (\phi,\phi,g)}^{(0)} {N_\phi\, s_{13} s_{23} \over s_{12} (s_{13} + s_{23}) } \pnt \label{rational-3pt}
\end{equation}
Since the computation in section \ref{sec:ffactor} is based on the four-dimensional unitarity method,
one might be concerned whether additional rational terms are missed in this approach. In the following, we show that the above result is actually complete by comparing with a Feynman diagram computation following the strategy of \cite{Xiao:2006vr}.

First, from the  power counting criterion given in \cite{Bern:1994cg}, a one-loop integral can generate rational-term contributions only if it has a high enough power of the loop momentum $l$ in the numerator of the loop integrand, which is given by
\begin{equation}
\begin{aligned}
&m>n-2 \quad  & &\text{ for } \quad I_n^{D}[(l)^m] \quad \text{ with } n>2 \col \\
&m>1   \quad  & &\text{ for } \quad I_2^{D}[(l)^m] \pnt
\end{aligned}
\end{equation}
Second, we can safely neglect Feynman diagrams that appear in the computation of the BPS form factor, since the sum of them is known to be free of rational terms \cite{Brandhuber:2010ad}. From these findings, it turns out that only two diagrams need to be considered, which are shown in figure \ref{fig:rt}.
\begin{figure}[tbp]
\begin{center}
\begin{equation*}
\begin{aligned}
\oneloopfftriang[
\fmf{plain}{vo,vl1}
\fmf{plain_arrow,label=$\scriptstyle l$,l.dist=4,l.side=left}{vl1,vl3}
\fmf{plain}{vl3,vo}
\fmf{photon}{vl3,v3}
\fmf{photon}{vl1,vv}
\fmf{plain}{vv,v1}
\fmf{plain}{vv,v2}
\fmfv{label=$\scriptstyle p_1$,l.dist=2,l.a=0}{v1}
\fmfv{label=$\scriptstyle p_2$,l.dist=2,l.a=0}{v2}
\fmfv{label=$\scriptstyle p_3$,l.dist=2,l.a=0}{v3}
]
&=
\oneloopfftriang[
\fmf{plain}{vo,vl1}
\fmf{plain_arrow,label=$\scriptstyle l$,l.dist=4,l.side=right}{vl3,vl1}
\fmf{plain}{vl3,vo}
\fmf{photon}{vl1,v1}
\fmf{photon}{vl3,vv}
\fmf{plain}{vv,v2}
\fmf{plain}{vv,v3}
\fmfv{label=$\scriptstyle p_3$,l.dist=2,l.a=0}{v1}
\fmfv{label=$\scriptstyle p_1$,l.dist=2,l.a=0}{v2}
\fmfv{label=$\scriptstyle p_2$,l.dist=2,l.a=0}{v3}
]
= {- i N_\phi g_\YM^3 \over 2 \sqrt{2} s_{12}} \int\frac{\de^Dl}{(2\pi)^D}\frac{(2l+p_1+p_2)\cdot(p_1-p_2)\,(2l-p_3)\cdot\epsilon_3^+}{l^2(l+p_1+p_2)^2(l-p_3)^2}
\end{aligned}
\end{equation*}

\caption{Feynman diagrams of the one-loop three-point Konishi form factor that contribute to the rational term.}
\label{fig:rt}

\end{center}
\end{figure}

Using standard color-ordered Feynman rules (see e.g. \cite{Mangano:1990by}), these two graphs give
\begin{equation}
F_{\mK, (\phi,\phi,g)}^{(0)} \left(-\frac{4 N_\phi}{\sqrt{2} s_{12} }\right) {\langle 2\,3\rangle \langle 3\,1\rangle \over \langle 1\,2\rangle  } {(\e^{\gamma_\E}\mu^2)^{\peps}} \int {\de^D l \over i \pi^{\frac{D}{2}}} {(l \cdot \epsilon^+_3) [l \cdot (p_1-p_2)] \over l^2(l - p_3)^2 (l+p_{12})^2} \col \label{rt-graph}
\end{equation}
where the polarization vector is given by $\epsilon_3^+ = {\sqrt{2} \xi \tilde\lambda_3 \over \langle \xi \lambda_3 \rangle}$ and $\xi$ is an arbitrary reference spinor. Then, applying the identity%
\footnote{This identity can be obtained using PV reduction and \eqref{rationaltermtriangle}. Formulae for more general cases can be found in \cite{Xiao:2006vr}.}
\begin{equation}
{\cal R} \bigg[ \int {\de^D l \over i \pi^{\frac{D}{2}}} {(l \cdot k_1) (l \cdot k_2) \over l^2(l+q_1)^2 (l+q_2)^2} \bigg] = { k_1 \cdot k_2 \over2} - { (k_1 \cdot k_2) (k_2 \cdot q_1) \over 2 q_1\cdot q_2} \,, \qquad k_1 \cdot q_1 = q_1^2 = 0 \col
\end{equation}
we can extract the rational term of \eqref{rt-graph} immediately, which, after some simple spinor algebra, turns out to be identical to that given in \eqref{rational-3pt}. Thus, we have shown that the unitarity method gives the complete rational terms. 

\subsubsection*{Spurious poles}

The coefficients of the integrals in \eqref{oneloop3Konscalar} and \eqref{oneloop3Konfermion} contain unphysical poles, such as the pole ${1\over s_{13}+s_{23}}={1\over s_{123}-s_{12}}$.  Such poles cannot originate from propagators in the underlying Feynman diagrams. Physical consistency requires that they must cancel when multiplying the coefficients with the respective integrals and summing all contributions.%
\footnote{This is a common feature for one-loop QCD amplitudes, see e.g.\ \cite{Bern:2007dw}.} Here, we check that this is indeed the case. We focus on the pole ${1\over s_{13}+s_{23}}$; the other poles can be treated in a similar way.

Let us first consider the case of $F_{\cal K}^{(1)}(1_\psi, 2_\psi, 3_\phi)$. Only the coefficients of the bubble integrals contain spurious poles. Summing over all bubble integrals, the $1\over \peps$ term is free of the pole, and at finite order we find
\begin{equation}
- 3 {s_{12} s_{13} \over s_{12} + s_{13} } \log \Big({s_{123} \over s_{23}} \Big) \pnt
\end{equation}
This is indeed finite for $s_{12} + s_{13}\rightarrow0$, as can be seen from the expansion
\begin{equation}
{\log(1+x) \over x} = 1 - {x\over2} + {x^2 \over3} +  {\cal O}(x^3)
\end{equation}
with $x=\frac{s_{12}+s_{13}}{s_{23}}$.

The $F_{\cal K}^{(1)}(1_\phi, 2_\phi, 3_{g})$ case is a little more complicated. In this case, both the bubble and triangle integral contain the pole ${1\over s_{13}+s_{23}}$ in their coefficients. Expanding to finite order and extracting the terms that contain this pole, we find 
\begin{equation}
- 6 {s_{13} s_{23} \over (s_{13} + s_{23})^2 } \log \Big({s_{123} \over s_{12}} \Big) + 6 {s_{13} s_{23} \over s_{12} (s_{13} + s_{23}) }\,,
\end{equation}
where the first term stems from the sum of bubble integrals and the second term is the rational term. Each term itself is divergent when taking the limit $s_{13}+s_{23}\rightarrow0$; however, the sum of the two terms is finite in this limit. 


\section{Phase-space parametrization}
\label{App:phase-space}

In this appendix, we provide formulae for the parametrization of the phase-space integrals. Furthermore, we give details on the non-trivial three-particle phase-space integration encountered in section \ref{sec:CS}.

The $n$-particle phase-space integral is defined as
\begin{equation} 
\int \de {\rm PS}_n \, (\bullet) =  \int \bigg( \prod_{j=1}^n {\de^D p_j \over (2\pi)^D} \, 2\pi \, \delta_+(p_j^2) \bigg)  (2\pi)^D \delta^{(D)} \Big(q -\sum_{j=1}^n p_j \Big) \, (\bullet) \col
\end{equation}
where $(\bullet)$ denotes the integrand, i.e.\ the squared matrix element.

When $n=2$, the squared matrix element depends only on $q^2$, and we can evaluate the two-particle phase-space integral independently:
\begin{equation} 
\label{PS-2pt}
\int \de {\rm PS}_2 \, (\bullet) =  f_{\rm PS2} \, (\bullet) \col \qquad f_{\rm PS2} = { (q^2)^{-\peps} \over 4 (16\pi)^{{1\over2}-\peps} \, \Gamma({3\over2}-\peps)} \pnt
\end{equation}

The three-particle phase space can be parametrized as
\begin{align} 
\label{PS-3pt}
\int \de {\rm PS}_3 \, (\bullet) &= f_{\rm PS3}  \int_0^1\de x \, x^{1-2\peps} (1-x)^{-\peps} \int_0^1\de y \,[y(1-y)]^{-\peps} \ (\bullet) \col \intertext{with} f_{\rm PS3} &= {(q^2)^{1-2\peps} \over 2(4\pi)^{3-2\peps} \Gamma(2-2\peps)} \pnt
\end{align}
The ratios of Mandelstam variables occurring in the squared matrix element are pa\-ram\-e\-trized as
\begin{equation} 
\Big\{ {s_{ij} \over q^2} \col {s_{jk} \over q^2} \col {s_{ki} \over q^2} \Big\} =  \big\{ x(1-y) \col  1- x \col  xy \big\} \col
\end{equation}
in which $(i,j,k)$ can be any permutation of $(1,2,3)$, since the phase-space measure is totally symmetric in $p_1, p_2, p_3$.

\subsection*{Some details about the three-particle phase-space integral}

The phase-space integration becomes non-trivial for the squared matrix element involving the three-point one-loop form factor. It contains the finite part of the box integral \eqref{eq:finite-Box}, which involves the hypergeometric functions $_2 F_1$.

The corresponding phase-space integrals, which are necessary to evaluate 
\eqref{threeparticlestwoloop}, are
\begin{equation}
 \begin{aligned}
& \int \de {\rm PS}_3 \, {1\over s_{13}} \Big(\frac{\mu^2}{-s_{23}}\Big)^{\peps} {1 \over \peps^2} ~_2 F_1\Big(1, -\peps, 1-\peps, - {s_{13} \over s_{12}} \Big) \\
& = {f_{\rm PS3} \over q^2}\Big(\frac{\mu^2}{-q^2}\Big)^{\peps}  \, {1\over\peps^2} \int_0^1\de x \, x^{-2\peps} (1-x)^{-2\peps} \int_0^1\de y \,y^{-1-\peps} (1-y)^{-\peps} ~_2 F_1\Big(1, -\peps, 1-\peps, - {y \over 1-y} \Big) \\
& = {f_{\rm PS3} \over q^2}\Big(\frac{\mu^2}{-q^2}\Big)^{\peps}\, {\Gamma(1-2\peps) \Gamma(-\peps)^2 \over 4 \peps^2 \Gamma(2-4\peps)} \bigg[ (1+6\peps)~_2 F_1\Big(1, 1, 1-2\peps, 1 \Big) \\
&\hphantom{{} = {f_{\rm PS3} \over q^2}\Big(\frac{\mu^2}{-q^2}\Big)^{\peps}\, {\Gamma(1-2\peps) \Gamma(-\peps)^2 \over 4 \peps^2 \Gamma(2-4\peps)} \bigg[ }- (8 \peps) ~_3F_2\Big(1,1,-\peps; 1-\peps, -2\peps; 1 \Big) \bigg]
\end{aligned}
\end{equation}
and
\begin{equation}
 \begin{aligned}
& \int \de {\rm PS}_3 \, {1\over s_{13}} \Big(\frac{\mu^2}{-q^2}\Big)^{\peps}{1 \over \peps^2}~_2 F_1\Big(1, -\peps, 1-\peps, - {q^2\over s_{12}s_{23}} \Big) \\
& = {f_{\rm PS3} \over q^2}\Big(\frac{\mu^2}{-q^2}\Big)^{\peps} \, {1\over\peps^2} \int_0^1 \de x \, x^{-2\peps} (1-x)^{-\peps} \int_0^1\de y \,y^{-1-\peps} (1-y)^{-\peps} \\
&\phantom{{} = {f_{\rm PS3} \over q^2}\Big(\frac{\mu^2}{-q^2}\Big)^{\peps} \, {1\over\peps^2} \int_0^1 \de x \, x^{-2\peps} (1-x)^{-\peps}}
~_2 F_1\Big(1, -\peps, 1-\peps, - {1\over 1-x} {y \over 1-y} \Big) \\
& = {f_{\rm PS3} \over q^2}\Big(\frac{\mu^2}{-q^2}\Big)^{\peps} \,  \bigg[ -{1\over\peps^3} - {3\over\peps^2} - {9 - {5\pi^2 \over6} \over \peps}  + \Big(-27 + {17\pi^2 \over6} + 21 \zeta_3 \Big) + {\cal O}(\peps) \bigg]
\pnt
\end{aligned}
\end{equation}

\section{Anomalous dimensions via two-point form factors}
 \label{app:anomalous-from-2pt}
In the main part of this paper, we have determined the anomalous dimension of the Konishi operator from its cross section, i.e.\ from the imaginary part of its two-point function.
It is also possible to determine the anomalous dimension of the Konishi operator from its two-point form factor alone. As seen throughout this paper, form factors of non-protected operators 
contain both UV and IR divergences. To extract the UV divergences, one needs to subtract the IR divergences. The computation of the cross section, as done in section \ref{sec:CS}, is one of the safest ways to do so. 
On the other hand, the IR divergences, in particular for Sudakov form factors, have an universal structure \cite{Mueller:1979ih, Collins:1980ih, Sen:1981sd}. This allows us to subtract the IR divergences directly from the form factors.
The remaining divergences are purely UV divergences, from which we can read off the anomalous dimension of the operator.\footnote{This route was also taken in the unpublished notes of Boucher-Veronneau, Dixon, and Pennington \cite{BDP-notes}.}

In terms of the effective planar coupling constant \eqref{gdef}, the logarithm of any (renormalized and diagonally renormalizing) Sudakov form factor in ${\cal N}=4$ SYM theory has the following structure; see e.g.\ \cite{Bern:2005iz}:\footnote{This form was checked for the minimal form factor of the BPS operator $\tr(\phi_{12}^2)$ up to the third loop order \cite{Gehrmann:2011xn}, for the minimal form factor of the BPS operator $\tr(\phi_{12}^k)$ up to the second loop order \cite{Brandhuber:2014ica}, for the $n$-point MHV form factor of the BPS operator $\tr(\phi_{12}^2)$ up to the first loop order \cite{Brandhuber:2010ad} and for the $3$-point MHV form factor of the BPS operator $\tr(\phi_{12}^2)$ up to the second loop order \cite{Brandhuber:2012vm}.}
\begin{equation}
\begin{aligned}
\log f_{{\cal O},\ren} &= \sum_{\ell=1}^{\infty} g^{2\ell} \, \left(\log f_{{\cal O},\ren} \right)^{(\ell)} 
= \sum_{\ell=1}^{\infty} g^{2\ell} \Big(\frac{\mu^2}{-q^2}\Big)^{\ell\peps} \bigg( - {\gamma_{\rm cusp}^{(\ell)} \over (2\ell\peps)^2} - {\mathcal{G}_0^{(\ell)} \over 2\ell\peps} + {\rm Fin}^{(\ell)} \bigg) + {\cal O}(\peps) ,
\end{aligned} 
\end{equation}
where the pole terms originate from the IR divergences and are determined by the universal cusp and collinear anomalous dimensions
\begin{equation}
\begin{aligned}\label{gammacuspcolldef}
\gamma_{\rm cusp}(g) & = \sum_{\ell=1}^\infty \gamma_{\rm cusp}^{(\ell)} g^{2\ell} =  8g^2 - 16 \zeta_2 g^4 + 176 \zeta_4 g^6 + {\cal O}(g^8) \col \\
\mathcal{G}_0(g) & = \sum_{\ell=1}^\infty \mathcal{G}_0^{(\ell)} g^{2\ell} = - 4\zeta_3 g^4 + 16\Big( 2 \zeta_5 + {5\over 3} \zeta_2 \zeta_3 \Big) g^6 + {\cal O}(g^8) \pnt 
\end{aligned}
\end{equation}
The finite terms of the logarithm of the form factor depend on the specific properties of the form factor such as the choice of the operator. In particular, they contain a remainder function, which was studied in \cite{Brandhuber:2012vm,Brandhuber:2014ica}.

For non-protected operators, renormalization is required. The renormalized form factor is given by
\begin{equation}
f_{{\cal O},\ren}^{(L)} = \sum_{\ell=1}^L {\cal Z}^{(\ell)} f^{(L-\ell)}_{{\cal O},\bare} \col
\end{equation}
where the renormalization constant is related to the anomalous dimension as shown in \eqref{eq: relation of calZ and gamma}.

The universal structure of IR divergences, together with the bare Konishi form factor, allow us to determine the renormalization constant and therefore the anomalous dimension.  In the following, we employ the two-loop Konishi form factor to reproduce the Konishi anomalous dimension \eqref{gammaK} up to two-loop order. Reversing the logic, we then give a prediction for the bare three-loop two-point Konishi form factor up to and including ${\cal O}(\peps^{-1})$ order by using the known three-loop anomalous dimension.

\subsection*{One-loop form factor}
The one-loop bare form factor is given in \eqref{eq:final-2pt-1loop}. 
From the universal IR structure, we know that
\begin{equation}
\begin{aligned}
\left(\log f_{\cal K,\ren} \right)^{(1)}  
&= f_{\cal K,\ren}^{(1)}  = f_{{\cal K},\bare}^{(1)} + {\cal Z}_{\cal K}^{(1)} =\Big(\frac{\mu^2}{-q^2}\Big)^{\peps} \Big( -\frac{\gamma_{\rm cusp}^{(1)}}{4\peps^2}- {\mathcal{G}_0^{(1)} \over 2\peps} \Big)+  {\cal O}(\peps^0)
\col
\end{aligned}
\end{equation}
where the one-loop cusp and collinear anomalous dimensions are given in \eqref{gammacuspcolldef}.
The simple pole in $f_{{\cal K},\bare}^{(1)}$ has to be canceled by the one-loop term in the operator renormalization constant, which yields\begin{equation}
{\cal Z}_{\cal K}^{(1)} = {6\over\peps}  \col 
\end{equation}
in agreement with \eqref{eq:calZ-1loop} and the known one-loop anomalous dimension. 
Thus, the one-loop renormalized form factor is
\begin{equation}
f_{{\cal K},\ren}^{(1)} = \Big(\frac{\mu^2}{-q^2}\Big)^{\peps} \, { 2 (1+\peps+\peps^2) e^{\peps\gamma_E} \Gamma(-\peps)^2 \Gamma(1+\peps) \over (-1+2\peps) \Gamma(1-2\peps)} +  {6\over\peps}  \pnt 
\end{equation}

\subsection*{Two-loop form factor}

The two-loop bare Konishi form factor is given in \eqref{eq:final-2pt-2loop}. 
From the universal IR structure, we know that
\begin{equation}
\begin{aligned}
\left(\log f_{\cal K,\ren} \right)^{(2)} & = f_{\cal K,\ren}^{(2)} - {1\over2} \left( f_{\cal K,\ren}^{(1)} \right)^2
= \left( f_{{\cal K},\bare}^{(2)} +  {\cal Z}_{\cal K}^{(1)} f_{{\cal K},\bare}^{(1)} +  {\cal Z}_{\cal K}^{(2)} \right) - {1\over2} \left( f_{\cal K,\ren}^{(1)} \right)^2 \\
&=
 \Big(\frac{\mu^2}{-q^2}\Big)^{2\peps} \Big( -\frac{\gamma_{\rm cusp}^{(2)}}{16\peps^2}- {\mathcal{G}_0^{(2)} \over 4\peps} \Big)+  {\cal O}(\peps^0)
\col
\end{aligned}
\end{equation}
where the two-loop cusp and collinear anomalous dimensions are given in \eqref{gammacuspcolldef}.
This determines the two-loop term of the renormalization constant as
\begin{equation}
{\cal Z}_{\cal K}^{(2)} =  {18 \over \peps^2} - {12 \over \peps}  \col
\end{equation}
which perfectly agrees with \eqref{gammaK12} and yields the known two-loop anomalous dimension.
Hence, the two-loop renormalized form factor is 
\begin{equation}
\begin{aligned}
f_{{\cal K},\ren}^{(2)} 
&= \Big(\frac{\mu^2}{-q^2}\Big)^{2\peps} 
\bigg[ {2\over\peps^4}  + {28 - {\pi^2 \over6}\over\peps^2} + {56 - \pi^2 - {25 \zeta_3 \over3}  \over\peps} + \Big( {316 - {26\pi^2 \over 3} - {28\zeta_3} - {7\pi^4 \over60}} \Big) 
\\ 
&\hphantom{{}={}\Big(\frac{\mu^2}{-q^2}\Big)^{2\peps} \bigg[}
+ \Big( {1172 - {131\pi^2 \over 3} - {572\zeta_3 \over3} - {53\pi^4 \over120}} + {23 \pi^2\zeta_3 \over18} +{71\zeta_5 \over 5} \Big)\peps \bigg]  + {\cal O}(\peps^2) \pnt 
\end{aligned}
\end{equation}

\subsection*{Prediction for the three-loop bare Konishi form factor}

Now, we reverse the logic.
From the universal IR structure, we know that
\begin{equation}
\begin{aligned}
\left(\log f_{\cal K,\ren} \right)^{(3)} & = f_{\cal K,\ren}^{(3)} -  f_{\cal K,\ren}^{(2)}\, f_{\cal K,\ren}^{(1)} + {1\over3} \left( f_{\cal K,\ren}^{(1)} \right)^3 \\
 &= \left( f_{{\cal K},\bare}^{(3)} + {\cal Z}_{\cal K}^{(1)} f_{{\cal K},\bare}^{(2)} +  {\cal Z}_{\cal K}^{(2)} f_{{\cal K},\bare}^{(1)} +  {\cal Z}_{\cal K}^{(3)}  \right) -  f_{\cal K,\ren}^{(2)}\, f_{\cal K,\ren}^{(1)} + {1\over3} \left( f_{\cal K,\ren}^{(1)} \right)^3 \\
 &=
 \Big(\frac{\mu^2}{-q^2}\Big)^{3\peps} \Big( -\frac{\gamma_{\rm cusp}^{(3)}}{36\peps^2}- {\mathcal{G}_0^{(3)} \over 6\peps} \Big)+  {\cal O}(\peps^0)
\col
\end{aligned}
\end{equation}
where the three-loop cusp and collinear anomalous dimensions are given in \eqref{gammacuspcolldef}.
Using the known one- and two-loop form factors, and together with the renormalization constant up to three loops obtained from \eqref{eq: relation of calZ and gamma},
\eqref{gammaK}
and given by
\begin{equation}
\mathcal{Z}_{\mK}^{(3)}
=
\frac{(\gamma_{\mK}^{(1)})^3}{48\peps^3}+\frac{\gamma_{\mK}^{(1)}\gamma_{\mK}^{(2)}}{8\peps^2}+\frac{\gamma_{\mK}^{(3)}}{6\peps}
=\frac{36}{\peps^3}-\frac{72}{\peps^2}+\frac{56}{\peps}
\col
\end{equation}
we can predict the three-loop bare Konishi form factor as:
\begin{equation}
\begin{aligned}
f_{{\cal K},\bare}^{(3)} 
&= \Big(\frac{\mu^2}{-q^2}\Big)^{3\peps} \bigg[ -{4\over 3\peps^6}- {12\over\peps^5} - {64\over \peps^4} - {284 - 2\pi^2 - {22\zeta_3\over3} \over\peps^3}   - {1180 - {65\pi^2 \over 3} - {78 \zeta_3} - {247 \pi^4 \over 3240}  \over\peps^2} \\
&\hphantom{{}={}\Big(\frac{\mu^2}{-q^2}\Big)^{3\peps}\bigg[}  
- \, {4744 - {141\pi^2} - {554\zeta_3} - {51\pi^4 \over40} + {85 \pi^2 \zeta_3 \over 54} + {878 \zeta_5 \over 15} \over \peps} \bigg]   + {\cal O}(\peps^0)\pnt 
\end{aligned}
\end{equation}
This should be compared with a direct computation.

\section{Renormalization-scheme transformations}
 \label{app:scheme}

In this appendix, we review transformations between different mass-independent
renormalization schemes and derive the behavior of the cross section \eqref{logsigmarengeneral} under such transformations.

A renormalization scheme specifies a regularization procedure for 
the UV divergences encountered in 
perturbation theory beyond tree-level 
and a prescription for the subtraction of these
divergences into renormalized fields, coupling constants and composite operators.
The subtraction prescription specifies how the UV divergences are removed from the perturbation series. In particular, it has to be indicated which finite parts 
are absorbed together with the UV divergences into the counter terms or --- equivalently --- the renormalization constants determining the relations between the bare and renormalized quantities.

A modified renormalization scheme,
which contains a different prescription for subtracting the UV divergences 
from the perturbation series in $g$,
 can be described by applying the subtraction of the original scheme but to
the perturbation series in a modified coupling constant
$g_\rho=g\e^{\varrho\peps}$. Thereby, 
the parameter $\varrho$ specifies the finite terms that are subtracted together with 
the UV divergences. Since the combination $g\mu^\peps$ of the coupling constant $g$ and 't Hooft mass $\mu$ is the 
expansion parameter of the perturbation series, 
the change between schemes, i.e.\ between $g$ and $g_\varrho$,  
can easily be implemented by changing $\mu$. 
If we demand $g_\varrho\mu_\varrho^\peps=g\mu^\peps$, the transformation of 
the perturbation series to the scheme $\varrho$, but written in terms of the 
original coupling constant $g$, is given by replacing $\mu\to\mu_\varrho=\mu\e^{-\varrho}$.

A widely used renormalization scheme is the dimensional reduction (DR) 
scheme, which combines regularization by dimensional reduction with
a minimal subtraction of the divergences into counter terms or --- equivalently ---
renormalization constants. Minimal subtraction means that no finite terms are subtracted. In the DR scheme, minimal subtraction 
is applied to the perturbation series in the 
coupling constant $\frac{\sqrt{\lambda}}{4\pi}$, $\lambda=g_\YM^2N_{\text{c}}$.

In this paper, we work in the modified dimensional reduction 
($\overline{\text{DR}}$) scheme. 
It employs dimensional reduction as a
regularization procedure, but the subtraction is non-minimal
in terms of the coupling constant $\frac{\sqrt{\lambda}}{4\pi}$, $\lambda=g_\YM^2N_{\text{c}}$. It is, however, minimal in terms of the coupling constant $g$ defined in \eqref{gdef}. Hence, the subtraction procedures of the 
DR and $\overline{\text{DR}}$ scheme are related in the same way as those
of the famous MS and  $\overline{\text{MS}}$ schemes defined in \cite{'tHooft:1973mm}
and \cite{Bardeen:1978yd}, respectively, 
that employ dimensional 
regularization as a regularization procedure.
The expressions in the former schemes are obtained from the ones in the latter 
schemes by replacing $\mu\to\mu\e^{-\varrho}$, where 
$\varrho=\frac{1}{2}(\log4\pi-\gamma_\E)$.

Consider the renormalized cross section $\sigma_\ren$. 
Inserting \eqref{opren} into \eqref{ImPi-sigma}, it is given as the product 
of the squared operator renormalization constant $\mathcal{Z}_{\mathcal{O}}$ 
introduced in \eqref{opren} and
the bare cross section $\sigma_\bare$:
\begin{equation}\label{sigmaRinsigmaB}
\sigma_\ren=\mathcal{Z}_{\mathcal{O}}(g,\peps)^2\sigma_\bare\pnt
\end{equation}
The logarithm of the ratio of the bare and the tree-level cross section $\sigma^{(0)}$ then has the following expansion up to two-loop order:
\begin{equation}\label{logsigmaBexpansion}
\log\frac{\sigma_\bare}{\sigma^{(0)}}
=g^2\Big(\frac{\mu^2}{q^2}\Big)^\peps\bigg(-\frac{\gamma_{\mathcal{O}}^{(1)}}{\peps}+s_0^{(1)}\bigg)
+g^4\Big(\frac{\mu^2}{q^2}\Big)^{2\peps}\bigg(
-\frac{\gamma_{\mathcal{O}}^{(2)}}{2\peps}
+s_0^{(2)}-\frac{(s_0^{(1)})^2}{2}\bigg)+\mathcal{O}(g^6,\peps)\pnt
\end{equation}
The finite terms $s_0^{(\ell)}$ become the coefficients of the 
perturbative expansion of the ratio of the renormalized and the tree-level 
cross section 
\begin{equation}\label{sigmaRexpansion}
\frac{\sigma_\ren}{\sigma^{(0)}}
=\Big(\frac{q^2}{4\mu^2}\Big)^{\gamma_{\mathcal{O}}}\big[1+g^2s_0^{(1)}+g^4s_0^{(2)}
+\mathcal{O}(g^6,\peps)\big]
\end{equation}
in the $\overline{\text{DR}}$ scheme where the coupling constant is 
\eqref{gdef}.

The condition $g_\varrho\mu_\varrho^\peps=g\mu^\peps$
implies that the expression \eqref{logsigmaBexpansion} is 
the same in all schemes. However, only the subtraction prescription of the 
$\overline{\text{DR}}$ scheme leads to the expression \eqref{sigmaRexpansion}
for the renormalized cross section.

The renormalization constant $\mathcal{Z}_{\mathcal{O}}(g,\peps)$ of the 
$\overline{\text{DR}}$ scheme 
obtained by performing minimal subtraction at the coupling constant $g$ can 
be expressed as the renormalization constant
 $\mathcal{Z}_{\mathcal{O},\varrho}=\mathcal{Z}_{\mathcal{O}}(g_\varrho,\peps)$ in the scheme $\varrho$ obtained by performing minimal subtraction at the coupling constant 
$g_\varrho$ times a factor without poles in $\peps$. 
Hence, the difference of the logarithms of these constants is finite and 
given by
\begin{equation}
\label{logZOrelation-general}
\log\mathcal{Z}_{\mathcal{O}}(g,\peps)=\log\mathcal{Z}_{\mathcal{O},\varrho}+g_\varrho^2\Delta\mathcal{Z}_{\mathcal{O}}^{(1)}+g_\varrho^4\Delta\mathcal{Z}_{\mathcal{O}}^{(2)}+\mathcal{O}(g_\varrho^6)\col
\end{equation}
where
\begin{equation}
\begin{aligned}
\label{logZOrelation}
\Delta\mathcal{Z}_{\mathcal{O}}^{(1)}
&=\Big(\Big(\frac{g}{g_\varrho}\Big)^2-1\Big)\mathcal{Z}_{\mathcal{O}}^{(1)}
=-\gamma_{\mathcal{O}}^{(1)}\varrho(1-\varrho\peps)+\mathcal{O}(\peps^2)
\col\\
\Delta\mathcal{Z}_{\mathcal{O}}^{(2)}
&=\Big(\Big(\frac{g}{g_\varrho}\Big)^4-1\Big)\mathcal{Z}_{\mathcal{O}}^{(2)}
-\frac{(\Delta\mathcal{Z}_{\mathcal{O}}^{(1)})^2}{2}
-\mathcal{Z}_{\mathcal{O}}^{(1)}\Delta\mathcal{Z}_{\mathcal{O}}^{(1)}
=
-\gamma_{\mathcal{O}}^{(2)}\varrho+\mathcal{O}(\peps)
\pnt
\end{aligned}
\end{equation}
We have used the expansion given in \eqref{eq: relation of calZ and gamma}.  In the expression $\Delta\mathcal{Z}_{\mathcal{O}}^{(1)}$ we have kept the term linear in $\peps$, since it leads to a finite term in $\peps$ in the expression $\Delta\mathcal{Z}_{\mathcal{O}}^{(2)}$, when it is multiplied by $\mathcal{Z}_{\mathcal{O}}^{(1)}$.

Adding \eqref{logsigmaBexpansion} and \eqref{logZOrelation} leads to the 
following relation of the renormalized cross sections in both schemes
\begin{equation}
\log\frac{\sigma_{\ren,\varrho}}{\sigma^{(0)}}=\log\frac{\sigma_\ren}{\sigma^{(0)}}
-2\big(g^2\Delta\mathcal{Z}_{\mathcal{O}}^{(1)}+g^4\Delta\mathcal{Z}_{\mathcal{O}}^{(2)}\big)+\mathcal{O}(g^6,\peps)
=\log\frac{\sigma_\ren}{\sigma^{(0)}}+2\gamma_{\mathcal{O}}\varrho+\mathcal{O}(g^6,\peps)
\col
\end{equation}
where we have inserted $g_\varrho=g\e^{\varrho\peps}$ and neglected terms that 
vanish when $\peps\to0$.
This relation can be interpreted in two ways, as follows.

First, one can 
insert the expansion \eqref{sigmaRexpansion} for $\sigma_\ren$ and the same 
expression for $\sigma_{\ren,\varrho}$ but with $\mu$ replaced by $\mu_\varrho$.
Then, one obtains the relation $\mu_\varrho=\mu\e^{-\varrho}$ mentioned already
at the beginning of this appendix. This shows that a scheme change can be
performed by changing $\mu$.

Second, one can 
insert the expansion \eqref{sigmaRexpansion} for $\sigma_\ren$
and a similar expression for $\sigma_{\ren,\varrho}$ but with the finite
expansion coefficients $s_{0}^{(\ell)}$ replaced by $s_{0,\varrho}^{(\ell)}$.
Then, one obtains the behavior of the finite terms under
a scheme change, given by the relations
\begin{equation}
\label{eq:s-scheme-change}
s_{0,\varrho}^{(\ell)}=s_0^{(\ell)}+2\gamma_{\mathcal{O}}^{(\ell)}\varrho
\pnt
\end{equation}


%% file: feyndiag.tex
\section{Feynman diagrams}
\label{app:feyndiag}

In this appendix, we compute the unrenormalized form factors of section \ref{sec:ffactor} to two-loop order via Feynman diagrams. See e.g.\ \cite{Fokken:2013aea} for the Feynman rules of the ${\cal N}=4$ SYM theory in our conventions.
In particular, we demonstrate how the analysis of section \ref{sec: subtleties in unitarity} works for the concrete diagrams and that we did not miss any rational terms in section \ref{sec:ffactor}.


\subsection*{One-loop self energies}

For the calculation of the unrenormalized two-loop form factors, we need the 
one-loop self-energies of the gauge and scalar fields. They occur as subdiagrams in certain two-loop diagrams.

The one-loop self-energy of the gauge field is determined from diagrams in which the scalar fields, the fermion fields, the gauge field itself or the ghost field propagates in the loop. They evaluate to
\begin{equation}
\begin{aligned}\label{seAdiags}
\swfone{photon}{plain,left=1}{plain,left=1}
&=\frac{g^2}{2}N_\phi\delta^{ab}I_{\text{s}\,\mu\nu}
\col\qquad
&\swfone{photon}{dashes,left=1}{dashes,left=1}
&=g^2 N_\psi\delta^{ab}I_{\text{f}\,\mu\nu}
\col\\
\swfone{photon}{photon,left=1}{photon,left=1}
&=\frac{g^2}{2}\delta^{ab}(DI_{\text{s}\,\mu\nu}+I_{\text{ph}\,\mu\nu}+2I_{\text{gh}\,\mu\nu})
\col\qquad
&\swfone{photon}{dots,left=1}{dots,left=1}
&=-g^2\delta^{ab}I_{\text{gh}\,\mu\nu}
\col\\
\end{aligned}
\end{equation}
where $g$ is the coupling in the $\overline{\text{DR}}$ scheme 
defined in \eqref{gdef}, and besides the number of scalar flavors
$N_\phi=6+2\peps$ we have also introduced the number of fermion flavors
$N_\psi=4$ of $\mathcal{N}=4$ SYM theory.
Moreover, we have split the contribution from the gauge loop into the 
tensor integrals $I_{\text{s}\,\mu\nu}$, $I_{\text{gh}\,\mu\nu}$ 
occurring in case of the scalar- and ghost-loop contribution, respectively,
and into $I_{\text{ph}\,\mu\nu}$, which is associated with the 
remaining physical degrees of freedom of the gauge-field polarizations.
The occurring integrals are expressed in terms of the simple 
bubble integral in \eqref{oneloopint:bubble} as
\begin{equation}
\begin{aligned}
I_{\text{s}\,\mu\nu}
&=(\e^{\gamma_\E}\mu^2)^\peps\int\frac{\de^Dl}{i\pi^{\frac{D}{2}}}\frac{(q-2l)_\mu(q-2l)_\nu}{l^2(q-l)^2}
=
\frac{1}{D-1}(q_\mu q_\nu-q^2g_{\mu\nu})
\swfone{double}{plain,left=1}{plain,left=1}
\col\\
I_{\text{f}\,\mu\nu}
&=(\e^{\gamma_\E}\mu^2)^\peps\int\frac{\de^Dl}{i\pi^{\frac{D}{2}}}\frac{\tr\tilde\sigma_\mu(q-l)\tilde\sigma_\nu l}{l^2(q-l)^2}
=\frac{D-2}{D-1}
(q_\mu q_\nu-q^2g_{\mu\nu})\swfone{double}{plain,left=1}{plain,left=1}
\col\\
I_{\text{ph}\,\mu\nu}
&=(\e^{\gamma_\E}\mu^2)^\peps\int\frac{\de^Dl}{i\pi^{\frac{D}{2}}}\frac{1}{l^2(q-l)^2}
(-6q_\mu q_\nu+4q_\mu(q-l)_\nu+4(q-l)_\mu q_\nu-8(q-l)_\mu(q-l)_\nu)\\
&\hphantom{{}={}(\e^{\gamma_\E}\mu^2)^\peps\int\frac{\de^Dl}{i\pi^{\frac{D}{2}}}\frac{1}{l^2(q-l)^2}(}
+(5q^2-2q\cdot(q-l)+2(q-l)^2)g_{\mu\nu})\\
&=
-\frac{4D-2}{D-1}(q_\mu q_\nu-q^2g_{\mu\nu})\swfone{double}{plain,left=1}{plain,left=1}
\col\\
I_{\text{gh}\,\mu\nu}
&=(\e^{\gamma_\E}\mu^2)^\peps\int\frac{\de^Dl}{i\pi^{\frac{D}{2}}}\frac{(l-q)_\mu l_\nu}{l^2(q-l)^2}
=\frac{1}{4(D-1)}
(-(D-2)q_\mu q_\nu-q^2g_{\mu\nu})\swfone{double}{plain,left=1}{plain,left=1}
\pnt
\end{aligned}
\end{equation}

Inserting the results for the tensor integrals into \eqref{seAdiags} and summing all contributions, we obtain
\begin{equation}
\begin{aligned}\label{seA}
\settoheight{\eqoff}{$\times$}%
\setlength{\eqoff}{0.5\eqoff}%
\addtolength{\eqoff}{-6\unitlength}%
\raisebox{\eqoff}{%
\fmfframe(1,1)(1,1){%
\begin{fmfchar*}(14,10)
\fmfleft{v1}
\fmfright{v2}
\fmf{photon}{v1,v2}
\fmfcmd{pair vert[]; vert1 = vloc(__v1);}
\vacpolp[0.5]{vert1--vloc(__v2)}
\end{fmfchar*}}}
&=\frac{g^2}{2}\delta^{ab}((N_\phi+D)I_{\text{s}\,\mu\nu}+2N_\psi I_{\text{f}\,\mu\nu}+I_{\text{ph}\,\mu\nu})\\
&=\frac{g^2}{2}\delta^{ab}
\frac{N_\phi+D+2(D-2)N_\psi-(4D-2)}{D-1}(q_\mu q_\nu-q^2g_{\mu\nu})
\swfone{double}{plain,left=1}{plain,left=1}\\
&=2g^2\delta^{ab}(q_\mu q_\nu-q^2g_{\mu\nu})\swfone{double}{plain,left=1}{plain,left=1}
\pnt
\end{aligned}
\end{equation}
The first line shows that our decomposition of the 
gauge-loop contribution in \eqref{seAdiags} is advantageous:
$D$ and $N_\phi$ only appear in the combination $D+N_\phi$
which is insensitive to the simultaneous continuation of 
$D$ and $N_\phi$ as prescribed by the $\overline{\text{DR}}$ scheme, cf.\ 
the discussion in section \ref{sec: subtleties in unitarity}.
We note that when inserting the 
appropriate numbers flavors in the second line, the dependence on $D$ originating from the tensor integrals is also canceled.

The remaining one-loop self energies for the scalar and fermion fields
read
\begin{equation}
\begin{aligned}\label{sephipsi}
\settoheight{\eqoff}{$\times$}%
\setlength{\eqoff}{0.5\eqoff}%
\addtolength{\eqoff}{-6\unitlength}%
\raisebox{\eqoff}{%
\fmfframe(1,0)(1,0){%
\begin{fmfchar*}(14,10)
\fmfleft{v1}
\fmfright{v2}
\fmf{plain}{v1,v2}
\fmfcmd{pair vert[]; vert1 = vloc(__v1);}
\vacpolp[0.5]{vert1--vloc(__v2)}
\end{fmfchar*}}}
&=-2g^2\delta^{ab}\delta_I^Jq^2\swfone{double}{plain,left=1}{plain,left=1}\col\\
\settoheight{\eqoff}{$\times$}%
\setlength{\eqoff}{0.5\eqoff}%
\addtolength{\eqoff}{-6\unitlength}%
\raisebox{\eqoff}{%
\fmfframe(1,1)(1,1){%
\begin{fmfchar*}(14,10)
\fmfleft{v1}
\fmfright{v2}
\fmf{dashes}{v1,v2}
\fmfcmd{pair vert[]; vert1 = vloc(__v1);}
\vacpolp[0.5]{vert1--vloc(__v2)}
\end{fmfchar*}}}
&=-4g^2\delta^{ab}\delta_I^Jq_{\dot\alpha}{}^\alpha\swfone{double}{plain,left=1}{plain,left=1}\pnt\\
\end{aligned}
\end{equation}

\subsection*{One-loop form factors}

In the Feynman-diagram approach, the one-loop form factors for the BPS operator \eqref{BPSinso6} and the Konishi operator \eqref{Kinso6} are
obtained from the two diagrams given in table \ref{tab:oneloopFFdiags}.
\begin{table}[htbp]
\begin{center}
\begin{tabular}{c|c|c|c|@{}c@{}}
diagram & \multicolumn{2}{c|}{$g^2(A\unitmatrix+B\tracematrix)$} & $f(l)$ & \\ 
& $A$ & $B$ &  & \\
\hline
\twoloopfftriang[%
\fmf{plain,left=1}{vo,vmc}
\fmf{plain,left=1}{vmc,vo}
\fmf{plain}{vmc,v1}
\fmf{plain}{vmc,v2}
]
& $1$ & $-1$ 
& $(l+p_2)^2$ & \multirow{2}{*}{$\left.\rule[-8\unitlength]{0pt}{18\unitlength}\right\}f(l)\twoloopfftriang[%
\fmf{plain}{vo,vr1}
\fmf{plain}{vr1,v1}
\fmf{plain}{vo,vr2}
\fmf{plain}{vr2,v2}
\fmf{plain}{vr1,vr2}
]$} \\
\cline{1-4}
\twoloopfftriang[%
\fmf{plain}{vo,vr1}
\fmf{plain}{vr1,v1}
\fmf{plain}{vo,vr2}
\fmf{plain}{vr2,v2}
\fmf{photon}{vr1,vr2}
]
& $1$ & $0$ 
& $(l+2q-p_2)\cdot(l-p_2)$ & \\
\hline
\end{tabular}
\caption{Diagrams
for the unrenormalized one-loop form factors.
The prefactors $g^2(A\unitmatrix+B\tracematrix)$ of each diagram 
consist of the identity and trace operator in flavor space 
$\unitmatrix$ and $\tracematrix$, respectively.
For the BPS operator
\eqref{BPSinso6} and for the Konishi operator \eqref{Kinso6} the 
prefactors reduce to $g^2A$
and  $g^2(A+BN_\phi)$, respectively. They multiply the triangle integral which contains the numerator factor $f(l)$.
\label{tab:oneloopFFdiags}}
\end{center}
\end{table}
Completing the numerator of the second integral in table \ref{tab:oneloopFFdiags} to squared momenta occurring in the denominator, it can be transformed to the expression
\begin{equation}
\begin{aligned}\label{oneloopintreduction}
(l+2q-p_2)\cdot(l-p_2)
\twoloopfftriang[%
\fmf{plain}{vo,vr1}
\fmf{plain}{vr1,v1}
\fmf{plain}{vo,vr2}
\fmf{plain}{vr2,v2}
\fmf{plain}{vr1,vr2}
]
&=
-\twoloopfftriang[%
\fmf{plain,left=1}{vo,vmc}
\fmf{plain,left=1}{vmc,vo}
\fmf{plain}{vmc,v1}
\fmf{plain}{vmc,v2}
]
+\twoloopfftriang[%
\fmf{plain}{vo,vr1}
\fmf{plain}{vr1,v1}
\fmf{plain}{vo,v2}
\fmf{plain,left=0.4}{vr1,vo}
]
+\twoloopfftriang[%
\fmf{plain}{vo,v1}
\fmf{plain}{vo,vr2}
\fmf{plain}{vr2,v2}
\fmf{plain,right=0.4}{vr2,vo}
]
+(p_1^2+p_2^2-2q^2)
\twoloopfftriang[%
\fmf{plain}{vo,vr1}
\fmf{plain}{vr1,v1}
\fmf{plain}{vo,vr2}
\fmf{plain}{vr2,v2}
\fmf{plain}{vr1,vr2}
]
\pnt
\end{aligned}
\end{equation}
Only the first three integrals
are UV divergent.  Moreover, they develop IR divergences if 
the corresponding external momentum square
$q^2$, $p_1^2$ or $p_2^2$ vanishes. In this case, the respective integral 
vanishes identically in dimensional reduction 
since its IR pole and its UV pole cancel. 
The fourth integral is UV finite, but it becomes IR divergent if at least one of the three momentum 
squares vanishes.
In case that $p_1^2$, $p_2^2$ are not zero, also the self-energy corrections of 
the scalar fields contribute to the form factor. 
Using the expression for the one-loop scalar self energy given in \eqref{sephipsi},
the respective contribution can be written as
\begin{equation}
\frac{1}{2}\bigg[
\twoloopfftriang[%
\fmf{plain}{vo,v1}
\fmf{plain}{vo,v2}
\fmfipath{p[]}
\fmfiset{p1}{vpath(__vo,__v1)}
{\vacpolp[0.333]{p1}}
]
+
\twoloopfftriang[%
\fmf{plain}{vo,v1}
\fmf{plain}{vo,v2}
\fmfipath{p[]}
\fmfiset{p2}{vpath(__vo,__v2)}
{\vacpolp[0.333]{p2}}
]
\bigg]
=
-g^2\bigg[
\twoloopfftriang[%
\fmf{plain}{vo,vr1}
\fmf{plain}{vr1,v1}
\fmf{plain}{vo,v2}
\fmf{plain,left=0.4}{vr1,vo}
]
+\twoloopfftriang[%
\fmf{plain}{vo,v1}
\fmf{plain}{vo,vr2}
\fmf{plain}{vr2,v2}
\fmf{plain,right=0.4}{vr2,vo}
]
\bigg]
\col
\end{equation}
where the factor $\frac{1}{2}$ originates from the fact that 
the square root of the
renormalization constant determined from the self-energy contribution 
renormalizes the corresponding elementary field.
When added to the sum of the two 
diagrams given in table \ref{tab:oneloopFFdiags}, this
contribution exactly cancels the second and third term in the 
expansion of the second integral given in \eqref{oneloopintreduction}, 
irrespective of the vanishing or non-vanishing of $p_1^2$, $p_2^2$. 
In the case of the BPS operator, where both diagrams of table \ref{tab:oneloopFFdiags} only contribute with the coefficient $A$, the remaining UV divergence contained in the bubble integral cancels 
among the two diagrams given in table \ref{tab:oneloopFFdiags}. Hence, in the BPS case, there is only a contribution from the triangle integral of \eqref{oneloopintreduction}.
In the case of the Konishi operator, the contributions of the bubble integral
do not cancel for the flavor-trace contribution, which comes with the 
coefficient $B$.
The one-loop form factors for the BPS operator and the Konishi operator
hence read
\begin{equation}
\begin{aligned}
f_{\BPS,2}^{(1)}&=(p_1^2+p_2^2-2q^2)
\FDinline[triangle]
\col\\
f_{\mK,(\phi,\phi)}^{(1)}
&=
-N_\phi
\FDinline[bubble]
+(p_1^2+p_2^2-2q^2)
\FDinline[triangle]
\pnt
\end{aligned}
\end{equation}
We have calculated the above form factors for generic off-shell momenta $p_1^2\neq0$ and $p_2^2\neq0$. Hence, they are generalizations of the respective
expressions with $p_1^2=p_2^2$, given 
for the BPS operator in \eqref{F2-BPS-1loop} and for the operator $\mK_6$ in
\eqref{Konishi1loopfull} into which the factor $r_\phi$ has to be
introduced as prescribed in \eqref{eq: ftilde-correction-rule} in order to 
obtain the Konishi form factor.

The contribution ${\tilde f}_{\mK,n}^{(\ell)}$ in \eqref{eq: Kdecomposition}
is free of any contribution from the triangle integral, and it is 
in particular independent of $p_1^2$ and $p_2^2$.
This explicitly confirms that the IR divergence is universal, i.e.\ the same for the BPS and the Konishi operator. Moreover, the UV divergence of the Konishi operator can be extracted from the final $\peps$-expansions of the Konishi 
and the BPS form factor given in \eqref{eq:final-2pt-1loop} in the on-shell case 
$p_1^2=p_2^2=0$ where the the $\frac{1}{\peps}$-poles 
originate from both, the UV and the IR divergences.

\subsection*{Two-loop form factors}

The one-particle-irreducible (1PI) diagrams for the two-loop form factors of the BPS operator 
\eqref{BPSinso6} and the Konishi operator \eqref{Kinso6} are displayed in 
table \ref{tab:twoloopFFdiags}.
\afterpage{%
\begin{center}
\footnotesize
\begin{longtable}[p]{c|c|c|c|@{}c@{}}
diagram & \multicolumn{2}{c|}{$g^4(A\unitmatrix+B\tracematrix)$} &  $f(k,l)$ 
& \\ 
& $A$ & $B$ & & \\
\hline
\endfirsthead
\multicolumn{4}{c}%
{{Continued from previous page}} & \\
diagram & \multicolumn{2}{c|}{$g^4(A\unitmatrix+B\tracematrix)$} & $f(k,l)$  
& \\ 
& $A$ & $B$ & & \\
\hline
\endhead
\multicolumn{5}{l}{}\\[-0.1cm]
  \multicolumn{5}{l}{\normalsize {\bfseries \tablename\ \thetable{}} -- continued on next page} 
\endfoot
\caption{
Diagrams
for the unrenormalized two-loop form factors.
The prefactors $g^4(A\unitmatrix+B\tracematrix)$ of each diagram 
consist of the identity and the trace operator in flavor space, 
$\unitmatrix$ and $\tracematrix$, respectively.
For the BPS operator
\eqref{BPSinso6} and for the Konishi operator \eqref{Kinso6}, the 
prefactors reduce to $g^4A$
and  $g^4(A+BN_\phi)$, respectively. They multiply the corresponding integral $I_{\text{tb}}$, $I_{\text{bt}}$ or $I_{\text{bb}}$,  which are given in \eqref{ItbIbtIbbdef} and
contain the numerator factors $f(k,l)$.
For all diagrams which are not symmetric under a reflection at the horizontal axis, also the corresponding reflected version has to be considered.
\label{tab:twoloopFFdiags}
}
\endlastfoot
\nopagebreak\twoloopfftriang[%
\fmf{plain,left=1}{vo,vlc}
\fmf{plain,right=1}{vo,vlc}
\fmf{plain,left=1}{vlc,vrc}
\fmf{plain,right=1}{vlc,vrc}
\fmf{plain}{vrc,v1}
\fmf{plain}{vrc,v2}
]
& $1$ & $N_\phi-2$ 
& $(k-l)^2(l+p_2)^2$ 
& \multirow{16}{*}{$\left.\rule[-80\unitlength]{0pt}{152\unitlength}\right\}I_{\text{tb}}$} 
\\
\cline{1-4}
\nopagebreak\twoloopfftriang[%
\fmf{plain,left=1}{vo,vlc}
\fmf{plain,right=1}{vo,vlc}
\fmf{photon,left=1}{vlc,vrc}
\fmf{photon,right=1}{vlc,vrc}
\fmf{plain}{vrc,v1}
\fmf{plain}{vrc,v2}
]
& $0$ & $D$ 
& $(k-l)^2(l+p_2)^2$ & \\
\cline{1-4}
\nopagebreak\twoloopfftriang[%
\fmf{plain,left=1}{vo,vmc}
\fmf{plain,left=1}{vmc,vo}
\fmffreeze
\fmfposition
\fmf{plain}{vmc,v1}
\fmf{plain}{vmc,v2}
\fmfipair{vg[]}
\fmfipath{p[]}
\fmfiset{p1}{vpath(__vo,__vmc)}
\fmfiset{p2}{vpath(__vmc,__vo)}
\svertex{vg1}{p1}
\svertex{vg2}{p2}
\fmfi{photon}{vg1--vg2}
]
& $1$ & $-1$ 
& $(l+p_2)^2(k+l)\cdot(k+l+2q)$ & \\
\cline{1-4}
\nopagebreak\twoloopfftriang[%
\fmf{plain,left=1}{vo,vmc}
\fmf{plain,left=1}{vmc,vo}
\fmf{plain}{vmc,v1}
\fmf{plain}{vmc,v2}
\fmffreeze
\fmfposition
\fmfipair{vg[]}
\fmfipath{p[]}
\fmfiset{p1}{vpath(__vmc,__v2)}
\fmfiset{p2}{vpath(__vmc,__vo)}
\svertex{vg1}{p1}
\svertex{vg2}{p2}
\fmfi{photon}{vg1{dir -135}..{dir 135}vg2}
]
& $0$ & $0$ 
& $(l+q)^2(l-2k)\cdot(l+2p_2)$ & \\
\cline{1-4}
\nopagebreak\twoloopfftriang[%
\fmf{plain,left=1}{vo,vmc}
\fmf{plain,left=1}{vmc,vo}
\fmf{plain}{vmc,v1}
\fmf{plain}{vmc,v2}
\fmffreeze
\fmfposition
\fmfipair{vg[]}
\fmfipath{p[]}
\fmfiset{p1}{vpath(__vmc,__v1)}
\fmfiset{p2}{vpath(__vmc,__v2)}
\svertex{vg1}{p1}
\svertex{vg2}{p2}
\fmfi{photon}{vg1{dir -45}..{dir 225}vg2}
]
& $1$ & $-1$ 
& $(k-l)^2(l-p_2)\cdot(l+2q-p_2)$ & \\
\cline{1-4}
\nopagebreak\twoloopfftriang[%
\fmf{plain,left=1}{vo,vmc}
\fmf{plain,left=1}{vmc,vo}
\fmf{photon,left=0}{vmc,vg1}
\fmf{plain,tension=3}{vg1,v1}
\fmf{photon,right=0}{vmc,vg2}
\fmf{plain,tension=3}{vg2,v2}
\fmffreeze
\fmfposition
\fmf{plain}{vg1,vg2}
]
& $0$ & $1$ 
& $(k-l)^2(l+2p_2)\cdot(l+2p_2-q)$ & \\
\cline{1-4}
\nopagebreak\twoloopfftriang[%
\fmf{phantom,left=1}{vo,vmc}
\fmf{phantom,left=1}{vmc,vo}
\fmffreeze
\fmfposition
\fmf{plain}{vmc,v1}
\fmf{plain}{vmc,v2}
\fmfipair{vg[]}
\fmfipath{p[]}
\fmfiset{p1}{vpath(__vo,__vmc)}
\fmfiset{p2}{vpath(__vmc,__vo)}
\svertex{vg1}{p1}
\svertex{vg2}{p2}
\fmfi{plain}{subpath (0,length(p1)/2) of p1}
\fmfi{plain}{subpath (length(p2)/2,length(p2)) of p2}
\fmfi{photon}{subpath (length(p1)/2,length(p1)) of p1}
\fmfi{photon}{subpath (0,length(p2)/2) of p2}
\fmfi{plain}{vg1--vg2}
]
& $0$ & $-1$ 
& $(l+p_2)^2(l-2k)\cdot(l-2k-q)$ & \\
\cline{1-4}
\nopagebreak\twoloopfftriang[%
\fmf{plain,left=1}{vo,vmc}
\fmf{plain,left=1}{vmc,vo}
\fmf{plain}{vmc,v1}
\fmf{plain}{vmc,v2}
\fmffreeze
\fmfposition
\fmfipair{vg[]}
{\vacpolp[0.25]{vpath(__vmc,__vo)}}
]
& $-4$ & $4$ 
& $(l+q)^2(l+p_2)^2$ & \\
\cline{1-4}
\nopagebreak\twoloopfftriang[%
\fmf{plain}{vo,vr1}
\fmf{plain}{vr1,v1}
\fmf{plain}{vo,vr2}
\fmf{plain}{vr2,v2}
\fmf{plain,left=0.25}{vr1,vr2}
\fmf{plain,right=0.25}{vr1,vr2}
]
& $\frac{3}{2}(N_\phi-3)$ & $3$ 
& $l^2(l+q)^2$ & \\
\cline{1-4}
\nopagebreak\twoloopfftriang[%
\fmf{plain}{vo,vr1}
\fmf{plain}{vr1,v1}
\fmf{plain}{vo,vr2}
\fmf{plain}{vr2,v2}
\fmf{photon,left=0.25}{vr1,vr2}
\fmf{photon,right=0.25}{vr1,vr2}
]
& $\frac{3}{2}D$ & $0$ 
& $l^2(l+q)^2$ & \\
\cline{1-4}
\nopagebreak\twoloopfftriang[%
\fmf{plain}{vo,vr1}
\fmf{plain}{vr1,v1}
\fmf{plain}{vo,vl2}
\fmf{plain}{vl2,vr2}
\fmf{plain}{vr2,v2}
\fmf{photon}{vr1,vl2}
\fmf{photon}{vr1,vr2}
]
& $-\frac{3}{2}$ & $0$ 
& $(l+q)^2(k+l)\cdot(l-p_2)$ & \\
\cline{1-4}
\nopagebreak\twoloopfftriang[%
\fmf{plain}{vo,vl1}
\fmf{dashes}{vl1,vr1}
\fmf{plain}{vr1,v1}
\fmf{plain}{vo,vl2}
\fmf{dashes}{vl2,vr2}
\fmf{plain}{vr2,v2}
\fmf{dashes}{vl1,vl2}
\fmf{dashes}{vr1,vr2}
]
& $0$ & $-8$ 
& $\begin{aligned}&2(l-k)\cdot l(l+p_2)\cdot(l+q)-2(l-k)\cdot(l+p_2)l\cdot(l+q)\\[-1.33\baselineskip]
&+2(l-k)\cdot(l+q)l\cdot(l+p_2)\end{aligned}$ & \\
\cline{1-4}
\nopagebreak\twoloopfftriang[%
\fmf{plain}{vo,vl1}
\fmf{plain}{vl1,vr1}
\fmf{plain}{vr1,v1}
\fmf{plain}{vo,vl2}
\fmf{plain}{vl2,vr2}
\fmf{plain}{vr2,v2}
\fmf{photon}{vl1,vl2}
\fmf{photon}{vr1,vr2}
]
& $1$ & $0$ 
& $(k+l)\cdot(k+l+2q)(l-p_2)\cdot(l+2q-p_2)$ & \\
\cline{1-4}
\nopagebreak\twoloopfftriang[%
\fmf{plain}{vo,vl1}
\fmf{photon}{vl1,vr1}
\fmf{plain}{vr1,v1}
\fmf{plain}{vo,vl2}
\fmf{photon}{vl2,vr2}
\fmf{plain}{vr2,v2}
\fmf{plain}{vl1,vl2}
\fmf{plain}{vr1,vr2}
]
& $0$ & $1$ 
& $(l-2k-q)\cdot(l+p_2-q)(l-2k)\cdot(l+2p_2)$ & \\
\hline
\hline
\newpage
\twoloopfftriang[%
\fmf{plain}{vo,vr1}
\fmf{plain}{vr1,v1}
\fmf{plain}{vo,vl2}
\fmf{plain}{vl2,vr2}
\fmf{plain}{vr2,v2}
\fmf{photon}{vr1,vr2}
\fmf{photon,left=0.666}{vl2,vr2}
]
& $-\frac{3}{2}$ & $0$ 
& $(l+p_2)^2(k+l)\cdot(k+2q-p_2)$ \
& \multirow{7}{*}{$\left.\rule[-38\unitlength]{0pt}{76\unitlength}\right\}I_{\text{bt}}$} 
\\
\cline{1-4}
\nopagebreak\twoloopfftriang[%
\fmf{plain}{vo,vr1}
\fmf{plain}{vr1,v1}
\fmf{plain}{vo,vl2}
\fmf{plain}{vl2,vr2}
\fmf{plain}{vr2,v2}
\fmf{photon}{vr1,vl2}
\fmf{photon,left=0.666}{vl2,vr2}
]
& $\frac{3}{2}$ & $0$ 
& $(k-l)^2(l-p_2)\cdot(k+2q-p_2)$ & \\
\cline{1-4}
\nopagebreak\twoloopfftriang[%
\fmf{plain}{vo,vm1}
\fmf{plain}{vm1,v1}
\fmf{plain}{vo,vl2}
\fmf{plain}{vl2,vr2}
\fmf{plain}{vr2,v2}
\fmf{photon}{vm1,vm2}
\fmf{photon,right=0.666}{vl2,vr2}
]
& $\frac{1}{2}$ & $0$ 
& $(l-2k)\cdot(l+2p_2)(k-2l-p_2)\cdot(k+2q-p_2)$ & \\
\cline{1-4}
\nopagebreak\twoloopfftriang[%
\fmf{plain}{vo,vr1}
\fmf{plain}{vr1,v1}
\fmf{plain}{vo,vl2}
\fmf{plain}{vl2,vr2}
\fmf{plain}{vr2,v2}
\fmf{photon}{vr1,vrc}
\fmf{photon}{vrc,vr2}
\fmf{photon}{vrc,vl2}
]
& $-\frac{1}{2}$ & $0$ 
& $\begin{aligned}&(2k-l+p_2)\cdot(l-p_2)(k+l)\cdot(k+2q-p_2)\\[-1.33\baselineskip]
&\phantom{{}{}}-(k+l+2p_2)\cdot(k+l)(k+2q-p_2)\cdot(l-p_2)\\[-1.33\baselineskip]
&\phantom{{}{}}+(2l-k+p_2)\cdot(k+2q-p_2)(k+l)\cdot(l-p_2)\end{aligned}$ & \\
\cline{1-4}
\nopagebreak\twoloopfftriang[%
\fmf{plain}{vo,vr1}
\fmf{plain}{vr1,v1}
\fmf{plain}{vo,vl2}
\fmf{dashes}{vl2,vr2}
\fmf{plain}{vr2,v2}
\fmf{photon}{vr1,vrc}
\fmf{dashes}{vrc,vr2}
\fmf{dashes}{vrc,vl2}
]
& $-4$ & $0$ 
& $\begin{aligned}&2l\cdot(-l-p_2)(k+2q-p_2)\cdot(k-l)\\[-1.33\baselineskip]
&\phantom{{}{}}-2l\cdot(k+2q-p_2)(-l-p_2)\cdot(k-l)\\[-1.33\baselineskip]
&\phantom{{}{}}+2l\cdot(k-l)(k+2q-p_2)\cdot(-l-p_2)\end{aligned}$ & \\
\cline{1-4}
\nopagebreak\twoloopfftriang[%
\fmf{plain}{vo,vr1}
\fmf{plain}{vr1,v1}
\fmf{plain}{vo,vr2}
\fmf{plain}{vr2,v2}
\fmf{photon}{vr1,vr2}
\fmffreeze
\fmfposition
{\vacpolp[0.25]{vpath(__vr1,__vr2)}}
]
& $-2$ & $0$ 
& $\begin{aligned}&l^2(k-p_2)\cdot(k+2q-p_2)\\[-1.33\baselineskip]
&\phantom{{}{}}-\frac{l^2}{(k+p_2)^2}(k+p_2)\cdot(k-p_2)(k+p_2)\cdot(k+2q-p_2)\end{aligned}$ & \\
\cline{1-4}
\nopagebreak\twoloopfftriang[%
\fmf{plain}{vo,vr1}
\fmf{plain}{vr1,v1}
\fmf{plain}{vo,vr2}
\fmf{plain}{vr2,v2}
\fmf{photon}{vr1,vr2}
\fmffreeze
\fmfposition
{\vacpolp[0.25]{vpath(__vo,__vr2)}}
]
& $-2$ & $0$ 
& $(l+p_2)^2(k-p_2)\cdot(k+2q-p_2)$ & \\
\hline
\hline
\twoloopfftriang[%
\fmf{dashes}{vl1,vr2}
\fmf{dashes,rubout}{vr1,vl2}
\fmf{plain}{vo,vl1}
\fmf{dashes}{vl1,vr1}
\fmf{plain}{vr1,v1}
\fmf{plain}{vo,vl2}
\fmf{dashes}{vl2,vr2}
\fmf{plain}{vr2,v2}
]
& $-4$ & $2$ 
& $\begin{aligned}&2l\cdot(l+p_2)(k+q-p_2)\cdot k-2l\cdot(k+q-p_2)k\cdot(l+p_2)\\[-1.33\baselineskip]
&\phantom{{}{}}+2l\cdot k(l+p_2)\cdot(k+q-p_2)\end{aligned}$ 
& \multirow{2}{*}{$\left.\rule[-8\unitlength]{0pt}{18\unitlength}\right\}I_{\text{bb}}$} 
\\
\cline{1-4}
\nopagebreak\twoloopfftriang[%
\fmf{photon}{vl1,vr2}
\fmf{photon,rubout}{vr1,vl2}
\fmf{plain}{vo,vl1}
\fmf{plain}{vl1,vr1}
\fmf{plain}{vr1,v1}
\fmf{plain}{vo,vl2}
\fmf{plain}{vl2,vr2}
\fmf{plain}{vr2,v2}
]
& $\frac{1}{2}$ & $0$ 
& $(2k+l+2q-p_2)\cdot(l-p_2)(k+2l)\cdot(k+2q-2p_2)$ & \\
\hline
\label{tab: twoloopdiagrams}
\end{longtable}
\end{center}
}
The occurring integrals are given by
\begin{equation}
\begin{aligned}\label{ItbIbtIbbdef}
I_{\text{tb}}
&=f(k,l)\twoloopfftriang[%
\fmf{plain}{vo,vl1}
\fmf{plain}{vl1,vr1}
\fmf{plain}{vr1,v1}
\fmf{plain}{vo,vl2}
\fmf{plain}{vl2,vr2}
\fmf{plain}{vr2,v2}
\fmf{plain}{vl1,vl2}
\fmf{plain}{vr1,vr2}
]
=(\e^{\gamma_\E}\mu^2)^{2\peps}\int\frac{\de^Dk}{i\pi^{\frac{D}{2}}}\frac{\de^Dl}{i\pi^{\frac{D}{2}}}
\frac{f(k,l)}{k^2(k+q)^2(k-l)^2l^2(l+q)^2(l+p_1)^2}
\col\\
I_{\text{bt}}
&=
f(k,l)\twoloopfftriang[%
\fmf{plain}{vo,vr1}
\fmf{plain}{vr1,v1}
\fmf{plain}{vo,vl2}
\fmf{plain}{vl2,vr2}
\fmf{plain}{vr2,v2}
\fmf{plain}{vr1,vrc}
\fmf{plain}{vrc,vr2}
\fmf{plain}{vrc,vl2}
]
=(\e^{\gamma_\E}\mu^2)^{2\peps}\int\frac{\de^Dk}{i\pi^{\frac{D}{2}}}\frac{\de^Dl}{i\pi^{\frac{D}{2}}}
\frac{f(k,l)}{k^2(k+q)^2(k+p_1)^2(k-l)^2l^2(l+p_1)^2}
\col\\
I_{\text{bb}}
&=
f(k,l)\twoloopfftriang[%
\fmf{plain}{vl1,vr2}
\fmf{plain,rubout}{vr1,vl2}
\fmf{plain}{vo,vl1}
\fmf{plain}{vl1,vr1}
\fmf{plain}{vr1,v1}
\fmf{plain}{vo,vl2}
\fmf{plain}{vl2,vr2}
\fmf{plain}{vr2,v2}
]
=(\e^{\gamma_\E}\mu^2)^{2\peps}\int\frac{\de^Dk}{i\pi^{\frac{D}{2}}}\frac{\de^Dl}{i\pi^{\frac{D}{2}}}
\frac{f(k,l)}{(k+l)^2(k+l+q)^2k^2(k+p_1)^2l^2(l+p_2)^2}
\pnt
\end{aligned}
\end{equation}

For $p_1^2\neq0$, $p_2^2\neq0$, contributions from diagrams involving the two-loop self-energy of the scalar fields
have to be considered in addition to the 1PI diagrams shown in table \ref{tab: twoloopdiagrams}. Also, the second diagram coming with the integral
$I_{\text{bt}}$ yields a non-vanishing contribution, while it vanishes otherwise.
All graphs are then IR finite, and the UV divergence can easily be extracted
by setting e.g.\ one external momentum to zero and the other one to $q^2$ 
such that no new IR divergences are accidentally created. Moreover, since all 
integrals are superficially logarithmically divergent, one can neglect 
external momenta in the numerators as convenient for maximal simplifications.
We have checked that this produces the known result for the two-loop overall
UV divergence of the Konishi operator when subdivergences are subtracted by 
considering also the corresponding  counter-term diagrams. 
This also produces a vanishing result for the BPS operator.

For $p_1^2=p_2^2=0$, where the 1PI diagrams shown in table \ref{tab: twoloopdiagrams} are the only
contributions to the form factors, it is advantageous to express the scalar
products in the numerators in terms of squares of momenta found in the 
denominator from the propagators. Then, one can use IBP reduction as 
e.g.\ implemented in {\tt LiteRed} \cite{Lee:2012cn} in
order to further reduce the integrals to a set of master integrals. 
The results exactly match the ones given in \eqref{eq:final-2pt-2loop}.
This confirms the absence of further rational terms that might not have 
been detected in the unitarity-based approach.